\documentclass[prb,aps,nofootinbib,amssymb,twocolumn,superscriptaddress,10pt,floatfix]{revtex4-2}

\usepackage{graphicx}
\usepackage{amsmath}
\usepackage{dcolumn}
\usepackage{bm}
\usepackage{longtable}
\usepackage{subfigure}
\usepackage{graphicx}
\usepackage{chemformula}
\usepackage{physics}
\usepackage{mathptmx}
\usepackage{lipsum}
\usepackage[colorlinks=true,linkcolor=blue,citecolor=blue,urlcolor=blue]{hyperref}
\usepackage{cleveref}
\usepackage{url}
\usepackage{tabularray}
\usepackage{xcolor}
\usepackage{soul}
\usepackage{amsmath}
\usepackage{chemformula}
\usepackage{makecell}
\usepackage[version=4]{mhchem}
\usepackage{latexsym,amsmath,amssymb,bm,euscript}
\usepackage{array}
\usepackage{color}
\usepackage{comment}
\usepackage{siunitx}
\usepackage{tikz}
\usepackage{minitoc}

\usepackage{xfp}
\newcommand\SupplementalMaterials{%
  \xdef\presupfigures{\arabic{figure}}
  \xdef\presupsections{\arabic{section}}
  \renewcommand\thefigure{S\fpeval{\arabic{figure}-\presupfigures}}
  \renewcommand\thesection{S\fpeval{\arabic{section}-\presupsections}}
}

\crefname{appendix}{Appendix}{Appendices}
\crefname{equation}{Eq.}{Eqs.}
\crefname{figure}{Fig.}{Figs.}
\crefname{table}{Table}{Tables}
\crefname{section}{Section}{Sections}
\crefname{enumi}{Point}{Points}
\creflabelformat{appendix}{[#2#1#3]}
\crefmultiformat{figure}{Figs.~#2#1\xdef\mycreffirstarg{#1}#3}%
 { and~#2#1#3}{, #2#1#3}{ and~#2#1#3}

\usepackage{soul}

\newcommand{\TwoDDBNbrEntries}{8,872}

\newcommand{\TwoDDBMaterialsTIWithSOC}{905}
\newcommand{\TwoDDBPercentMaterialsTIWithSOC}{10.28}
\newcommand{\TwoDDBMaterialsSMWithSOC}{2,129}

\newcommand{\TwoDDBMaterialsNLCWithSOC}{258}

\newcommand{\TwoDDBMaterialsSEBRWithSOC}{647}

\newcommand{\TwoDDBMaterialsESWithSOC}{19}

\newcommand{\TwoDDBMaterialsESFDWithSOC}{2,110}

\newcommand{\TwoDDBMaterialsOAIWithSOC}{1,003}

\newcommand{\TwoDDBMaterialsOOAIWithSOC}{36}
\newcommand{\TwoDDBMaterialsNonTrivialOrObstructedWithSOC}{4,073}
\newcommand{\TwoDDBPercentMaterialsNonTrivialOrObstructedWithSOC}{46.27}

\newcommand{\TwoDDBMaterialsTIFullGapWithSOC}{139}

\newcommand{\TwoDDBMaterialsSMWithSOCAndEvenNbrElectrons}{602}

\newcommand{\webBCS}{\href{https://cryst.ehu.es/}{Bilbao Crystallographic Server}}
\newcommand{\webBCSAbbr}{\href{https://cryst.ehu.es/}{BCS}}
\newcommand{\webtwoDTQC}{\href{https://www.topologicalquantumchemistry.org/topo2d/index.html}{Topological 2D Materials Database}}
\newcommand{\webtwoDTQCAbbr}{\href{https://www.topologicalquantumchemistry.org/topo2d/index.html}{2D-TQCDB}}

\newcommand{\webTQCAbbr}{\href{https://www.topologicalquantumchemistry.org/}{TQC website}}

\newcommand{\webTQCphonon}{\href{https://www.topologicalquantumchemistry.org/topophonons/index.html}{Topological Phonon Database}}

\newcommand{\CtwoDB}{\href{https://c2db.fysik.dtu.dk/}{C2DB}}
\newcommand{\MCtwoD}{\href{https://mc2d.materialscloud.org/}{MC2D}}

\newcommand{\icsd}{\href{https://icsd.products.fiz-karlsruhe.de}{ICSD}}
\newcommand{\cod}{\href{http://www.crystallography.net/cod}{COD}}

\newcommand{\LBANDREP}{\href{https://cryst.ehu.es/cryst/lbandrep.html}{LBANDREP}}
\newcommand{\DLCOMPREL}{\href{https://cryst.ehu.es/cryst/dlcomprel.html}{DLCOMPREL}}

\newcommand{\CheckT}{\href{https://cryst.ehu.eus/cryst/checktopologicallayer.html}{Check Topological Layer Mat.}}
\newcommand{\LTOPOINDICES}{\href{https://cryst.ehu.eus/cryst/topoindiceslayer.html}{Topological Indices}}
\newcommand{\RepresentationsDLSG}{\href{https://cryst.ehu.eus/cryst/representationslayer.html}{Representations DLSG}}
\newcommand{\RSI}{\href{https://cryst.ehu.eus/cryst/RSIlayer.html}{RSI-layer}}

\newcommand{\siref}{SM}

\newcommand{\mctwodpath}[1]{\href{https://mc2d.materialscloud.org/\#/details/#1}{#1}}

\newcommand{\mctwodpathold}[1]{\href{https://www.materialscloud.org/discover/legacy-mc2d/details/#1}{#1}}

\newcommand{\ctwodbpath}[1]{\href{https://c2db.fysik.dtu.dk/material/#1}{#1}}

\newcommand{\serialidweb}[1]{\href{https://www.topologicalquantumchemistry.org/topo2d/index.html\#/?serialid=#1}{#1}}
\newcommand{\twodtqclink}[1]{\href{https://topologicalquantumchemistry.com/topo2d/index.html\#/?serialid=#1}{#1}}

\newcommand{\codpath}[1]{\href{http://www.crystallography.net/cod/#1.html}{#1}}

\newcommand{\lgsymb}[1]{\ifnum#1=1
$p1$\else
\ifnum#1=2
$p\bar{1}$\else
\ifnum#1=3
$p112$\else
\ifnum#1=4
$p11m$\else
\ifnum#1=5
$p11a$\else
\ifnum#1=6
$p112/m$\else
\ifnum#1=7
$p112/a$\else
\ifnum#1=8
$p211$\else
\ifnum#1=9
$p2_111$\else
\ifnum#1=10
$c211$\else
\ifnum#1=11
$pm11$\else
\ifnum#1=12
$pb11$\else
\ifnum#1=13
$cm11$\else
\ifnum#1=14
$p2/m11$\else
\ifnum#1=15
$p2_1/m11$\else
\ifnum#1=16
$p2/b11$\else
\ifnum#1=17
$p2_1/b11$\else
\ifnum#1=18
$c2/m11$\else
\ifnum#1=19
$p222$\else
\ifnum#1=20
$p2_122$\else
\ifnum#1=21
$p2_12_12$\else
\ifnum#1=22
$c222$\else
\ifnum#1=23
$pmm2$\else
\ifnum#1=24
$pma2$\else
\ifnum#1=25
$pba2$\else
\ifnum#1=26
$cmm2$\else
\ifnum#1=27
$pm2m$\else
\ifnum#1=28
$pm2_1b$\else
\ifnum#1=29
$pb2_1m$\else
\ifnum#1=30
$pb2b$\else
\ifnum#1=31
$pm2a$\else
\ifnum#1=32
$pm2_1n$\else
\ifnum#1=33
$pb2_1a$\else
\ifnum#1=34
$pb2n$\else
\ifnum#1=35
$cm2m$\else
\ifnum#1=36
$cm2e$\else
\ifnum#1=37
$pmmm$\else
\ifnum#1=38
$pmaa$\else
\ifnum#1=39
$pban$\else
\ifnum#1=40
$pmam$\else
\ifnum#1=41
$pmma$\else
\ifnum#1=42
$pman$\else
\ifnum#1=43
$pbaa$\else
\ifnum#1=44
$pbam$\else
\ifnum#1=45
$pbma$\else
\ifnum#1=46
$pmmn$\else
\ifnum#1=47
$cmmm$\else
\ifnum#1=48
$cmme$\else
\ifnum#1=49
$p4$\else
\ifnum#1=50
$p\bar{4}$\else
\ifnum#1=51
$p4/m$\else
\ifnum#1=52
$p4/n$\else
\ifnum#1=53
$p422$\else
\ifnum#1=54
$p42_12$\else
\ifnum#1=55
$p4mm$\else
\ifnum#1=56
$p4bm$\else
\ifnum#1=57
$p\bar{4}2m$\else
\ifnum#1=58
$p\bar{4}2_1m$\else
\ifnum#1=59
$p\bar{4}m2$\else
\ifnum#1=60
$p\bar{4}b2$\else
\ifnum#1=61
$p4/mmm$\else
\ifnum#1=62
$p4/nbm$\else
\ifnum#1=63
$p4/mbm$\else
\ifnum#1=64
$p4/nmm$\else
\ifnum#1=65
$p3$\else
\ifnum#1=66
$p\bar{3}$\else
\ifnum#1=67
$p312$\else
\ifnum#1=68
$p321$\else
\ifnum#1=69
$p3m1$\else
\ifnum#1=70
$p31m$\else
\ifnum#1=71
$p\bar{3}1m$\else
\ifnum#1=72
$p\bar{3}m1$\else
\ifnum#1=73
$p6$\else
\ifnum#1=74
$p\bar{6}$\else
\ifnum#1=75
$p6/m$\else
\ifnum#1=76
$p622$\else
\ifnum#1=77
$p6mm$\else
\ifnum#1=78
$p\bar{6}m2$\else
\ifnum#1=79
$p\bar{6}2m$\else
\ifnum#1=80
$p6/mmm$\else
{\color{red}{Invalid LG number}}
\fi
\fi
\fi
\fi
\fi
\fi
\fi
\fi
\fi
\fi
\fi
\fi
\fi
\fi
\fi
\fi
\fi
\fi
\fi
\fi
\fi
\fi
\fi
\fi
\fi
\fi
\fi
\fi
\fi
\fi
\fi
\fi
\fi
\fi
\fi
\fi
\fi
\fi
\fi
\fi
\fi
\fi
\fi
\fi
\fi
\fi
\fi
\fi
\fi
\fi
\fi
\fi
\fi
\fi
\fi
\fi
\fi
\fi
\fi
\fi
\fi
\fi
\fi
\fi
\fi
\fi
\fi
\fi
\fi
\fi
\fi
\fi
\fi
\fi
\fi
\fi
\fi
\fi
\fi
\fi}

\newcommand{\lgsymbnum}[1]{LG #1 (\lgsymb{#1})}

\newcommand{\titlePaper}{
Two-dimensional Topological Quantum Chemistry and Catalog of Topological Materials}

\newcommand{\paperAuthors}{
    \author{Urko Petralanda}
    \thanks{These authors contributed equally to this work.}
    \email{urko.petralanda@ehu.eus}
    \affiliation{Department of Applied Mathematics, University of the Basque Country UPV/EHU, Vitoria-Gasteiz, Spain}
    \affiliation{Department of Physics, University of the Basque Country UPV/EHU, Apartado 644, 48080 Bilbao, Spain}
    \author{Yi Jiang}
    \thanks{These authors contributed equally to this work.}
    \affiliation{Donostia International Physics Center (DIPC), Paseo Manuel de Lardizábal. 20018, San Sebastián, Spain}
	\author{Hanqi Pi}
	\affiliation{Donostia International Physics Center (DIPC), Paseo Manuel de Lardizábal. 20018, San Sebastián, Spain}
	\author{Maia G. Vergniory}
   	\affiliation{Donostia International Physics Center (DIPC), Paseo Manuel de Lardizábal. 20018, San Sebastián, Spain}
	\affiliation{Département de Physique et Institut Quantique, Université de Sherbrooke, Sherbrooke QC J1K 2R1, Canada}
    \affiliation{Regroupement Québécois sur les Matériaux de Pointe (RQMP), Quebec H3T 3J7, Canada}
	\author{B.~Andrei Bernevig}
	\email{bernevig@princeton.edu}
	\affiliation{Department of Physics, Princeton University, Princeton, New Jersey 08544, USA}
   	\affiliation{Donostia International Physics Center (DIPC), Paseo Manuel de Lardizábal. 20018, San Sebastián, Spain}
	\affiliation{IKERBASQUE, Basque Foundation for Science, Bilbao, Spain}
    \author{Nicolas Regnault}
    \email{regnault@princeton.edu}
    \affiliation{Center for Computational Quantum Physics, Flatiron Institute, 162 5th Avenue, New York, NY 10010, USA}
    \affiliation{Department of Physics, Princeton University, Princeton, New Jersey 08544, USA}
    \affiliation{Laboratoire de Physique de l'Ecole normale sup\'{e}rieure, ENS, Universit\'{e} PSL, CNRS, Sorbonne Universit\'{e}, Universit\'{e} Paris-Diderot, Sorbonne Paris Cit\'{e}, 75005 Paris, France}
    \author{Luis Elcoro}
    \email{luis.elcoro@ehu.eus}
    \affiliation{Department of Physics, University of the Basque Country UPV/EHU, Apartado 644, 48080 Bilbao, Spain}
}

\begin{document}
\title{\titlePaper}
\paperAuthors

\begin{abstract}

Two-dimensional topological materials host symmetry-protected, disorder-robust edge states central to quantum technologies. Yet, they remain scarce compared with their three-dimensional counterparts.
We adapt the topological quantum chemistry formalism to layer groups, and study the band topology of \TwoDDBNbrEntries\  entries from the computational materials databases C2DB and MC2D. We find \TwoDDBMaterialsNonTrivialOrObstructedWithSOC\  topological or obstructed entries, including \TwoDDBMaterialsTIWithSOC\  topological insulators and \TwoDDBMaterialsOAIWithSOC\ obstructed atomic insulators, greatly expanding the library of such materials in two dimensions. Detailed electronic and topological properties of all materials are compiled into a a Topological 2D Materials Database (2D-TQCDB), and the edge states of selected new materials are studied in detail. Our methodology is implemented in new publicly available programs, designed to study the topology of any non-magnetic monolayer or multilayer material.

\end{abstract}

\maketitle

\section*{}\label{sec:Intro}

Two decades ago, the theoretical prediction \cite{PhysRevLett.95.226801,PhysRevLett.95.146802,Bernevig2006} and the first experimental verification \cite{Konig2007} of the 2D quantum spin Hall (QSH) effect sparked strong interest in topological insulators, which continues to this day. During this time, various classes of topological insulators beyond QSH insulators \cite{tcli,hoti0,Schindler2018} have contributed to the expanding field of topology in non-magnetic materials.

Recently identified topological materials have mostly been three-dimensional, with symmetry-based methods such as topological quantum chemistry (TQC) and symmetry indicators being used to diagnose band topology and predict thousands of 3D topological insulators.\cite{Vergniory2019,Zhang2019,Vergniory2022,Wieder2021,bradlyn2017topological,Po2017,slager2017}.
However, 2D topological insulators possess fundamental features essential for future topological device architectures \cite{hasan-kane,qi-zhang}. The one-dimensional protected edge states in 2D-QSH insulators have been proposed \cite{PhysRevLett.100.096407,Strunz2019} as a platform to realize Majorana anyons when coupling to a superconductor \cite{majorana}. In addition, the low dimensionality of monolayers offers important advantages, such as the ability to tune physical properties through multilayer design, including adjusting the twisting angle formed by the components \cite{magicAngle1}. 

The library of experimentally fabricated 2D materials has steadily grown since graphene was first exfoliated \cite{Novoselov2004}. Computational exploration of these materials, especially by means of density functional theory (DFT) calculations, has, in parallel, made substantial progress during this time. This is reflected in the several 2D materials databases that collect thousands of monolayers relaxed and studied in detail by DFT \cite{Haastrup_2018,Gjerding_2021,Mounet2018,2DMatpedia}. 
This enables the prediction of new materials and the identification of monolayers for experimental investigation, particularly QSH insulators \cite{PhysRevMaterials.3.024005,Marrazzo2019,Wang2019,Choudhary2020,Schleder2021,Costa2021,soliton,tynerTI,yao2011,liu2011,ascl-topo-liu,Grassano2023,mc2d-database}. Topological crystalline and obstructed atomic insulators \cite{PhysRevMaterials.3.024005,oai-graphyidyne,Sdequist2022,yi-oai,oaihexa,oai58,oailigand,oai2023,oai2024}, as well as topological semimetals (TSMs) \cite{2dtoposemi,ben-layer,3toposemimetals}, have also been identified.
Nevertheless, the catalogue of topological or obstructed atomic insulator monolayers, numbering around 400 materials, remains small compared to their 3D counterparts.

Isolated groups of connected bands are trivial if induced from atomic orbitals. In topological insulators, however, the valence band lacks an atomic limit \cite{soluyanov2011}. In TQC, elementary band representations (EBRs) are defined as the (finite) basis of trivial bands induced from atomic orbitals,
enabling classification of gapped bands via DFT as trivial or topological depending on whether they decompose into EBRs.
Here, we adapt the TQC formalism to study the topology of materials with layer group (LG) symmetry. To accomplish this, we have developed a set of new programs in the Bilbao Crystallographic Server (BCS) \cite{aroyo2011crystallography,aroyo2006bilbao1, aroyo2006bilbao2}.
\nocite{Topo2DDB}%
We have applied the formalism to \TwoDDBNbrEntries\  entries of monolayers from the computational materials databases C2DB  \cite{Haastrup_2018,Gjerding_2021} and MC2D  \cite{Mounet2018,mc2d2}, finding \TwoDDBMaterialsNonTrivialOrObstructedWithSOC\ topological or obstructed entries.
With our results, we have built a comprehensive online database, Topological 2D Materials Database (2D-TQCDB) \cite{Topo2DDB}, which provides the band structures and topological properties of each material.

\section{Workflow}\label{sec:Workflow}
\begin{figure*}
    \includegraphics[width=0.9\linewidth]{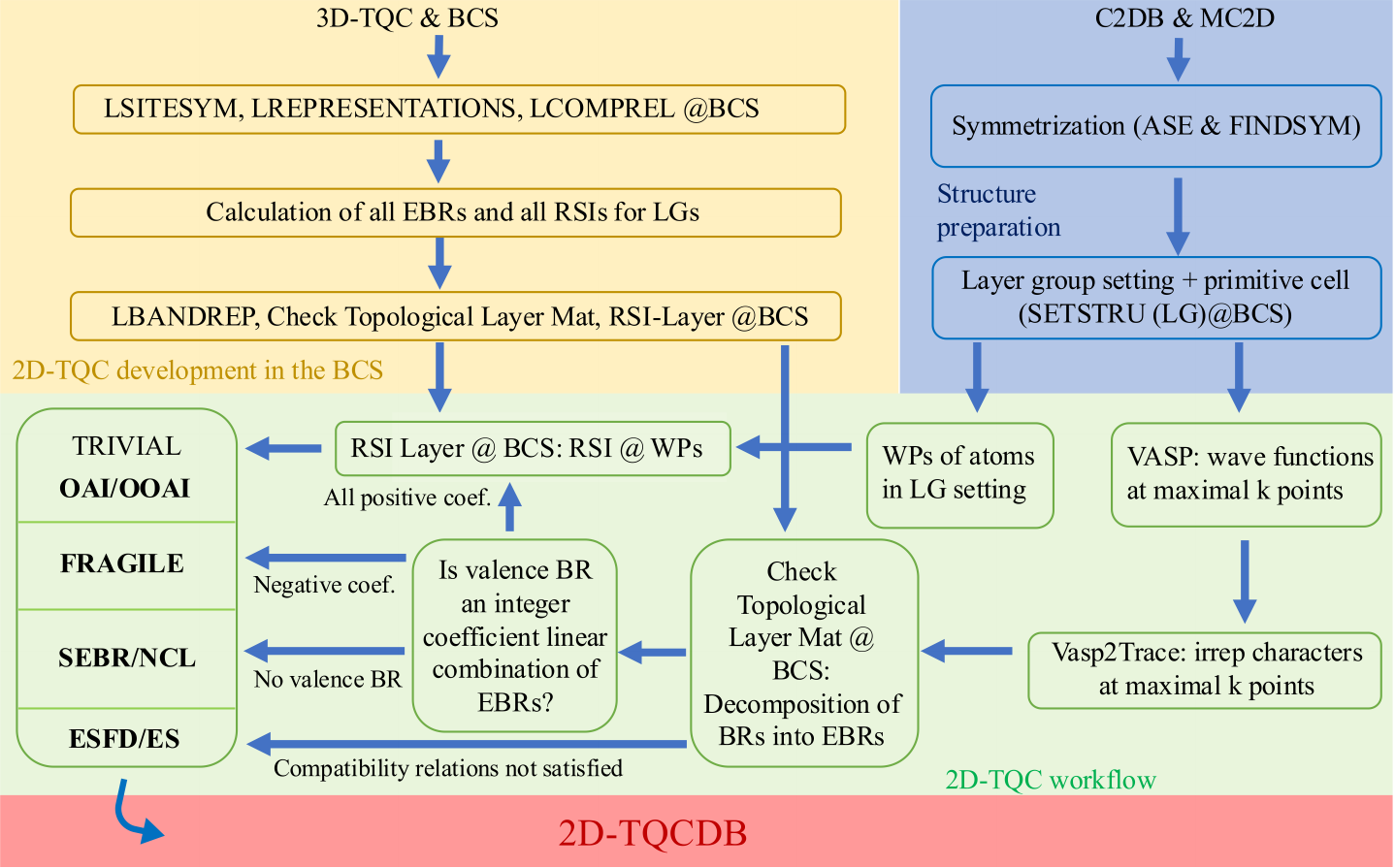}
    \caption{Flowchart of the process for the discovery of monolayer topological materials. The first stage involves the adaptation of Topological Quantum Chemistry (TQC) to layer groups (LGs). This includes the development of new programs in the BCS. During the structure preparation stage, entries from the 2D materials computational databases C2DB; and MC2D; are processed using the Atomic Simulation Environment (ASE) \cite{HjorthLarsen2017}, FINDSYM \cite{Stokes2005}, and the BCS. In the 2D-TQC workflow, a combination of VASP, Vasp2Trace, and BCS is used to classify structures based on their band topology. Finally, the data are added to the Topological 2D Materials Databse (2D-TQCDB).
    }
    \label{fig:workflow}
\end{figure*}

Our workflow is divided into three major phases: development of new BCS programs, structure preparation, and topological analysis (\cref{fig:workflow}).

\subsection{New BCS programs}
We first wrote a set of programs in the BCS.  
Among them, we highlight DLCOMPREL, LBANDREP, Check Topological Layer Mat., and RSI-layer. DLCOMPREL is a database that gives the compatibility relations between the irreducible representations (irreps) of the little groups of a pair of k-vectors with a group-subgroup relation. 
The output of LBANDREP shows, for a given (double) LG, the band representations induced from every irrep of the site-symmetry group of every Wyckoff position. The program also identifies the set of EBRs, and points out which EBRs are decomposable.
The knowledge of all the EBRs subsequently allowed us to write the Check Topological Layer Mat. program. This program determines whether, and if so, how, a set of bands can be decomposed into EBRs, thereby (partially) determining its topology. Finally, we have calculated all real space invariants (RSIs) for all LG at any Wyckoff position, now available in the RSI-layer\ program.
A detailed description of the new BCS programs introduced here is given in Section S3 of \cite{SM-note}.

\subsection{Structure preparation}
In the second phase, we extracted structural data from the computational databases C2DB and MC2D. ~From the C2DB, we selected two groups of structures. One consisted of all non-magnetic structures labeled as highly thermodynamically and/or dynamically stable. 
The second group consisted of all non-magnetic structures with fewer than 40 electrons per unit cell not included in the previous group. 
From the MC2D, we downloaded the 2D relaxed structures after retrieving their numerical identification codes.
We then assigned each structure a LG and transformed it to a primitive unit cell.
More details about structure acquisition and symmetrization are given in \cite{SM-note}.

\subsection{Topological analysis}
The third phase comprised screening the selected structures for band topology using the TQC formalism adapted to layer groups. Detailed descriptions of the TQC approach are provided in Refs. \cite{bradlyn2017topological,Vergniory2019,xu2021three} and Section S2 of \cite{SM-note} , while our adaptation to layer groups is presented in Section S3 of \cite{SM-note}. 
We first calculated the wavefunctions at high-symmetry (maximal) $\mathbf{k}$-points of the Brillouin zone using the DFT code VASP \cite{vasp1,vasp2} (see \cite{SM-note} for computational details). This allowed us to compute the traces of the representation matrices of each symmetry operation of the little group of every maximal $\mathbf{k}$-point, using the program vasp2trace \cite{Vergniory2019}. 
This was then used as input for the program Check Topological Layer Mat., which determines the band topology of the compound.
Using the traces and DLCOMPREL, we infer the possible connections between the eigenstates at all pairs of maximal $\mathbf{k}$-vectors in the valence bands. 
If the eigenstates cannot be all connected, the material is diagnosed as a topological semimetal [enforced semimetal (ES) or enforced semimetal with Fermi level degeneracy (ESFD)]. If they can be connected, the program identifies the material as a trivial insulator when the set of bands can be expressed as a linear combination of EBRs with positive integers [linear combination of EBR (LCEBR)], as Fragile when all possible linear combinations require at least a negative integer, or as a topological insulator [split EBR (SEBR), or no linear combination (NLC)] if there is no such linear combination. A detailed derivation and description of these cases is provided in \cite{SM-note}.

In non-magnetic 2D insulators with spin-orbit coupling (SOC), gapped stable topological states are classified as either QSH insulators (QSHI) or mirror topological crystalline insulators (TCI). QSHIs are protected solely by time-reversal symmetry (TRS), whereas mirror TCIs require both TRS and an in-plane mirror symmetry (i.e., $M_z$ when the system is placed on the $xy$ plane). 
These topological phases can be diagnosed through symmetry-based indices derived from crystalline symmetries~\cite{song2018diagnosis, Song2018, Po2017}. For example, inversion symmetry enables the definition of the $\mathbb{Z}_2$ index $z_{2w,1}$, whose non-zero value denotes a QSHI~\cite{fu2007topological}. When $C_{nz}$ rotation perpendicular to the structure plane is present in addition to $M_z$ symmetry, a mirror TCI can be diagnosed using the $z_{nm0}$ ($n=3,4,6$) indices. 
When multiple indices can be defined, they will not necessarily be independent. 
Check Topological Layer Mat. provides the values for topological indices (tabulated in all LGs in the new program TopologicalIndices) that characterize the topological class of a material.
In \cite{SM-note}, we provide an overview of the derivation of these topological indices and their physical implications.

Obstructed atomic insulators (OAIs)~\cite{xu2021three, oaienforced, gao2022unconventional} are characterized by Wannier centers that are pinned to empty sites and cannot be adiabatically moved to atomic positions without breaking crystal symmetry. In orbital-selected obstructed atomic insulators (OOAIs), the valence band representation can be expressed as LCEBRs induced from the occupied atomic positions, but in all combinations, at least one EBR does not correspond to any orbital of the outer shell of any atom. OAIs are identified through RSIs, which present at least a nonzero RSI at an unoccupied Wyckoff position 
(see Section S2 of \cite{SM-note} or Appendix C of Ref. \cite{xu2021three}). While RSIs may involve single or multiple Wyckoff positions, any nonzero value of RSIs indicates charge obstruction at interstitial sites when all relevant positions are empty.

The bulk-boundary correspondence has distinct manifestations for different topological phases. Both QSHI and mirror TCI exhibit helical edge states when confined to 1D ribbons. Topological semimetals present gapless edge states originating from the bulk gapless nodes. 
In contrast, OAIs display fundamentally different boundary phenomena: non-helical edge states emerge when ribbons are cut along obstructed Wyckoff positions, whereas zero-dimensional flakes preserving the symmetry of the monolayer reveal fractional corner charges at these positions~\cite{fraccharge1,indicators-oal,Li2022,oaienforced, gao2022unconventional}. 

\begin{figure*}
    \centering
    \includegraphics[width=\linewidth]{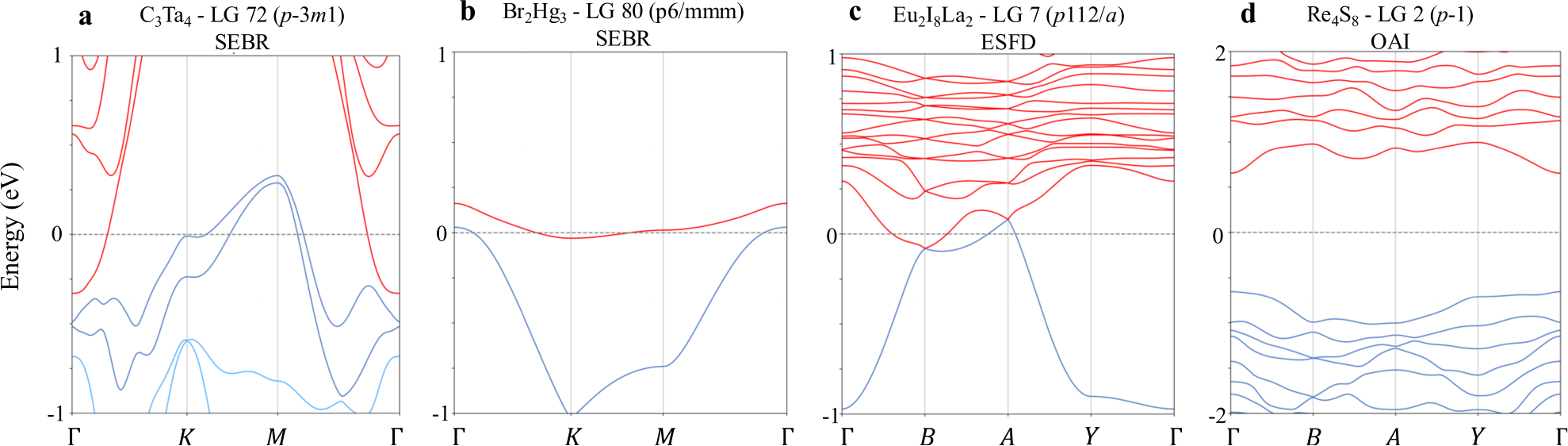}
    \caption{Materials selected from the high-throughput search. We show, in all cases, the formula, layer group (LG), and topological type of the material. Valence (conduction) bands are drawn in blue (red) unless the bands belong to a BR with fragile topology, in which case they are drawn in light blue (magenta). (A)
    MXene C$_3$Ta$_4$ in \lgsymbnum{72} (2D-TQC \serialidweb{1.1.12}), a SEBR QSH insulator that has been experimentally realized in monolayer form~\cite{cotton2017epitaxial}.
    (B) Br$_2$Hg$_3$ in \lgsymbnum{80} (2D-TQC \serialidweb{1.2.133}), a computationally exfoliated monolayer topological insulator with SEBR topology. (C)
    Eu$_2$I$_8$La$_2$ in \lgsymbnum{7}, (2D-TQC \serialidweb{2.2.30}) a computationally exfoliated Dirac semimetal. (D) Re$_4$S$_8$ in \lgsymbnum{2} (2D-TQC \serialidweb{3.1.1}), an experimentally existing~\cite{wolverson2014raman} obstructed atomic insulator. }
    \label{fig:materials}
\end{figure*}

\section{\label{sec:Mat}Materials statistics and examples}

We identified a total of \TwoDDBMaterialsNonTrivialOrObstructedWithSOC\  topological or obstructed 
entries, including \TwoDDBMaterialsTIWithSOC\  topological insulators (\TwoDDBMaterialsSEBRWithSOC\  SEBR and \TwoDDBMaterialsNLCWithSOC\  NLC, \TwoDDBMaterialsTIFullGapWithSOC\ with a non-zero global gap), \TwoDDBMaterialsSMWithSOC\  topological semimetals (\TwoDDBMaterialsESWithSOC\  ES and \TwoDDBMaterialsESFDWithSOC\  ESFD), \TwoDDBMaterialsOAIWithSOC\  OAIs, and \TwoDDBMaterialsOOAIWithSOC\ OOAIs. Therefore, over \TwoDDBPercentMaterialsNonTrivialOrObstructedWithSOC\% of the \TwoDDBNbrEntries\ materials studied exhibit non-trivial band topology or an obstructed atomic limit, about \TwoDDBPercentMaterialsTIWithSOC\% of them being topological insulators. 
Most of the studied 2D materials have yet to be realized experimentally. Through a literature review, we found about 170 entries corresponding to experimentally fabricated non-magnetic materials, either in monolayer or few-layer configurations. Among these, we identified 86 as topological or possessing an OAL.

We have tabulated all topological or obstructed materials in
Tables S4 to S27
in \cite{SM-note}. Materials in each class (TI, TSM, OAI, and OOAI) are further categorized as experimentally existing, computationally exfoliated (MC2D), stable (C2DB), and not-stable (C2DB).
In Section S10 of \cite{SM-note}, we have shortlisted all the topological insulators with a full gap.

In \cref{fig:materials}, we showcase four representative materials discovered here. \cref{fig:materials}A shows the band structure of an experimentally realized \cite{c3ta4exp} topological insulator of QSHI type, MXene C$_3$Ta$_4$ in \lgsymbnum{72} (2D-TQC \serialidweb{1.1.12}). It features a 100 meV direct gap at $\Gamma$, zero global gap, and SEBR topology. In \cref{fig:materials}B we display the band plot of Br$_2$Hg$_3$ in \lgsymbnum{80} (2D-TQC \serialidweb{1.2.133}), a computationally exfoliated monolayer from the MC2D, identified here as a SEBR TI. Its topological index is $z_{m6,0}=1$, corresponding to the mirror Chern number \cite{MCN}. 
In \cref{fig:materials}C we show the bands of Eu$_2$I$_8$La$_2$ in \lgsymbnum{7}, (2D-TQC \serialidweb{2.2.30}), a computationally exfoliated Dirac semimetal, displaying two symmetry-enforced crossings near $E_F$. 
Re$_4$S$_8$ in \lgsymbnum{2} (2D-TQC \serialidweb{3.1.1}, shown in \cref{fig:materials}D, is an experimentally realized \cite{exfoliation} monolayer OAI with a gap of 1.28 eV. Its RSIs can be found in Table S20 of \cite{SM-note}. 
While the atoms occupy $e$ Wyckoff positions, RSIs defined at $b$, $d$ are nonzero.
Obstructed edge states 
arising from charge filling anomaly \cite{filanomaly1,filanomaly2} are expected to emerge at both $b$ and $d$ when the material is cut into 1D ribbons passing through these positions. Nanoflakes of the material that preserve inversion symmetry are expected to show symmetry-protected, localized fractional corner charges.

Our analysis fully characterizes the topology of all computed bands, enabling a search for materials with fragile topology \cite{fragile0} at any energy. Fragile topological materials are scarce; to our knowledge, no crystalline material exhibits fragile cumulative topology at $E_F$ \cite{fragilemonoid}.
In Section S6 of \cite{SM-note} we show a material, Cu$_2$H$_4$ in \lgsymbnum{72} (2D-TQC \serialidweb{2.4.669}), presenting cumulative fragile topology at $N_e-1$ electrons ($N_e$ being the total number of electrons). 
Twisted bilayer graphene exhibits fragile topological bands at $E_F$ relevant to its correlated phases \cite{fragilegraphene,fragileMott}. 
We identified experimentally realized materials with isolated fragile bands near $E_F$, listed in Table S3 of \cite{SM-note}.

We also find a wide variety of topological semimetals.
In Section S7 of \cite{SM-note}, we present representative materials, including computationally exfoliated O$_6$P$_2$Sn$_2$ in \lgsymbnum{5} (2D-TQC \serialidweb{2.2.25}); dynamically stable Ir$_2$S$_6$ in \lgsymbnum{5} (2D-TQC \serialidweb{2.3.34}), containing \emph{hourglass} dispersion \cite{hourglass2016,hourgla2} at $E_F$; Cu$_2$F$_4$, in \lgsymbnum{63} (2D-TQC \serialidweb{2.2.203}), a computationally exfoliated nodal-line semimetal with a Dirac crossing near $E_F$; Ir$_2$O$_4$ in \lgsymbnum{15}, a dynamically stable, non-symmorphic semimetal with 4D crossings (2D-TQC \serialidweb{2.3.134}). In \cite{SM-note}, we 
discuss in detail all TSM highlighted in the main text.

Our results are available to the community through the 2D-TQCDB website, as a part of the Topological Quantum Chemistry website \cite{Vergniory2022}. Currently, it consists of around 9,000 entries and will be open to further enhancement as new 2D materials emerge. In Section S4 of \cite{SM-note} we provide a more detailed overview of the database.

\begin{figure*}
    \includegraphics[width=\linewidth]{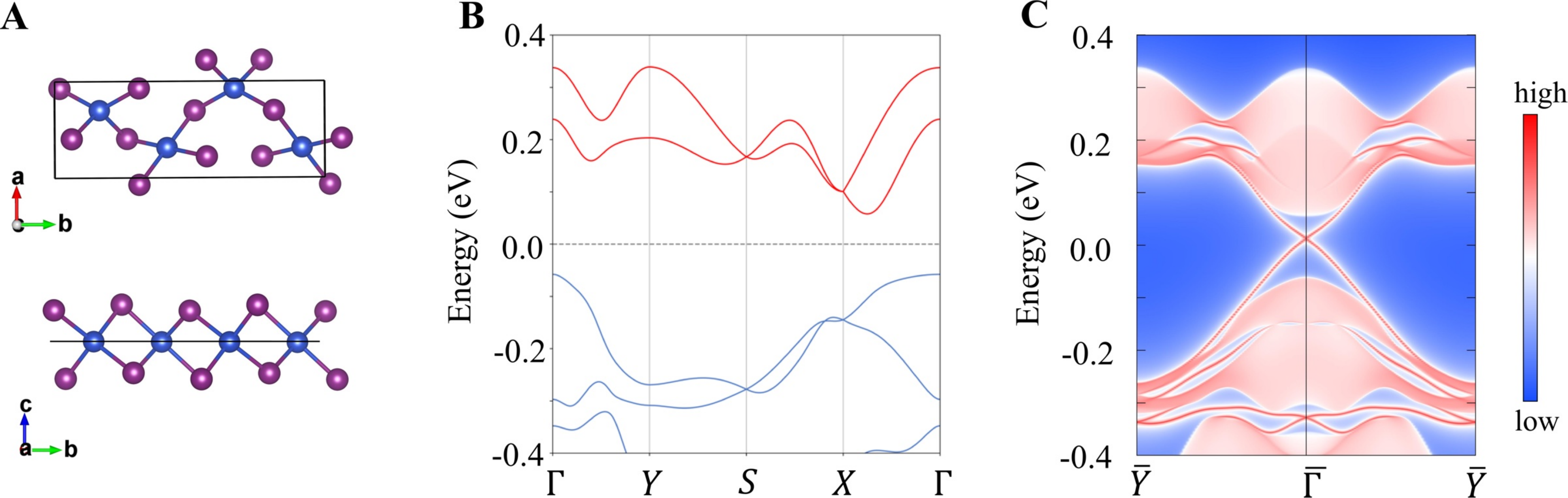}
    \caption{Atomic and electronic properties of monolayer Cu$_4$I$_8$ \lgsymbnum{16} (2D-TQC \serialidweb{1.3.26}). (A) Top and lateral views of the compound, where the blue (purple) spheres denote Cu (I) atoms. (B) Band structure calculated by DFT with SOC included, where valence bands are represented in red and conduction bands in blue. (C) Spectral function of the tight-binding model on a semi-infinite ribbon, cut along the $[10]$ direction of the structure. Red (blue) color indicates a higher (lower) contribution of edge atoms. 
    }
    \label{fig:ti}
\end{figure*}

\section{\label{sec:edgesates}Edge states in selected materials}

We have selected three thermodynamically stable topological insulator representatives, Cu$_4$I$_8$ in \lgsymbnum{16} (2D-TQC \serialidweb{1.3.26}),
Bi$_2$Br$_2$ in \lgsymbnum{15} (2D-TQC \serialidweb{1.3.6}), and As$_2$Cl$_2$ in \lgsymbnum{72} (2D-TQC \serialidweb{1.3.193}), and two OAIs, Re$_4$Se$_8$ in \lgsymbnum{2} (2D-TQC \serialidweb{3.1.3}), and C$_6$N$_2$ in \lgsymbnum{80} (2D-TQC \serialidweb{3.1.48}) to calculate their edge states. The two OAIs exist as experimental monolayers \cite{wolverson2014raman,c6n3exp}.

We first consider the topological insulator \ch{Cu4I8} in \lgsymbnum{16}. \cref{fig:ti} A and B show the crystal structure and the monolayer band structure with SOC, respectively. It exhibits a global band gap of 0.12 eV, which emerges owing to the SOC-induced gap opening at the symmetry-protected crossing point along the $\Gamma$-$X$ path in the absence of SOC. \ch{Cu4I8} has a topological index $z_{2w,1}=1$, corresponding to a time-reversal protected $\mathbb{Z}_2$ index $\nu=1$ QSHI (NLC). 
We consider a ribbon with edges parallel to the $a$ axis of the unit cell, and calculate its electronic properties based on a tight-binding model constructed from maximally localized Wannier functions \cite{Pizzi2020} (see details of tight-binding models in \cite{SM-note}). In \cref{fig:ti}C, we show the band structure of a semi-infinite length ribbon calculated using the Green's function method, as implemented in the WannierTools~\cite{Wu2018}. A surface Dirac cone centered at the $\Gamma$ point emerges, formed by topologically protected helical edge states. For the experimental validation of a topological insulator, it is essential to verify the presence of edge states in realistic conditions. We thus constructed a 1D ribbon cut along the $a$ axis to mimic the system simulated in the tight-binding model. After relaxing the ribbon in DFT, the band structure shows a similar Dirac cone at $\Gamma$ as in the tight-binding calculation \cite{SM-note}.  
For Bi$_2$Br$_2$ and As$_2$Cl$_2$, we have done a similar analysis, including the calculation of the spectral function of the tight-binding model on a semi-infinite ribbon and the ab initio band structure on a fully relaxed nanoribbon \cite{SM-note}. 

\begin{figure*}
    \includegraphics[width=\linewidth]{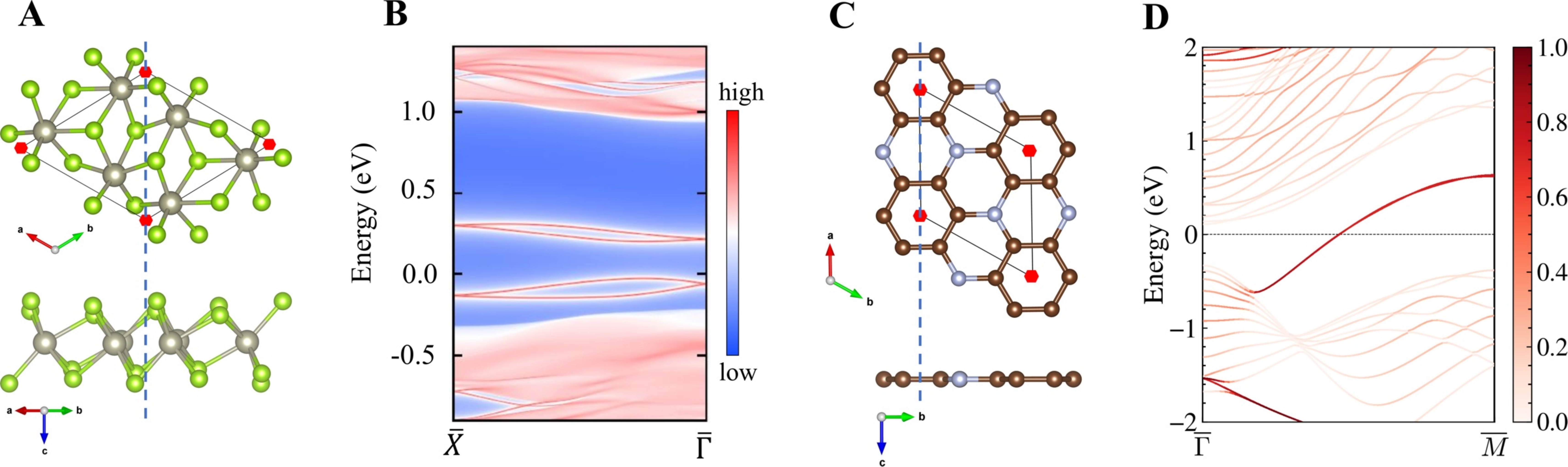}
    \caption{Atomic and electronic properties of monolayer Re$_4$Se$_8$ in \lgsymbnum{2} (2D-TQC \serialidweb{3.1.3}) and C$_6$N$_2$ in \lgsymbnum{80} (2D-TQC \serialidweb{3.1.48}). (A) Lateral and top views of Re$_4$Se$_8$. Re atoms, Se atoms, and Wyckoff positions $a$ are represented by grey spheres, green spheres and red hexagons, respectively. (B) Spectral function of the tight-binding model on a semi-infinite ribbon cut along the [11] direction of Re$_4$Se$_8$, where the cutting plane is indicated with a blue dashed line in \textbf{A}. Red (blue) color indicates a higher (lower) contribution of edge atoms. (C) Lateral and top views of C$_6$N$_2$. C atoms, N atoms, and Wyckoff position $a$ are represented by brown and grey spheres and red hexagons, respectively. 
    (D) Band structure of a ten-unit-cell-wide ribbon of \ch{C6N2}, cut along the [10] direction of C$_6$N$_2$, with the cutting plane indicated with a blue dashed line in (C). The edge projection highlights the in-gap obstructed edge states. 
    }
    \label{fig:oai}
\end{figure*}

We then discuss the two OAI materials. 
In Re$_4$Se$_8$, Se atoms are arranged in octahedral networks around Re atoms (\cref{fig:oai}A), forming the ``distorted diamond chain'' \cite{distorted-diamond}. The obstructed orbitals of \ch{Re4Re8} are located at the inversion-symmetric Wyckoff positions $a$ and $c$ of in \lgsymbnum{2}. The program RSI-layer\ shows that these obstructed orbitals can be identified using two $\mathbb{Z}$-type RSIs, both with value $\delta=1$. C$_6$N$_2$ in \lgsymbnum{80} has a graphene-like honeycomb lattice shown in \cref{fig:oai}C with distinct bonding environments: each carbon atom forms three $sp^2$ bonds from one nitrogen and two carbon neighbors, while each nitrogen atom bonds to three carbon atoms. As the atoms occupy only Wyckoff positions $b$ and $h$ in \ch{C6N2}, a nonzero RSI ($\delta=-1$) at $a$ (the carbon hexagon center) identifies \ch{C6N2} as an OAI.

To probe mid-gap obstructed edge states, we built tight-binding models for both materials. We then cut Re$_4$Se$_8$ into a ribbon in the $[11]$ direction, and C$_6$N$_2$ in $[10]$ direction. The cutting planes (dashed lines in \cref{fig:oai}A and \cref{fig:oai}C) pass through Wyckoff position $a$ (red hexagons) in both structures and avoid all atomic positions. In both cases, the ribbons preserve inversion symmetry. We then perform semi-infinite ribbon calculations. Re$_4$Se$_8$ presents four mid-gap bands that form two pairs of two-fold Kramers degenerate states at time-reversal invariant momenta, per ribbon edge (\cref{fig:oai}B). 
In Fig. S22D of \cite{SM-note}, we repeated the calculation for the $[10]$-cut ribbon of \ch{Re4Se8}, where the ribbon edge passes through both $a$ and $c$. As expected, the number of midgap states increases compared to the $[11]$-cut ribbon. The two cutting planes are shown in detail in Fig. S22E of \cite{SM-note}.
In the case of C$_6$N$_2$, two nearly degenerate mid-gap states are observed (see Fig. S21 of \cite{SM-note}, which intersect with bulk valence bands. These states originate from the obstructed Wyckoff position $a$, at the ribbon edge (\cref{fig:oai}C). Consequently, there are two spinful obstructed edge states per edge. 

We also performed ab initio calculations on Re$_4$Se$_8$ and C$_6$N$_2$ nanoribbons to validate tight-binding predictions. Both materials exhibit strong covalent bonds that would lead to significant edge reconstruction upon cleavage. To stabilize the edges while preserving intrinsic electronic properties, we employed hydrogen passivation—a well-established technique supported both theoretically \cite{louie-passi2006,jacobsen-passi2010} and experimentally \cite{Lawrence2022passi}. The hydrogen-passivated ribbons of both materials show clear mid-gap edge states, characteristic of OAIs. We present the results for Re$_4$Se$_8$ ribbons cut along the [10] direction (crossing Wyckoff positions $a$ and $c$) in Fig. S20 of \cite{SM-note}, showing four spinful bands per edge in the gap. This strongly indicates the presence of one obstructed atomic center (two spinful states) at each Wyckoff position $a$ and $c$. This suggests that two of the four midgap edge states in the $[11]$ ribbon in \cref{fig:oai}B originate from dangling bonds.
We also consider the C$_6$N$_2$ ribbon cut along the $[10]$ direction.
In the unhydrogenated ribbon (Fig. S21 of \cite{SM-note}), dangling-bond states from broken $sp^2$ bonds are pushed to high energies, leaving the obstructed surface states fully occupied. Hydrogenation saturates these bonds, and the associated bands are lowered into the valence manifold. This removes the extrinsic boundary compensation and exposes the intrinsic OAI filling anomaly: the inversion-related obstructed states, originating from the Wannier centers at $a$, are half-filled, as shown in \cref{fig:oai}D.

\section{Discussion and outlook}\label{Discussion}
We adapt the TQC formalism to account for symmetry-indicated topology and obstructed phases in non-magnetic materials with layer group symmetry, and implement it in several new programs of the Bilbao Crystallographic Server. We apply the formalism to \TwoDDBNbrEntries\  entries from computational 2D materials databases C2DB  and MC2D, including over 100 experimentally realized monolayers, and increase the pool of topological or OAL monolayers by an order of magnitude, whose edge states should be analyzed in the future. 
Key electronic properties of the band edges, including effective masses, vacuum-aligned energies, and twistability metrics~\cite{jiang2024twist}, are provided in the 2D-TQCDB database. A comprehensive table for hetero-bilayer band edge alignment is provided in Section S15 of \cite{SM-note}. 

Our work provides an atlas of monolayers to guide experimental synthesis. Following two decades of rapid progress in 2D material synthesis \cite{exforeview}, the field is now undergoing another revolution, driven by twisting different monolayers. Our database and computational tools enable advancement in this emerging area.
Our work should inspire continuous progress in building new 2D materials computationally, accompanied by a thorough analysis of their topology. Simultaneously, progress should be made in methods to accurately predict the synthesis feasibility of computationally predicted monolayers \cite{chemexfo}. Multilayer calculations should also be performed on our systems.

\begingroup
\renewcommand{\refname}{Acknowledgments}
\section*{\refname}
We thank Titus Neupert and Grigorii Skorupskii for fruitful discussions. 
The authors thank for technical support provided by the staff at SGIker (UPV/EHU/ ERDF, EU), and the Atlas supercomputer of the Donostia International Physics Center.
\\
\textbf{Funding:} U.P. acknowledges funding from the European Union’s Next Generation EU plan through the María Zambrano Programme. U.P. and L.E. were supported by the Government of the Basque Country (Project No. IT1458-22).
Y.J. was supported by the European Research Council (ERC) under the European Union’s Horizon 2020 research and innovation program (Grant Agreement No. 101020833), as well as by the IKUR Strategy under the collaboration agreement between Ikerbasque Foundation and DIPC on behalf of the Department of Education of the Basque Government. M.G.V. received financial support from the Canada Excellence Research Chairs Program for Topological Quantum Matter and PID2025-172166NB-C22 funded by MICIU/AEI/10.13039/501100011033 and “ERDF A way of making Europe”.
B.A.B. was supported by the Gordon and Betty Moore Foundation (grant GBMF8685) towards the Princeton theory program, the Gordon and Betty Moore Foundation’s EPiQS Initiative (grant GBMF11070), the Global Collaborative Network Grant at Princeton University, the Simons Investigator Grant No. 404513, the NSF MRSEC (grant DMR-2011750), the Simons Collaboration on New Frontiers in Superconductivity (grant SFI-MPS-NFS-00006741-01), the Princeton Catalysis Initiative (PCI), the Schmidt Transformative Technology Fund at Princeton University and the NSF through the AI Research Institutes (program award DMR-2433348). 
The Flatiron Institute is a division of the Simons Foundation.
\textbf{Author contributions:} U.P. and B.A.B. conceived the study. U.P. performed DFT calculations, with the help of Y.J.. Y.J., H.P., U.P., and M.G.V. performed edge state calculations of the highlighted materials. LE generated the crystallographic tables for layer groups, wrote all the new Bilbao Crystallographic Server programs, and performed the topological classification of the materials. U.P. and L.E. wrote the manuscript, with inputs from B.A.B., Y.J. and N.R.. N.R. curated the DFT data with help of U.P. and Y.J., and created the 2D-TQCDB database. B.A.B. supervised the whole project. All authors contributed to the review and editing of the final draft.
\\
\textbf{Competing interests: }
The authors declare that they have no competing interests.
\\
\textbf{Data and materials availability: }
The crystal structures analyzed in this work are available in the C2DB (https://c2db.fysik.dtu.dk) and MC2D (https://mc2d.materialscloud.org). Programs DLCOMPREL, LBANDREP, Check Topological Layer Mat., RSI-layer, and TopologicalIndices can be accessed from https://cryst.ehu.es. WannierTools package is available at https://github.com/quanshengwu/wanniertools. First-principles calculations were performed using version 5.4.4 of the Vienna Ab-initio Simulation Package (VASP), a commercial software package available at https://www.vasp.at/. All data are available in the supplementary materials, through our public website \cite{Topo2DDB} and in Zenodo \cite{zenodoref}.

\endgroup

\bibliography{2d-tqc}

\addtocontents{toc}{\protect\setcounter{tocdepth}{2}}

\clearpage
\newpage
\onecolumngrid

\begin{center}
\textbf{Supplementary Material: 
Two-dimensional Topological Quantum Chemistry and Catalog of Topological Materials}
\end{center}

\SupplementalMaterials

\setcounter{section}{0}
\newcommand{\toclesssection}[1]{\section*{#1}\addtocounter{section}{1}}

\renewcommand{\thefigure}{S\arabic{figure}}
\renewcommand{\thetable}{S\arabic{table}}
\renewcommand{\thesection}{S\arabic{section}}
\renewcommand{\theequation}{S\arabic{equation}}

\tableofcontents

\clearpage
\newpage

\renewcommand{\thesection}{S\arabic{section}}
\renewcommand{\thefigure}{S\arabic{figure}}
\renewcommand{\thetable}{S\arabic{table}}
\renewcommand{\theequation}{S\arabic{equation}}
\renewcommand{\thepage}{S\arabic{page}}

\setcounter{figure}{0}
\setcounter{table}{0}
\setcounter{equation}{0}
\setcounter{page}{1} 

\section{Materials and methods}\label{comp}
\addcontentsline{stoc}{section}{\protect\numberline{\thesection}Materials and methods}

\subsection{Structure acquisition and processing}
\addcontentsline{stoc}{subsection}{\protect\numberline{\thesubsection}Structure acquisition and processing}

We extracted structural data from the computational databases \CtwoDB\ \cite{Haastrup_2018,Gjerding_2021} and \MCtwoD\ \cite{Mounet2018,Grassano2023}. From the \CtwoDB, we selected two groups of structures. One group consists of all non-magnetic structures labeled as highly thermodynamically and/or dynamically stable. In the database, a structure is defined as highly thermodynamically stable if it lies less than 0.2 eV/atom above the convex hull minimum. The requirement for a structure to be labeled as dynamically stable is that neither the stiffness tensor nor the $\Gamma$ point hessian matrix of the $2\times 2$ supercell of the material presents negative eigenvalues. This data was previously calculated in \CtwoDB. Under these criteria, 4462 stable structures were found. The second group of \CtwoDB\ structures consisted of all non-magnetic structures with fewer than 40 electrons per unit cell, which were not included in the previous group. 2095 entries were found to satisfy these conditions. In total, 6557 \CtwoDB\  entries were thus collected for further analysis. 
As the \CtwoDB\  is continuously enhanced, the number of structures satisfying these criteria is expected to increase over time. From the \MCtwoD, we retrieved the numerical identification codes for 2763 relaxed 2D structures from the \emph{Raw Data} section within the website's API and subsequently downloaded all the structures associated with those codes. We note that the magnetization of only a small fraction of the \MCtwoD\  structures has been studied. In our approach, we treated materials whose magnetization has not been studied in \MCtwoD\  as non-magnetic. We then processed all the structures collected from the two databases to identify the symmetry operations that keep the set of atomic positions invariant and, thus, to assign an LG to the structure. It should be stressed that, in most cases, the structure data are given in P1 space group and the structure file contains the whole set of atomic positions in a primitive unit cell with no reference to the symmetry operations. We initially converted the original files into standard CIF files using the Atomic Simulation Environment (ASE) \cite{HjorthLarsen2017} and then symmetrized them using the FINDSYM utility \cite{Stokes2005}. We then performed a transformation of the structure description into the standard setting of the corresponding LG. Finally, in those LGs whose standard setting is c-centered, we transformed the structure description to a primitive unit cell defined by {\bf a}-{\bf b}, {\bf a}+{\bf b} and {\bf c} unit cell vectors.

\subsection{High-throughput calculations}
\addcontentsline{stoc}{subsection}{\protect\numberline{\thesubsection}High-throughput calculations}

Density functional theory (DFT) calculations were performed using VASP code \cite{vasp1,vasp2}, version 5.4.4. Interactions of electrons with ion cores were represented using projected atomic wave (PAW) formalism \cite{blochl1994paw}
, and the exchange-correlation functional was parameterized with the General Gradient approximation (GGA) 
\cite{perdew1996pbe}. The POTCAR files were built using PBE PAW datasets v.54.
The plane wave cutoff (ENCUT) in VASP was chosen to be 30 \% larger than the maximal cutoff (ENMAX) among the elements in the POTCAR file, and the PREC=accurate mode was set.
For the high-throughput part, we performed eight types of calculations per compound: a self-consistent field (scf) run, a run to obtaining wave functions at maximal $\mathbf{k}$ points, a density of states (DOS) calculation, and a band structure calculation for systems both with and without spin-orbit coupling (SOC).
For the scf run, used to obtain the charge density of the system, we set a $\mathbf{k}$ point grid such that $n_{k_i}=\frac{75}{a_i}$, where $n_{k_i}$ is the $\mathbf{k}$ point amount in $i$ direction and $a_i$ is the unit cell length in $i$ direction in \r{A}. This way, we kept the $\mathbf{k}$ grid density constant in all compounds. For the DOS calculation, we typically increased the $\mathbf{k}$ point density by 50 \%, except for compounds with many electrons, where we kept the scf settings. For band plots, we used 80 $\mathbf{k}$ points at each segment separating maximal $k$ points in the path.
All POSCAR files used for calculations are available in the \webtwoDTQCAbbr\; database.

For finite-size ribbon calculations, we kept the same $\mathbf{k}$ point density mentioned above in the periodic direction, and a single $\mathbf{k}$ point in the non-periodic directions. We did not include spin-orbit coupling effects for the relaxation, and set a criterion of 0.01 eV/angstrom for the largest allowed atomic force. The relaxation was carried out by setting ISIF=8 (fixed cell-shape, variable positions and volume) in VASP, and only relaxing the lattice along the periodic direction. The DFT-D2 method of Grimme~\cite{grimme2006semiempirical} is used as the Van der Waals (vdW) correction in the relaxation. 
For band structure calculations, spin-orbit coupling was taken into account. At least 15 \r{A} of vacuum was included in the non-periodic directions to avoid spurious interactions due to periodic boundary conditions. 

We employed the software VESTA \cite{vesta}, ASE \cite{HjorthLarsen2017}, pymatgen \cite{pymatgen}, and pyprocar \cite{pyprocar} for structure visualization and manipulation, DFT post-processing, and projected band structure plotting of the finite ribbons, respectively.

\subsection{Treatment of materials containing atoms with a negligible magnetization}
\addcontentsline{stoc}{subsection}{\protect\numberline{\thesubsection}Treatment of materials containing atoms with a negligible magnetization}

In VASP, the direction of the magnetization in a material can be constrained only partially, through the addition of an energy penalty if the magnetization deviates from a user-specified direction and/or value. However, to date, magnetic symmetry cannot be absolutely enforced in calculations. Hence, VASP cannot restrict the magnetic moments of atoms to zero when spin-orbit coupling is included, and residual magnetizations emerge in most materials, due to the finite threshold value used for energy convergence.

In the literature, it is standard practice to consider materials where the modulus of the largest magnetic moment of any atom ($m$) is smaller than 0.1$\mu_B$ where $\mu_B$ is the Bohr magneton, as non-magnetic \cite{Gjerding_2021}. 

Thus, although a small $m$ in VASP does not imply that the material is magnetic, it is still present in the calculation. This, through spin-orbit coupling, affects the eigenvalue spectrum of the material in such a way that the spectrum respects the magnetic symmetry of the material, not necessarily the space group symmetry. When this happens, the degeneracies that would be expected from the irreps of the space group can appear to be broken in the calculation. The value of this (artificial) degeneracy-breaking splitting is material-dependent and is correlated, but not totally determined, by the value of $m$. In this work, we have set an arbitrary threshold of 0.015 eV for this artificial splitting and discarded all materials exceeding it.

\subsection{Wannierization and tight-binding models}
\addcontentsline{stoc}{subsection}{\protect\numberline{\thesubsection}Wannierization and tight-binding models}

For the four compounds whose edge state study is presented in the main text, Bi$_2$Br$_2$ in \lgsymbnum{15} (2D-TQC \serialidweb{1.3.6}), Re$_4$Se$_8$ in \lgsymbnum{2} (2D-TQC \serialidweb{3.1.1}), C$_6$N$_2$, in \lgsymbnum{80} (2D-TQC \serialidweb{3.1.48}), and As2Cl2 in  \lgsymbnum{72} (2D-TQC \serialidweb{1.3.193}), we first did a wannierization, using Wannier90 code \cite{Pizzi2020}. We used spinor-Wannier functions to account for spin-orbit coupling in both cases. For Bi$_2$Br$_2$ we constructed a tight-binding (TB) model in the space of $p$ orbitals of Bi and Br. For Re$_4$Se$_8$, we built a TB model with $d$ orbitals of Re and $p$ orbitals of Se. For C$_6$N$_2$ we used $s$ and $p$ orbitals of both C and N atoms in order to build the TB model. Finally, for As2Cl2, we used $p$ orbitals of As and Cl.
For the spectral function calculations on the models of each of the two materials, a number of principal layers of 3 was chosen, after the pertinent convergence tests were performed.

\section{Overview of the topological quantum chemistry workflow}\label{tqc-summ}
\addcontentsline{stoc}{section}{\protect\numberline{\thesection}Overview of the topological quantum chemistry workflow}

Topological Quantum Chemistry (TQC) is a formalism that allows a symmetry-based classification of the band topology in crystals. In essence, the formalism relies on the so-called elementary band representations (EBRs), which are Fourier-transformed irreps of site symmetry groups. 

Detailed descriptions of the TQC approach are provided in Refs. \cite{bradlyn2017topological,Vergniory2019,xu2021three}. In this work we have first adapted the TQC formalism to layer groups, as detailed in \cref{adap}. To do this, we have written a series of new Bilbao Crystallographic Server programs, listed in the main text and described in detail in \cref{adap}. In our workflow, we have integrated all these programs. The workflow is described in detail below.

We first calculated the wave functions (WFs) at high symmetry (maximal) $\mathbf{k}$ points of the Brillouin Zone (BZ) using the DFT code VASP \cite{vasp1,vasp2} (see \cref{comp} in the \siref\, for details of the computational parameters of VASP calculations). Under a symmetry operation of the little group of $\mathbf{k}$, a set of degenerate Bloch eigenstates transforms into itself, i.e., the symmetry operation transforms a given eigenstate into a linear combination of eigenstates with the same energy. We can calculate, thus, a square matrix that represents the way this subset of states transforms under the given symmetry operation. The matrices calculated for every operation of the little group of $\mathbf{k}$ form a representation of the little group. We have used the program Vasp2trace \cite{Vergniory2019, Vergniory2022} to calculate, for every symmetry operation of the little group, the traces of the matrices. The list of traces of all the calculated bands at every maximal $\mathbf{k}$ are introduced into the program \CheckT\  to identify the representations making use of the previously tabulated irreps of the little groups in the database \RepresentationsDLSG. 

In the next step, the program uses the number $N_e$ of electrons in the primitive unit cell to check whether these electrons fully occupy subsets of degenerate states at every maximal $\mathbf{k}$ point. If at some maximal $\mathbf{k}$ points a set of degenerate states is partially occupied, we find three different cases: (a) the partially occupied degenerate states correspond to a single irrep, i.e., the whole set of degenerate states transform as a single irrep at every $\mathbf{k}$ point with partially occupied sets. The compound is tagged as ESFD. (b) The sets of degenerate states transform, at every $\mathbf{k}$ point, as the direct sum of two or more irreps. The compound is tagged as \emph{AccidentalDegeneracy}. (c) We find a mixed result: at some $\mathbf{k}$ point(s) we have the (a) case and at other $\mathbf{k}$ point(s) we have the (b) case. These compounds are tagged as \emph{ESFDplusAccidental} because the compound is undoubtedly ESFD, but the band structure close to the Fermi level has accidental degeneracies. If the program finds that there are no partially occupied degenerate states in any maximal $\mathbf{k}$ point, it constructs the so-called symmetry data vector (SDV) that contains the multiplicities of every irrep at every maximal $\mathbf{k}$ point in the decomposition of the complete set of occupied bands at $\mathbf{k}$.
Making use of the compatibility relations gathered in \DLCOMPREL, the program checks whether the sets of $N_e$ eigenstates at every pair of maximal $\mathbf{k}$ points can be interconnected, i.e., the subduction of the whole set of irreps into the common line that joins the two $\mathbf{k}$ points and whose little group is a subgroup of the little groups of both $\mathbf{k}$ points, gives as a result exactly the same multiplicities for all the irreps in the line. If for some pair of $\mathbf{k}$ points the multiplicities are not the same, bands at each point must be connected with bands at the other point but located in de conduction band. The program tags this compound as enforced semimetal (ES). In other words, ESFD emerges when the sum of irrep dimensions below the Fermi level at high-symmetry $\mathbf{k}$ points fails to match the electron count of the system, necessarily resulting in Fermi level crossings at maximal $\mathbf{k}$ points. In contrast, ES exhibit an enforced band crossing along intermediate $\mathbf{k}$ points or lines due to the specific energy ordering of irreducible representations at maximal k-points that breaks the compatibility relations.

If all the pairs of multiplicities match, there is a gap between the first $N_e$ (valence) bands and the (conduction) rest of bands. The compound has the band structure of an insulator. In this case, the program tries to perform
a linear decomposition of the symmetry data vector into EBRs. If a linear combination of EBRs with positive integers equates to the symmetry data vector, then the material is labeled at this stage as a trivial insulator (LCEBR); If the SDV can be expressed as an integer combination of EBRs with at least one negative coefficient, the material is considered to have fragile topology (FRAGILE).

When all the previous options have been discarded, the compound is a topological insulator. In the next step the program tries to express the whole set of bands as integer linear combination of EBRs and parts of decomposable EBRs, if there are such EBRs in the LG. When this decomposition is possible, the compound is tagged as Split EBR (SEBR). Stanene and silicene are a prototypes of this type of material \cite{stanenePRL,liu2011,yao2011}. 
Finally, if the symmetry data vector cannot be expressed as a linear combination of EBRs, or split parts of decomposable EBRs, the material is also a topological insulator and is classified as NLC. A notable example of this is monolayer 1T'-WTe$_2$ \cite{Tang2017} in \lgsymbnum{15} (2D-TQC \serialidweb{1.1.6}). The \CheckT\ program provides the values for topological indices that characterize the topological class of the material, and in the new \LTOPOINDICES\  program we have tabulated all possible independent topological indices of all LGs. In \cref{topoindices} of the \siref, we provide an overview of the derivation of these topological indices and their physical implications.

In addition, an accidental Fermi degeneracy can occur. In this type of system, a small perturbation (or another DFT pseudopotential choice) will lift the accidental degeneracy between distinct irreps at the Fermi level, pushing it into one of the main topological categories. It can also happen that a material has a band gap that is too small for DFT to distinguish it, so the irreps appear to be accidentally degenerate.

In LCEBR materials, we account for obstructed and orbital-selected obstructed atomic insulators (OAI and OOAI, respectively). For this, we first extract the atomic Wyckoff positions from cif files and the populated orbitals of elements in their atomic configurations from the POTCAR file of VASP. In an OAI, at least one Wannier center exists at an empty site that cannot be adiabatically moved to an atomic site without breaking the crystal symmetry. The OAIs are identified through the RSIs (see Appendix C of the Supplementary Information in Ref. \cite{xu2021three}). In LGs, the derivation of the RSI is again analogous to that in space groups. We consider an orbital at a Wyckoff position $W$ with site-symmetry group (SSG) $\mathcal{G}^W$, which transforms as an irrep of $\mathcal{G}^W$. If this orbital can be adiabatically moved through a Wyckoff position $w$ with SSG $\mathcal{G}^w<\mathcal{G}^W$ while preserving the symmetry of the system, it implies that the irrep of $\mathcal{G}^W$ can be induced from irreps of $\mathcal{G}^w$. Based on this principle, we can establish conditions for any physical orbital located at any Wyckoff position (of any space group) to remain pinned.
We have adapted the procedure developed in Ref. \cite{xu2021three} for SG, to layer group and obtained all RSI for all Wyckoff position of all (double) LGs. They are tabulated in the new \webBCSAbbr\  program \RSI. A material is classified as an OAI if there exists at least a non-vanishing RSI at one empty Wyckoff position and, moreover, all other Wyckoff positions of higher symmetry connected to it are also empty. If all non-zero RSIs correspond to occupied Wyckoff positions (or to Wyckoff positions connected to occupied Wyckoff positions of higher symmetry), the program finally checks whether the whole set of bands described by the SDV can be expressed as a linear integer combination of BRs induced from orbitals in the outer shell of the atoms in the corresponding Wyckoff positions. If this set of linear diophantine equations has no solution, the compound is tagged as OOAI. Otherwise it remains LCEBR.\color{black}.

\section{Adaptation of TQC to layer groups: tools implemented in the Bilbao Crystallographic Server}\label{adap}
\addcontentsline{stoc}{section}{\protect\numberline{\thesection}Adaptation of TQC to layer groups: tools implemented in the Bilbao Crystallographic Server}

The analysis of the electronic bands and their classification according to the topological properties under the TQC approach (\cite{bradlyn2017topological,elcoro2017,cano2018,vergniory2017}) has required the adaptation of the entire TQC machinery to layer groups. The tools adapted or developed include the tabulation of the irreducible representations of the layer groups with and without time-reversal symmetry, the tabulation of the compatibility relations, the tabulation of the band representations induced from the irreps of the site-symmetry group of every Wyckoff position (and the identification of the elementary band representations (EBRs)). Finally, once these tables are available, we have written a code to check the connectivity of the electronic bands and, depending on this connectivity, the program identifies the topological features of a given band structure. In the next sections we describe the databases and tools developed for this work.

\subsection{Symmetry operations, Wyckoff positions and {\bf k}-vectors in the layer groups}
\addcontentsline{stoc}{subsection}{\protect\numberline{\thesubsection}Symmetry operations, Wyckoff positions and {\bf k}-vectors in the layer groups}

\label{sec:wyckoff}
To construct the tables of irreducible representations, compatibility relations and the band representations of the 80 layer groups we have adopted the standard setting defined in the International Tables for Crystallography, vol. E \cite{ite2010}, abbreviated as ITE in the following. For most layer groups, the ITE tables include both the list of symmetry operations and the list of Wyckoff positions in a single standard basis but, for some groups, they include two different possible choices of the unit cell (\emph{cell choice 1 or 2} or two different choices for the origin of the unit cell (\emph{origin choice 1 or 2}). Recently, these tables have been implemented in the Bilbao Crystallographic Server (BCS) \cite{delaflor2021}, where it has been assumed a single standard description for every group. For the two layer groups with possible different cell choices (layer groups N. 5 and 7) it has been selected \emph{cell choice 1}: $p11a$ and $p112/a$, respectively, instead of the alternative $p11b$ and $p112/b$ (\emph{cell choice 2}). The origin choice 1 or 2 makes reference to the two possible choices of the origin in those layer (or space) groups which, together with an inversion center, they also include a roto-inversion of type $\overline{3}$, $\overline{4}$ or $\overline{6}$ whose center (single point that remains invariant under such roto-inversion) does not coincide with an inversion center. There are only three such layer groups, $p4/n$ (N. 52), $p4/nbm$ (N. 62) and $p4/nmm$ (N. 64). In those cases the \emph{origin choice 2}, i.e., origin at the inversion center, has been assumed.
We have adopted these choices for the development of the tables or representations and of the band representations. The symmetry operations and lists of Wyckoff positions used in this work have been thus obtained from the tools GENPOS and WYCKPOS, respectively, hosted in the section \emph{Subperiodic Groups: Layer, Rod and Frieze Groups} in the BCS (\href{https://cryst.ehu.es}{cryst.ehu.es}) \cite{delaflor2021}.

The 80 layer group types are 3-dimensional groups with 2-dimensional translations. In the standard or conventional setting defined in ITE the translations are parallel to the $a_1a_2$ plane. The point group of a layer group is a 3-dimensional point group whose operations, on the one hand, keep invariant the plane $a_1a_2$ and, on the other hand, keep invariant or invert the direction perpendicular to the plane of translations. This condition restricts the possible point group of a layer group to one of the non-cubic 27 crystallographic point group types. Although the crystallographic basis of a layer group includes three basis vectors, denoted as ${\bf a}_i$, $i=1,2,3$, only the first two are lattice vectors and they define the 2-dimensional periodicity of the system. In general, we will denote the symmetry operations of a layer group as $\{R|t_1t_20\}$. The rotational part $R$ has a block diagonal form with blocks of dimensions 2 and 1 with $R_{33}=\pm1$.

In principle one could construct the tables of Wyckoff positions, {\bf k}-vectors and irreducible representations for each layer group from scratch, following exactly the same procedure used in the determination of these magnitudes in the 3-dimensional space groups (see for instance the detailed explanation of the procedure in Ref.~\cite{aroyo2006} for single groups and Ref.~\cite{elcoro2017} for double groups). However, the calculation can be simplified due to the existence, for each layer group  $\mathcal{L}$, of a space group $\mathcal{G}$ which can be expressed as the semi-direct product of $\mathcal{L}$ and $\mathcal{T}_3$,
\begin{equation}
	\label{groupG}
	\mathcal{G}=\mathcal{L}\rtimes\mathcal{T}_3
\end{equation}
where $\mathcal{T}_3$ is the 1-dimensional translation group whose elements are $\{E|0,0,n\}$ with $n\in\mathbb{Z}$.

The space group type $\mathcal{G}$ is unique for each layer group type $\mathcal{L}$. Therefore, we can make use of the available tabulated magnitudes (Wyckoff positions, {\bf k}-vectors, irreducible representations,\ldots) of $\mathcal{G}$ to obtain the corresponding magnitudes of $\mathcal{L}$. In the following we refer to $\mathcal{G}$ as the \emph{reference} space group type of $\mathcal{L}$. It is important to note that, although $\mathcal{G}$ is unique for every $\mathcal{L}$, different layer group types $\mathcal{L}_1$ and $\mathcal{L}_2$ can have the same reference group $\mathcal{G}$. For instance, the layer group types $p112$ (N. 3) and $p211$ (N. 8) have as reference group the space group type $P2$ (N. 3). In the first case the binary ($C_2$) axis is perpendicular to the 2-dimensional lattice of the layer group and in the second case, it is parallel to the lattice. Due to the different orientations of the binary axis with respect to the lattice, these layer groups must be considered as different layer group types. This example also helps to realize that, together with the reference space group $\mathcal{G}$ that satisfies the coset decomposition (\cref{groupG}), it is necessary to provide a transformation matrix that transforms the symmetry operations of $\mathcal{G}$ expressed in its standard setting into the symmetry operations of $\mathcal{L}$ expressed in its own standard setting to completely identify a unique layer group. In general, this transformation matrix contains a change of the orientation of the crystallographic axes and/or an origin shift. It is obvious that, in our example, although the reference space group type for the layer groups $p112$ (N. 3) and $p211$ (N. 8) is the same, the transformation matrix is different in both cases. \cref{referenceG} gives the reference space group type for each layer group type and a transformation matrix. It is worth noting that only some monoclinic and orthorhombic layer groups have a transformation matrix whose rotational part is different from the identity and that only one layer group (N. 54) has an origin shift different from 0. Finally, it is also important to stress that the transformation matrix for each ($\mathcal{G},\mathcal{L}$) pair is not unique and, therefore, the matrices in \cref{referenceG} represent a particular choice. Any other transformation matrix obtained as the product of the matrix in \cref{referenceG} and a member of the normalizer of $\mathcal{L}$ is also a valid transformation matrix for the pair ($\mathcal{G},\mathcal{L})$. 

The rotational part $P$ of the transformation matrix $(P,\mathbf{p})$ relates the crystallographic basis $(\mathbf{a}^L,\mathbf{b}^L,\mathbf{c}^L)$ of the layer group in its standard setting with the crystallographic basis $\mathbf{a}_i$ of the corresponding space group in its standard setting,
\begin{equation}
	\mathbf{a}_i^L=\mathbf{a}_j\cdot P_{ji}
\end{equation}
and $\mathbf{p}$ is the origin shift (set of three coordinates after the semicolon in the third and sixth column of  \cref{referenceG}.

\begin{table}
\centering
\begin{tabular}{|lll|lll|}
    \hline
    LG ($\mathcal{L}$)&SG ($\mathcal{G}$)&Tr. Matrix&LG ($\mathcal{L}$)&SG ($\mathcal{G}$)&Tr. Matrix\\
    \hline
    $p1$\,(N.1)&$P1$\,(N.1)&\textbf{a},\textbf{b},\textbf{c};0,0,0&$p\bar{1}$\,(N.2)&$P\bar{1}$\,(N.2)&\textbf{a},\textbf{b},\textbf{c};0,0,0\\
$p112$\,(N.3)&$P2$\,(N.3)&\textbf{a},\textbf{c},-\textbf{b};0,0,0&$p11m$\,(N.4)&$Pm$\,(N.6)&\textbf{a},\textbf{c},-\textbf{b};0,0,0\\
$p11a$\,(N.5)&$Pc$\,(N.7)&\textbf{c},-\textbf{a},-\textbf{b};0,0,0&$p112/m$\,(N.6)&$P2/m$\,(N.10)&\textbf{a},\textbf{c},-\textbf{b};0,0,0\\
$p112/a$\,(N.7)&$P2/c$\,(N.13)&\textbf{c},-\textbf{a},-\textbf{b};0,0,0&$p211$\,(N.8)&$P2$\,(N.3)&\textbf{b},-\textbf{a},\textbf{c};0,0,0\\
$p2_1$\,(N.9)&$P2_1$\,(N.4)&\textbf{b},-\textbf{a},\textbf{c};0,0,0&$c211$\,(N.10)&$C2$\,(N.5)&\textbf{b},-\textbf{a},\textbf{c};0,0,0\\
$pm11$\,(N.11)&$Pm$\,(N.6)&\textbf{b},-\textbf{a},\textbf{c};0,0,0&$pb11$\,(N.12)&$Pc$\,(N.7)&\textbf{b},-\textbf{c},-\textbf{a};0,0,0\\
$cm11$\,(N.13)&$Cm$\,(N.8)&\textbf{b},-\textbf{a},\textbf{c};0,0,0&$p2/m11$\,(N.14)&$P2/m$\,(N.10)&\textbf{b},-\textbf{a},\textbf{c};0,0,0\\
$p2_1/m11$\,(N.15)&$P2_1/m$\,(N.11)&\textbf{b},-\textbf{a},\textbf{c};0,0,0&$p2/b11$\,(N.16)&$P2/c$\,(N.13)&\textbf{b},-\textbf{c},-\textbf{a};0,0,0\\
$p2_1/b11$\,(N.17)&$P2_1/c$\,(N.14)&\textbf{b},-\textbf{c},-\textbf{a};0,0,0&$c2/m11$\,(N.18)&$C2/m$\,(N.12)&\textbf{b},-\textbf{a},\textbf{c};0,0,0\\
$p222$\,(N.19)&$P222$\,(N.16)&\textbf{a},\textbf{b},\textbf{c};0,0,0&$p2_122$\,(N.20)&$P222_1$\,(N.17)&\textbf{c},\textbf{b},-\textbf{a};0,0,0\\
$p2_12_12$\,(N.21)&$P2_12_12$\,(N.18)&\textbf{a},\textbf{b},\textbf{c};0,0,0&$c222$\,(N.22)&$C222$\,(N.21)&\textbf{a},\textbf{b},\textbf{c};0,0,0\\
$pmm2$\,(N.23)&$Pmm2$\,(N.25)&\textbf{a},\textbf{b},\textbf{c};0,0,0&$pma2$\,(N.24)&$Pma2$\,(N.28)&\textbf{a},\textbf{b},\textbf{c};0,0,0\\
$pba2$\,(N.25)&$Pba2$\,(N.32)&\textbf{a},\textbf{b},\textbf{c};0,0,0&$cmm2$\,(N.26)&$Cmm2$\,(N.35)&\textbf{a},\textbf{b},\textbf{c};0,0,0\\
$pm2m$\,(N.27)&$Pmm2$\,(N.25)&\textbf{a},\textbf{c},-\textbf{b};0,0,0&$pm2_1b$\,(N.28)&$Pmc2_1$\,(N.26)&\textbf{a},\textbf{c},-\textbf{b};0,0,0\\
$pb2_1m$\,(N.29)&$Pmc2_1$\,(N.26)&-\textbf{b},\textbf{c},-\textbf{a};0,0,0&$pb2b$\,(N.30)&$Pcc2$\,(N.27)&\textbf{a},\textbf{c},-\textbf{b};0,0,0\\
$pm2a$\,(N.31)&$Pma2$\,(N.28)&\textbf{a},\textbf{c},-\textbf{b};0,0,0&$pm2_1n$\,(N.32)&$Pmn2_1$\,(N.31)&\textbf{a},\textbf{c},-\textbf{b};0,0,0\\
$pb2_1a$\,(N.33)&$Pca2_1$\,(N.29)&\textbf{a},\textbf{c},-\textbf{b};0,0,0&$pb2n$\,(N.34)&$Pnc2$\,(N.30)&-\textbf{b},\textbf{c},-\textbf{a};0,0,0\\
$cm2m$\,(N.35)&$Amm2$\,(N.38)&\textbf{b},\textbf{c},\textbf{a};0,0,0&$cm2e$\,(N.36)&$Aem2$\,(N.39)&\textbf{b},\textbf{c},\textbf{a};0,0,0\\
$pmmm$\,(N.37)&$Pmmm$\,(N.47)&\textbf{a},\textbf{b},\textbf{c};0,0,0&$pmaa$\,(N.38)&$Pccm$\,(N.49)&-\textbf{c},\textbf{b},\textbf{a};0,0,0\\
$pban$\,(N.39)&$Pban$\,(N.50)&\textbf{a},\textbf{b},\textbf{c};0,0,0&$pmam$\,(N.40)&$Pmma$\,(N.51)&\textbf{a},-\textbf{c},\textbf{b};0,0,0\\
$pmma$\,(N.41)&$Pmma$\,(N.51)&\textbf{a},\textbf{b},\textbf{c};0,0,0&$pman$\,(N.42)&$Pmna$\,(N.53)&\textbf{a},-\textbf{c},\textbf{b};0,0,0\\
$pbaa$\,(N.43)&$Pcca$\,(N.54)&-\textbf{c},-\textbf{a},\textbf{b};0,0,0&$pbam$\,(N.44)&$Pbam$\,(N.55)&\textbf{a},\textbf{b},\textbf{c};0,0,0\\
$pbma$\,(N.45)&$Pbcm$\,(N.57)&\textbf{b},\textbf{c},\textbf{a};0,0,0&$pmmn$\,(N.46)&$Pmmn$\,(N.59)&\textbf{a},\textbf{b},\textbf{c};0,0,0\\
$cmmm$\,(N.47)&$Cmmm$\,(N.65)&\textbf{a},\textbf{b},\textbf{c};0,0,0&$cmme$\,(N.48)&$Cmme$\,(N.67)&\textbf{a},\textbf{b},\textbf{c};0,0,0\\
$p4$\,(N.49)&$P4$\,(N.75)&\textbf{a},\textbf{b},\textbf{c};0,0,0&$p\bar{4}$\,(N.50)&$P\bar{4}$\,(N.81)&\textbf{a},\textbf{b},\textbf{c};0,0,0\\
$p4/m$\,(N.51)&$P4/m$\,(N.83)&\textbf{a},\textbf{b},\textbf{c};0,0,0&$p4/n$\,(N.52)&$P4/n$\,(N.85)&\textbf{a},\textbf{b},\textbf{c};0,0,0\\
$p422$\,(N.53)&$P422$\,(N.89)&\textbf{a},\textbf{b},\textbf{c};0,0,0&$p42_12$\,(N.54)&$P42_12$\,(N.90)&\textbf{a},\textbf{b},\textbf{c};0,1/2,0\\
$p4mm$\,(N.55)&$P4mm$\,(N.99)&\textbf{a},\textbf{b},\textbf{c};0,0,0&$p4bm$\,(N.56)&$P4bm$\,(N.100)&\textbf{a},\textbf{b},\textbf{c};0,0,0\\
$p\bar{4}2m$\,(N.57)&$P\bar{4}2m$\,(N.111)&\textbf{a},\textbf{b},\textbf{c};0,0,0&$p\bar{4}2_1m$\,(N.58)&$P\bar{4}2_1m$\,(N.113)&\textbf{a},\textbf{b},\textbf{c};0,0,0\\
$p\bar{4}m2$\,(N.59)&$P\bar{4}m2$\,(N.115)&\textbf{a},\textbf{b},\textbf{c};0,0,0&$p\bar{4}b2$\,(N.60)&$P\bar{4}b2$\,(N.117)&\textbf{a},\textbf{b},\textbf{c};0,0,0\\
$p4/mmm$\,(N.61)&$P4/mmm$\,(N.123)&\textbf{a},\textbf{b},\textbf{c};0,0,0&$p4/nbm$\,(N.62)&$P4/nbm$\,(N.125)&\textbf{a},\textbf{b},\textbf{c};0,0,0\\
$p4/mbm$\,(N.63)&$P4/mbm$\,(N.127)&\textbf{a},\textbf{b},\textbf{c};0,0,0&$p4/nmm$\,(N.64)&$P4/nmm$\,(N.129)&\textbf{a},\textbf{b},\textbf{c};0,0,0\\
$p3$\,(N.65)&$P3$\,(N.143)&\textbf{a},\textbf{b},\textbf{c};0,0,0&$p\bar{3}$\,(N.66)&$P\bar{3}$\,(N.147)&\textbf{a},\textbf{b},\textbf{c};0,0,0\\
$p312$\,(N.67)&$P312$\,(N.149)&\textbf{a},\textbf{b},\textbf{c};0,0,0&$p321$\,(N.68)&$P321$\,(N.150)&\textbf{a},\textbf{b},\textbf{c};0,0,0\\
$p3m1$\,(N.69)&$P3m1$\,(N.156)&\textbf{a},\textbf{b},\textbf{c};0,0,0&$p31m$\,(N.70)&$P31m$\,(N.157)&\textbf{a},\textbf{b},\textbf{c};0,0,0\\
$p\bar{3}1m$\,(N.71)&$P\bar{3}1m$\,(N.162)&\textbf{a},\textbf{b},\textbf{c};0,0,0&$p\bar{3}m1$\,(N.72)&$P\bar{3}m1$\,(N.164)&\textbf{a},\textbf{b},\textbf{c};0,0,0\\
$p6$\,(N.73)&$P6$\,(N.168)&\textbf{a},\textbf{b},\textbf{c};0,0,0&$p\bar{6}$\,(N.74)&$P\bar{6}$\,(N.174)&\textbf{a},\textbf{b},\textbf{c};0,0,0\\
$p6/m$\,(N.75)&$P6/m$\,(N.175)&\textbf{a},\textbf{b},\textbf{c};0,0,0&$p622$\,(N.76)&$P622$\,(N.177)&\textbf{a},\textbf{b},\textbf{c};0,0,0\\
$p6mm$\,(N.77)&$P6mm$\,(N.183)&\textbf{a},\textbf{b},\textbf{c};0,0,0&$p\bar{6}m2$\,(N.78)&$P\bar{6}m2$\,(N.187)&\textbf{a},\textbf{b},\textbf{c};0,0,0\\
$p\bar{6}2m$\,(N.79)&$P\bar{6}2m$\,(N.189)&\textbf{a},\textbf{b},\textbf{c};0,0,0&$p6/mmm$\,(N.80)&$P6/mmm$\,(N.191)&\textbf{a},\textbf{b},\textbf{c};0,0,0\\
    \hline
\end{tabular}
\caption{\label{referenceG} Reference space group (SG) type (columns 2 and 5) for each layer group (LG) type (columns 1 and 4). The third and sixth columns show the transformation matrix $(P,\mathbf{p})$ that relates the two sets of symmetry operations expressed in the corresponding standard or conventional setting.}
\end{table}

Once the reference space group and the transformation matrix have been identified, we transform the operations $\{R_{G}|{\bf t}_G\}$ of the space group using the transformation matrix,
\begin{equation}\label{transf}
	\{R|{\bf t}\}\equiv\{R_L|{\bf t}_L\}=\{P^{-1}R_GP|P^{-1}\cdot({\bf t}_G-{\bf p}+R_G{\bf p})\}
\end{equation}
The symmetry operations of the layer group are those operations $\{R|{\bf t}\}$ in \cref{transf} with $t_3=0$. The list of operations for each layer group can be found at the BCS, (\href{https://cryst.ehu.es/subperiodic/get_sub_gen.html}{cryst.ehu.es/subperiodic/get\_sub\_gen.html}) \cite{delaflor2021}. 

As an example of application of  \cref{referenceG} we will derive the symmetry operations of the layer groups $p112$ (N. 3) and $p211$ (N. 8) from the symmetry operations of the common reference space group $P2$ (N. 3).

In the first case, $p112$ (N. 3), \cref{referenceG} shows that the transformation matrix is ${\bf a},{\bf c},-{\bf b};0,0,0$, so that,
\begin{equation}\label{pmatrix1}
	P=\left(\begin{array}{rrr}
		1&0&0\\0&0&-1\\0&1&0
		\end{array}
	\right)\hspace{1cm}\textrm{and}\hspace{1cm}{\bf p}=(0,0,0)
\end{equation}
in \cref{transf}. Together with the three symmetry operations that generate the lattice, we can take as generator of the space group $P2$ (N. 3) the 2-fold axis $\{R_G|{\bf t_G}\}=\{2_{010}|0,0,0\}$ with
\begin{equation}\label{twofoldaxis}
2_{010}=\left(\begin{array}{rrr}-1&0&0\\0&1&0\\0&0&-1\end{array}\right)
\end{equation}
According to \cref{transf}, \cref{pmatrix1} and \cref{twofoldaxis} the 2-fold axis transforms into,
\begin{equation}
	\{R_G|{\bf t_G}\}=\{2_{010}|0,0,0\}\rightarrow\{R_L|{\bf t_L}\}=\{2_{001}|0,0,0\}
\end{equation}
The 2-fold axis in layer group $p112$ (N. 3) is parallel to ${\bf a}_3$ and perpendicular to the $a_1a_2$ plane.

In the second case, $p211$ (N. 8), according to \cref{referenceG}, the transformation matrix is ${\bf b},-{\bf a},{\bf c};0,0,0$, so that,
\begin{equation}\label{pmatrix2}
	P=\left(\begin{array}{rrr}
		0&-1&0\\1&0&0\\0&0&1
	\end{array}
	\right)\hspace{1cm}\textrm{and}\hspace{1cm}{\bf p}=(0,0,0)
\end{equation}
The 2-fold axis $\{R_G|{\bf t_G}\}=\{2_{010}|0,0,0\}$ of space group $P2$ (N. 3) transforms into,
\begin{equation}
	\{R_G|{\bf t_G}\}=\{2_{010}|0,0,0\}\rightarrow\{R_L|{\bf t_L}\}=\{2_{100}|0,0,0\}
\end{equation}
The 2-fold axis in layer group $p211$ (N. 8) is parallel to ${\bf a}_1$.

The list of Wyckoff positions of the layer group can also be obtained from the list of Wyckoff positions of the reference group. If ${\bf w}_G=(w_1,w_2,w_3)$ are the coordinates of a representative of a  Wyckoff position in the standard setting of the space group $\mathcal{G}$, its coordinates in the standard setting of the layer group $\mathcal{L}$ are,
\begin{equation}\label{wyckoff}
	{\bf w}_L=P^{-1}\cdot({\bf w}_G-{\bf p})
\end{equation}
The relation given by \cref{wyckoff} can be applied to every set of coordinates of every Wyckoff position of the reference space group $\mathcal{G}$ to get the corresponding coordinates in the setting of the layer group. However, only those Wyckoff positions that intersect the $z=0$ plane are meaningful in the layer group. Therefore, from the sets of Wyckoff positions of \cref{wyckoff}, those that have a fixed $z$ value different from 0 must be rejected. Finally, the labels of the allowed set of Wyckoff positions were assigned according to the standard definitions in ITE \cite{ite2010} and implemented afterward in the BCS (\href{www.cryst.ehu.es/subperiodic/get_sub_wp.html}{ cryst.ehu.es/subperiodic/get\_sub\_wp.html}) by de la Flor \emph{et al.} \cite{delaflor2021}. The site-symmetry group (SSG) type of every Wyckoff position of the layer group (and the point group isomorphic with the SSG) is exactly the same as the SSG type of the corresponding Wyckoff position in the reference group. Therefore, the set of irreps of the SSG is also exactly the same. This relation will allow us to simplify the calculations of the band representations in \cref{sec:lbandrep}.

In the reciprocal space, it is possible to follow the same procedure explained above for the Wyckoff positions to get the list of {\bf k}-vectors in the first Brillouin zone. Due to the lack of periodicity along ${\bf a}_3$, the reciprocal space of a layer group is a 2-dimensional space spanned by ${\bf a}_1^{*}$ and ${\bf a}_2^{*}$, ${\bf k}=k_1{\bf a}_1^{*}+k_2{\bf a}_2^{*}$, where the three basis vectors of the reciprocal space ${\bf a}_i^{*}$ are defined as ${\bf a}_i\cdot{\bf a}_j^{*}=\delta_{ij}$. If ${\bf k}=(k_1,k_2,k_3)$ are the coordinates in the standard (reciprocal) basis  ${\bf a}_ i^{*}$ of a representative of a {\bf k}-vector of the space group $\mathcal{G}$, its coordinates in the standard setting of the layer group $\mathcal{L}$ are,
\begin{equation}\label{kvecs}
	(k_1^L,k_2^L,k_3^L)^T=P^{T}\cdot(k_1,k_2,k_3)^T
\end{equation}
The relation given by \cref{kvecs} can be applied to every set of coordinates of every (already tabulated) {\bf k}-vector in the reference group $\mathcal{G}$, but only those vectors for which $k_3^L=0$ will correspond to an allowed {\bf k}-vector in the layer group, i.e., we can establish an isomorphism between the {\bf k}-vector types of the layer group $\mathcal{L}$ and the subset of {\bf k}-vectors types in the reference group $\mathcal{G}$ that fulfill the condition $k_3^L=0$ in \cref{kvecs}. The tables of {\bf k}-vector types for the layer groups have been recently implemented by de la Flor \emph{et al.}\cite{delaflor2021} in the BCS (program LKVEC, \href{https://cryst.ehu.es/subperiodic/get_layer_kvec.html}{www.cryst.ehu.es/subperiodic/get\_layer\_kvec.html}) following the tabulation and label assignation made by Litvin and Wike \cite{litvin1991}.

Finally, it is important to note, on the one hand, that the little co-group and the little group $\mathcal{G}^{\bf k}$ of a {\bf k}-vector type of the layer group are isomorphic with the little co-group and little group, respectively, of the corresponding {\bf k}-vector type in the reference space group through \cref{kvecs} with $k_3^L=0$. On the other hand, as the symmetry operations of the layer group keep invariant the $a_1a_2$ plane (and also the reciprocal $a_1^{*}a_2^{*}$ plane), the star of vectors of a given {\bf k}-vector (the set of symmetry equivalent vectors of {\bf k} under the operations of the point group) of the layer group has as many arms (vectors) as the number of arms of the corresponding star of the {\bf k}-vector in the reference group. Therefore, there exists a 1:1 correspondence between the arms of both wave-vector stars. 

\subsection{Irreducible representations of the layer (double) groups with and without time-reversal symmetry}
\addcontentsline{stoc}{subsection}{\protect\numberline{\thesubsection}Irreducible representations of the layer (double) groups with and without time-reversal symmetry}

\label{sec:irreps}
The existence of a 1:1 mapping between the {\bf k}-vector types in the layer group $\mathcal{L}$ and the {\bf k}-vector types of the reference group $\mathcal{G}$ for which $k_3^L=0$ in \cref{kvecs} and the existence of an isomorphism between their little (double) co-groups and little (double) groups allows us to establish a 1:1 relation between the irreps of both {\bf k}-vector types. Therefore, if $K$ is the label of a {\bf k}-vector type in the reference group $\mathcal{G}$ that corresponds to the {\bf k}-vector type $K^L$ in the layer group $\mathcal{L}$, we assume as matrices of the (single or double) irrep $K_i^L$ for the symmetry operations $\{R|{\bf t}\}_L$ with $t_3=0$, those matrices of the (single or double) irrep $K_i$ for the symmetry operations $\{R_G|{\bf t}_G\}$ of $\mathcal{G}$ that fulfill \cref{transf}. Therefore, the matrices of the irreps of the little groups for every {\bf k}-vector type in the layer group can be immediately obtained, without further calculations, from the corresponding {\bf k}-vector types of the reference space group. 

This assignation can be done with or without time-reversal symmetry. The reference group of a unitary layer group $\mathcal{L}_U$. i.e., a layer group that does not contain anti-unitary operations, is a unitary space group $\mathcal{G}_U$. If time-reversal symmetry is added to $\mathcal{L}_U$ (and all other operations of $\mathcal{L}_U$ combined with time-reversal) its reference group is the result of adding time-reversal symmetry to $\mathcal{G}_U$.

Note that, for some layer groups, the choice made above can represent a different labeling of the irreps compared to the table of irreducible representations calculated by Litvin and Wike \cite{litvin1991}. However, we think that the simple mapping used in our work has advantages in the analysis of real 2-dimensional materials. Usually, these materials are experimentally obtained through the exfoliation of a few-layer nanocrystals from the parent (or native) three-dimensional crystal. It is thus advisable to have a direct relationship between the irrep content of the electronic band configuration of the 3D native crystal and the configuration of the 2D crystal. Using the transformation matrix in \cref{referenceG} to establish the mapping between both sets of {\bf k}-vectors and the assumed relations between the labels of both subsets of irreps, the comparison between the irrep content in both crystals is immediate.

To construct the tables of the full representations (with and without time-reversal symmetry) of the layer group $\mathcal{L}$ from the irreps of the little groups, we can make use of the fact that there is a 1:1 correspondence between the arms in the stars of $K$ and $K^L$. 
Therefore, we can also assume as matrices of the (single or double) full representation $^*K_i^L$  for the symmetry operations $\{R|{\bf t}\}$ with $t_3=0$, those matrices of the (single or double) full representation $^*K_i$ for the symmetry operations $\{R_G|{\bf t}_G\}$ of $\mathcal{G}$ that fulfill \cref{transf}.

\begin{figure}
	\centering
	\includegraphics[scale=0.7]{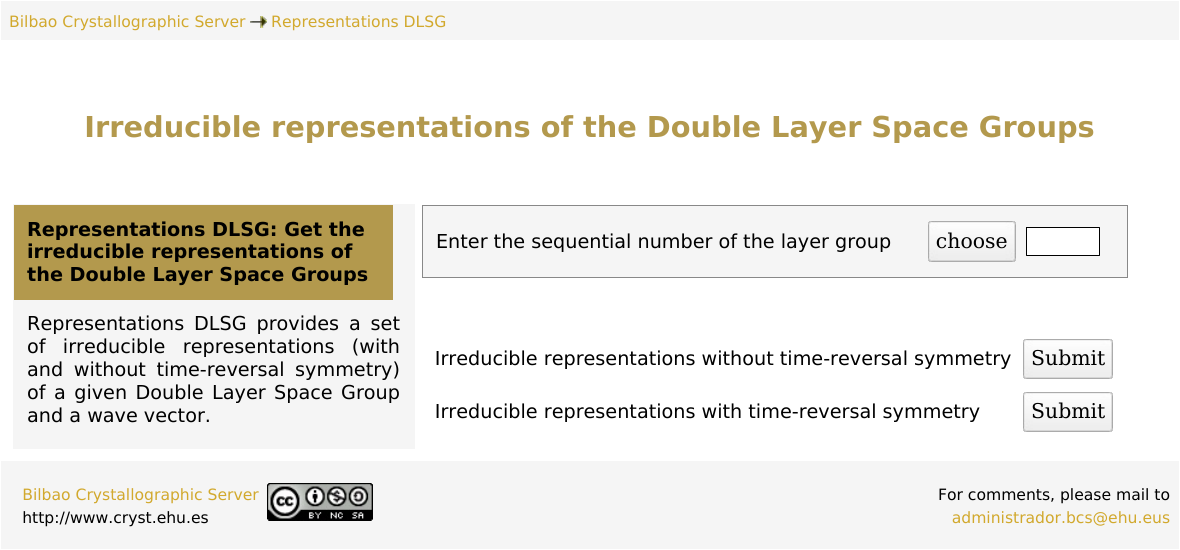}
	\caption{\label{toolrepres} Main menu of the tool REPRESENTATIONS DLSG hosted in the BCS. The user introduces the sequential number of the layer group in the white box or, clicking on the \emph{choose} button, a table with the Hermann-Mauguin symbols of the 80 layer groups is presented. Once the group has been chosen the user can ask for the irreps of the layer group with or without time-reversal symmetry.}
\end{figure}

The single and double irreducible representations of the little group of a given {\bf k}-vector type and the full representations induced from these irreps in the layer groups can be obtained using the newly available REPRESENTATIONS DLSG tool on the BCS (\href{https://cryst.ehu.es/cryst/representationslayer.html}{cryst.ehu.es/cryst/representationslayer.html}). \cref{toolrepres} shows the input page of REPRESENTATIONS DLSG. The user has to enter the sequential number (between 1 and 80) of the layer group according to ITE or, clicking on the \emph{choose} button, the user can select the layer group from a table where the groups are identified through their Hermann-Mauguin symbols. Once the layer group has been selected, it is possible to ask for the irreducible representations with and without time-reversal symmetry. The irreps of the little groups and the full irreps without time-reversal symmetry have been obtained from the unitary double crystallographic or Fedorov groups (type-I magnetic double space groups). The irreps with time-reversal symmetry (or irreducible co-representations) have been obtained from the corresponding type-II magnetic double space groups or \emph{gray} groups, which are derived from the Fedorov groups after the addition of the (anti-unitary) time-reversal symmetry operation. For the present work, where non-magnetic 2D compounds have been investigated, the relevant layer groups are those that include time-reversal as a symmetry operation and then the relevant irreps are the \emph{irreps with time-reversal symmetry} or co-irreps.
The matrices of the irreps and co-irreps of the magnetic double space groups of type-I and type-II used for the tabulation of the irreps of the layer groups have been obtained from the tool COREPRESENTATIONS  (\href{https://cryst.ehu.es/cryst/corepresentations.html}{cryst.ehu.es/cryst/corepresentations.html}) in the BCS \cite{elcoro2021}.

\begin{figure}
	\centering
	\includegraphics[scale=0.7]{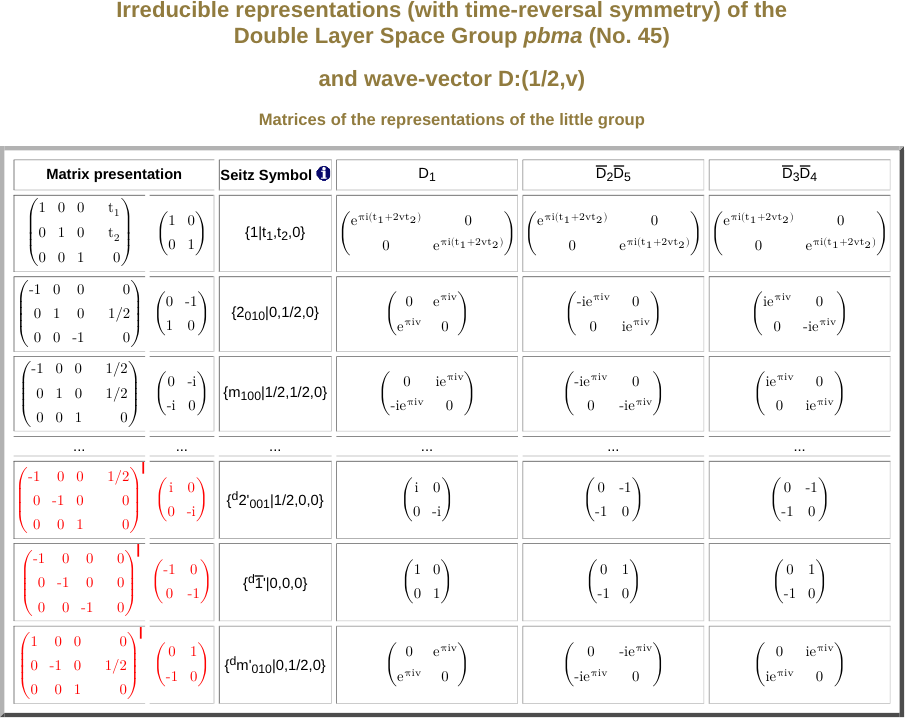}
	\caption{\label{littlegroupirreps} Matrices of the representations of the little group $\mathcal{L}^{\bf k}$ of the {\bf k}-vector D$:(1/2,v)$ in the double layer group $pbma$ (N. 45) with time-reversal symmetry. The first column gives the rotational and spin components of the symmetry operations. The second column shows the Seitz symbol of the symmetry operation. To distinguish between the two operations with the same rotational part but with different spin components the superscript $d$ has been added to one of them. The next columns show the matrices of the representations whose labels are indicated in the first row. The output has been cut to reduce the size of the figure and only the matrices of 6 (out of 16) symmetry operations are shown. The anti-unitary symmetry operations are colored in red.}
\end{figure}

Apart from general information about the group and the {\bf k}-vector, the output of REPRESENTATIONS DLSG consists basically of two tables. The first table gives the matrices of the representations for the symmetry operations of the little group $\mathcal{L}^{\bf k}$ of the {\bf k}-vector chosen. Together with the general translation $(t_1,t_2)$ one operation for each member of the point group is included in the list.
\cref{littlegroupirreps} shows the matrices of the little group of the {\bf k}-vector D$:(1/2,v)$ in the double layer group $pbma$ (N. 45) with time-reversal symmetry. The program gives the matrices of the single- and double-valued irreps. In the analysis of electronic bands done in this work, the single-valued irreps are the relevant ones when the Hamiltonian of the system does not depend on the spin of the electron (for instance when the spin-orbit coupling (SOC) is not considered) and the double-valued irreps are relevant when the hamiltonian depends on the spin of the electron (for instance when SOC is not negligible). The labels of the single-valued irreps follow the notation introduced by Miller and Love (\cite{miller1967}) and consist of the {\bf k}-vector letter and a sequential index. The labels of the double-valued irreps include a bar over the {\bf k}-vector letter. In \cref{littlegroupirreps}, $D_1$ is the unique single-valued irrep and both $\overline{D}_2\overline{D}_5$ and $\overline{D}_3\overline{D}_4$ are double-valued irreps.

\begin{figure}
	\centering
	\includegraphics[scale=0.7]{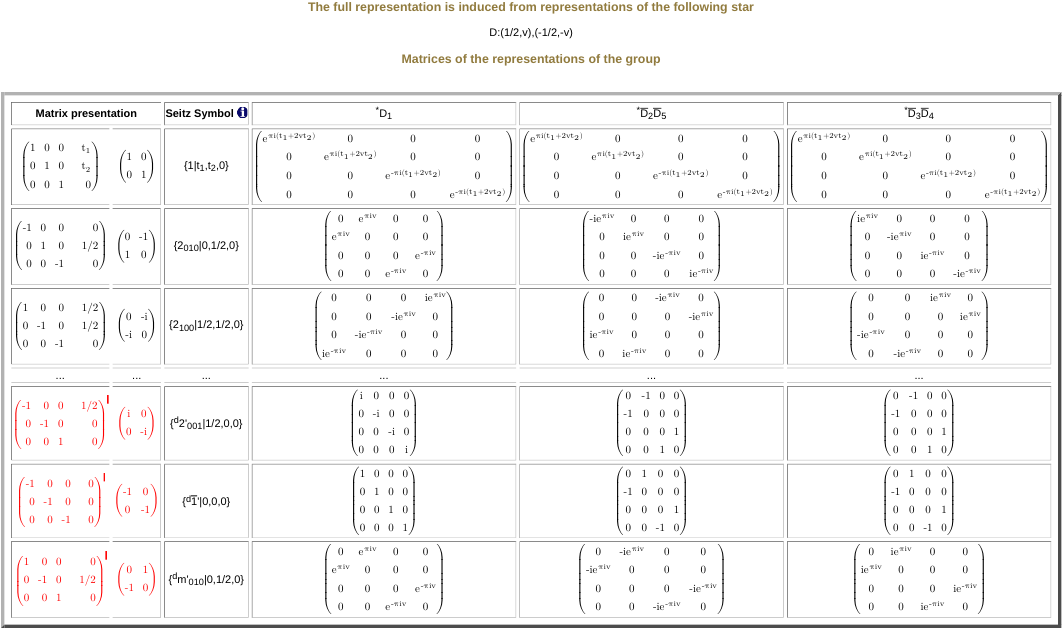}
	\caption{\label{fullirreps} Matrices of the full representations of the double layer group $pbma$ (N. 45) with time-reversal symmetry and wavevector D$:(1/2,v)$. The first column gives the rotational and spin components of the symmetry operations. The second column shows the Seitz symbol of the symmetry operation. To distinguish between the two operations with the same rotational part but with different spin components the superscript $d$ has been added to one of them. The next columns show the matrices of the representations whose labels are indicated in the first row. The output has been cut to reduce the size of the figure and only the matrices of 6 (out of 16) symmetry operations are shown. The anti-unitary symmetry operations are colored in red.}
\end{figure}

Immediately after the irreps of the little group of the chosen {\bf k}-vector, a second table gives the full representations of $\mathcal{L}$. The matrices of the full representations are represented in the standard form when the dimension of the matrix is less than 5. However, for matrices whose dimension is larger than 5, the output contains a list of elements with the format $(i,j):m_{ij}$ which states that the $i,j$ component of the matrix has value $m_{ij}$. Only those elements with $m_{ij}\neq0$ are included in the list. \cref{fullirreps} shows the matrices of the full representations for the D$:(1/2,v)$ wavevector and double layer group
$pbma$ (N. 45) with time-reversal symmetry.

\subsection{Compatibility relations of the layer (double) groups with and without time-reversal symmetry}
\addcontentsline{stoc}{subsection}{\protect\numberline{\thesubsection}Compatibility relations of the layer (double) groups with and without time-reversal symmetry}

\label{comprel}
The compatibility relations are crucial in the analysis of connectivity of the energy bands in the \emph{Topological Quantun Chemistry} (TQC) framework \cite{bradlyn2017topological,cano2018,elcoro2017,vergniory2017}. As we continuously move along a given electronic band represented in the reciprocal space and the path goes through {\bf k}-vectors of different symmetry (but keeping a group-subgroup relation between the little groups of adjacent vectors), the compatibility relations tell us how the sets of degenerate states at points of higher symmetry split when these states move to {\bf k}-vectors of lower symmetry. In the opposite sense, they tell us how states of different energy at {\bf k}-vectors of low symmetry merge at vectors of higher symmetry and become degenerate.

In group theory analysis, these relations are calculated through the \emph{subduction} process. If ${\bf k}_g$ and ${\bf k}_h$ are two vectors whose little groups $\mathcal{L}^{\bf k}_g$ and $\mathcal{L}^{\bf k}_h$ have a group-subgroup relation $\mathcal{L}^{\bf k}_g>\mathcal{L}^{\bf k}_h$ (for instance, ${\bf k}_g$ is a special point in a line ${\bf k}_h$ or ${\bf k}_g$ is a special line in a plane ${\bf k}_h$), the subset of matrices of an irrep of $\mathcal{L}^{\bf k}_g$, restricted to those elements of $\mathcal{L}^{\bf k}_g$ that also belong to $\mathcal{L}^{\bf k}_h$, form a representation, in general reducible, of $\mathcal{L}^{\bf k}_h$. The compatibility relations give the decomposition of this representation into irreps of $\mathcal{L}^{\bf k}_h$. Details about the calculation of the compatibility relations and their application in the TQC framework can be found in refs. \cite{elcoro2017,vergniory2017}.

\begin{figure}
	\centering
	\includegraphics[scale=0.7]{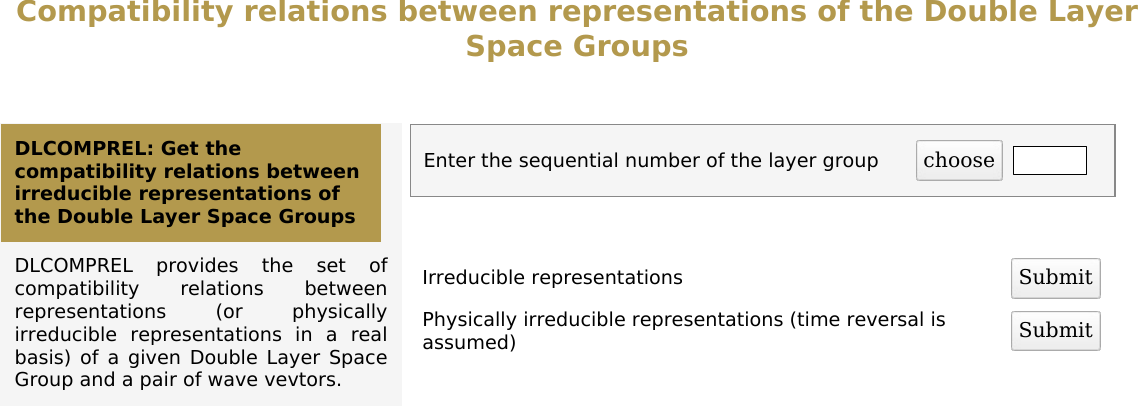}
	\caption{\label{dlcomprelmain} Main menu of the tool DLCOMPREL hosted in the BCS. The user introduces the sequential number of the layer group in the white box or, clicking on the \emph{choose} button, a table with the Hermann-Mauguin symbols of the 80 layer groups is presented. Once the group has been chosen the user can ask for the compatibility relations of irreps of the layer group with or without time-reversal symmetry.}
\end{figure}

The compatibility relations for layer groups can be obtained from the corresponding compatibility relations in 3D space groups assuming the mapping explained in \cref{sec:wyckoff} and \cref{sec:irreps}, based on the existence of the reference group. In \cref{sec:irreps} we have established a mapping between the irreps of a layer group and the irreps of its reference group. This mapping can also be translated to the compatibility relations. We have implemented the tool \DLCOMPREL (\href{https://cryst.ehu.es/cryst/dlcomprel.html}{cryst.ehu.es/cryst/dlcomprel.html}) in the BCS to retrieve the compatibility relations between pairs of {\bf k}-vectors connected in the reciprocal space, i.e., pairs whose little groups have a group-subgroup relation. \cref{dlcomprelmain} shows the main menu of DLCOMPREL. Once the layer group has been chosen, the user can ask for the compatibility relations between the irreps with or without time-reversal symmetry. 

\begin{figure}
	\centering
	\includegraphics[scale=0.7]{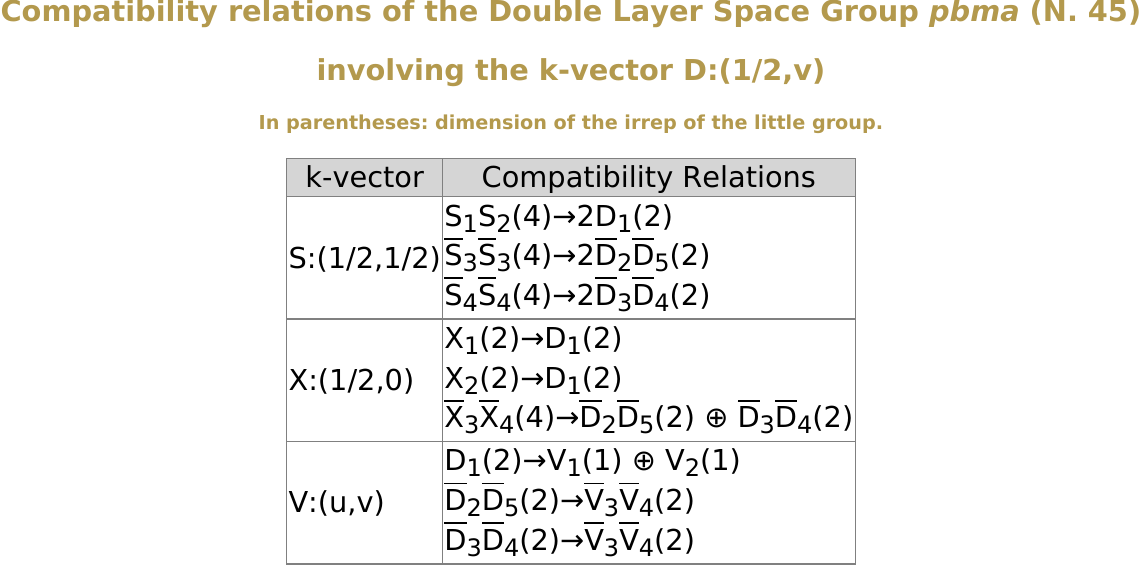}
	\caption{\label{dlcomprelout} Compatibility relations between the (co)irreps with time-reversal symmetry of the little group of {\bf k}-vector D$:(1/2,v)$ in the double layer group $pbma$ (N. 45) and the (co)irreps of all the {\bf k}-vectors connected to $D$ in this layer group, $S:(1/2,1/2)$, $X:(1/2,0)$ and the general position $V:(u,v)$. The second column shows the compatibility relations.}
\end{figure}

In the next step the user chooses one of the {\bf k}-vector types of the layer group. The program then shows  shows the list of {\bf k}-vectors connected to it in the chosen layer group. This list, in general, contains {\bf k}-vectors of higher symmetry (if any) and vectors of lower symmetry (if any). In this step the user must select one vector of the list to get the compatibility relations between the irreps of the little group of this vector and the irreps of the little group of ${\bf k}_1$. Optionally it is also possible to get the compatibility relations between the irreps of the little group of ${\bf k}_1$ and the irreps of the little group of every {\bf k}-vector connected to ${\bf k}_1$. \cref{dlcomprelout} shows an example of the output of DLCOMPREL for the layer group $pbma$ (N.45) with time-reversal symmetry and first {\bf k}-vector $D:(1/2,v)$. In this example we have asked for the compatibility relations between the irreps of the little group of $D$ and the little group of every {\bf k}-vector connected to $D$, i.e., $S:(1/2,1/2)$, $X:(1/2,0)$ and $V:(u,v)$ in the first column of the figure. The second column shows the compatibility relations. The first two {\bf k}-vectors $S$ and $X$ represent isolated points in the reciprocal space that sit on the line $D$ and, therefore, the little groups $\mathcal{L}^S$ and $\mathcal{L}^X$ of $S$ and $X$, respectively, are supergroups of the little group $\mathcal{L}^D$ of $D$. The irreps of  $\mathcal{L}^S$ and $\mathcal{L}^X$ subduce into representations of $\mathcal{L}^D$. The irrep $\overline{X}_3\overline{X}_4$ of $\mathcal{L}^X$, for instance, subduces into a reducible representation of $\mathcal{L}^D$ which can be expressed as the direct sum of the irreps $\overline{D}_2\overline{D}_5$ and $\overline{D}_3\overline{D}_4$. The {\bf k}-vector $V:(u,v)$ represents the general position where the line $D:(1/2,v)$ sits. The little group $\mathcal{L}^V$ of $V$ is thus a subgroup of $\mathcal{L}^D$ and the irreps of $\mathcal{L}^D$ subduce into representations of $\mathcal{L}^V$.

\subsection{Band representations in the layer groups}
\addcontentsline{stoc}{subsection}{\protect\numberline{\thesubsection}Band representations in the layer groups}

\label{sec:lbandrep}
The concept of \emph{band representation} was introduced by Zak \cite{Zak1980,ZakBandrep1,ZakBandrep2} for spinless systems with and without time-reversal symmetry and was later extended to spinful systems \cite{bradlyn2017topological,elcoro2017}. They represent a connection between magnitudes defined locally in direct space (as the electronic orbitals around a center of charges in this work) and its representation in the momentum or reciprocal space through extended states (Bloch wave-functions or electronic bands in this work). The symmetry of the extended states over the Brillouin zone is determined by the symmetry of the localized states around a Wyckoff position in the crystal.
The relation between the two descriptions based on representation theory is usually done through the so-called \emph{site-symmetry approach} \cite{EvarestovBook}. As the SSG of a given Wyckoff position is a (finite) subgroup of the (infinite) space group $\mathcal{G}$, a representation of the SSG induces a representation (band representation) in $\mathcal{G}$ whose dimension is infinite but can be expressed as a direct sum of irreducible representations of $\mathcal{G}$. The site-symmetry approach allows the calculation of the multiplicities of this decomposition of the band representation into irreps of $\mathcal{G}$, i.e., into irreps of the little group of every {\bf k}-vector in the first Brillouin zone. The apparent complexity of the determination of the multiplicities of the  (infinitely many) irreps of the little groups is much simplified by making use of the Frobenius reciprocity theorem. According to this theorem, if $\rho^H$ and $\rho^M$ are two representations of two subgroups $\mathcal{H}$ and $\mathcal{M}$, respectively, of $\mathcal{G}$, the multiplicity of $\rho^M$ of the induced representation into $\mathcal{G}$ from $\rho^H$ and the multiplicity of $\rho^H$ of the induced representation into $\mathcal{G}$ from $\rho^M$ are exactly the same. As the irreps of the little group of every {\bf k}-vector in all the space groups and the corresponding induced representations into the space group (the full representations) have been already tabulated, the systematic application of the site-symmetry approach allowed the tabulation of the multiplicities, as an intermediate step to characterize the band representations. The programs SITESYM (\href{https://cryst.ehu.es/cryst/}{cryst.ehu.es}) and DSITESYM (\href{https://cryst.ehu.es/cryst/sitesym.html}{cryst.ehu.es/cryst/sitesym.html}) in the BCS give these multiplicities in all the space groups and all the double space groups, respectively. More details about the site-symmetry approach can be found in Ref.~\cite{EvarestovBook} and about its implementation in the BCS in Ref.~\cite{elcoro2017}.

We have recently adapted DSITESYM to layer groups and implemented the tool DLSITESYM (\href{https://cryst.ehu.es/cryst/dlsitesym.html}{cryst.ehu.es/cryst/dlsitesym.html}) in the BCS. As it has been stressed in \cref{sec:wyckoff} and \cref{sec:irreps}, there is no need to apply the site-symmetry approach from scratch in the layer groups due to the existence of the reference space group $\mathcal{G}$ for each group $\mathcal{L}$ (see \cref{sec:wyckoff}). As we have established a correspondence between the Wyckoff positions, {\bf k}-vectors and irreps of $\mathcal{G}$ and $\mathcal{L}$, we can extend the correspondence to the multiplicities calculated in the site-symmetry approach without any further calculations, as we have done in the calculation of the compatibility relations in \cref{comprel}.

\begin{figure}
	\centering
	\includegraphics[scale=0.7]{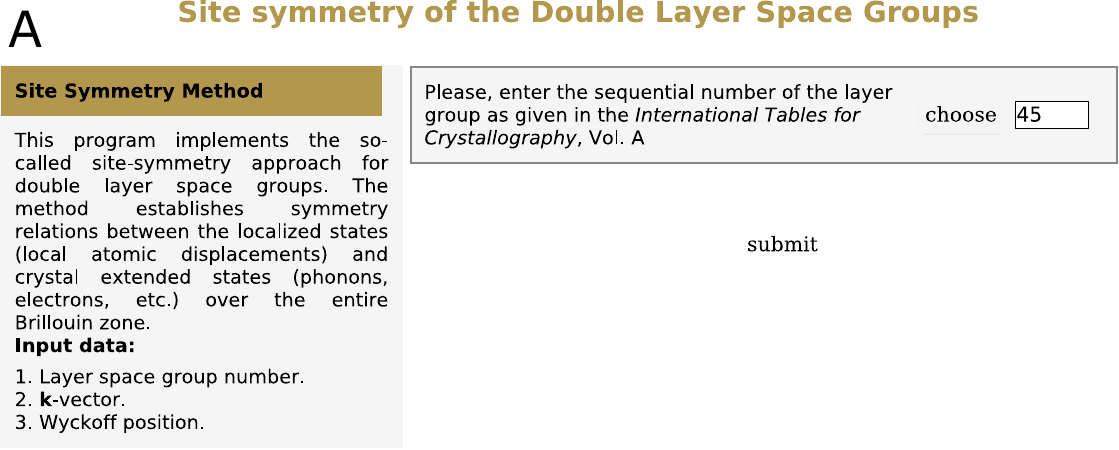}
	\includegraphics[scale=0.7]{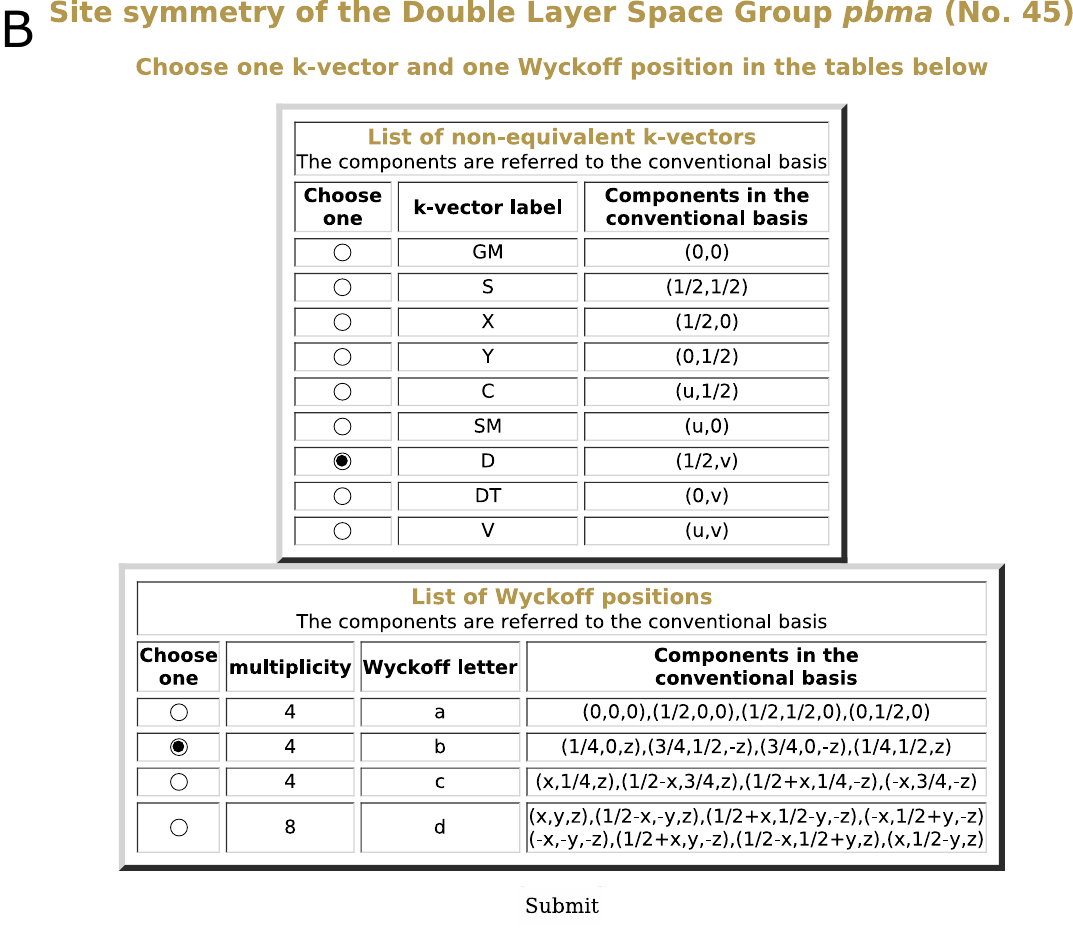}
	\caption{\label{dsitesymmenu} (A) First page of the menu of the tool DLSITESYM where the user selects the layer group. (B) Second page of the menu, once the user has selected the layer group ($pbma$ (N. 45) in the example of the figure). In this step the user chooses a {\bf k}-vector in the momentum space and a Wyckoff position ($D$ and $b$, respectively, in the example of the figure).}
\end{figure}

\begin{figure}
	\centering
	\includegraphics[scale=0.7]{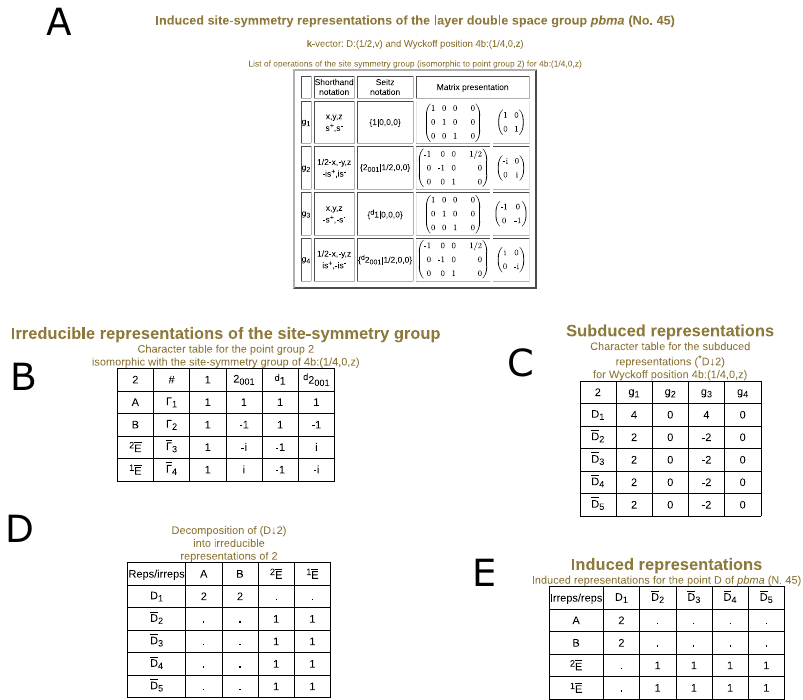}
	\caption{\label{dsitesymout} Output of the program DLSITESYM after the introduction of the input data of \cref{dsitesymmenu}. (A) Symmetry operations of the SSG of the Wyckoff position $4b$ in the double layer group $pbma$ (N. 45). (B) Character table of the point group isomorphic with the SSG of $4b$. (C) Character table of the subduced representations from the full representations at point $D$ into the SSG of $4b$. (D) Decomposition of the subduced representations of (C) into irreps of the SSG of $4b$. (E) Induced representations from the irreps of the SSG of $4b$ into $\mathcal{L}$ for the {\bf k}-vector $D$.}
\end{figure}

\cref{dsitesymmenu} shows the two steps of the input DLSITESYM. The user chooses sequentially the layer group (menu of \cref{dsitesymmenu}A), and a {\bf k}-vector and Wyckoff position (menu of \cref{dsitesymmenu}B). The output of the program consists of 5 tables, as shown in \cref{dsitesymout}.
The program gives (A) the symmetry operations of the SSG of the Wyckoff position chosen, (B) the character table of the irreps of the point group isomorphic with the SSG of (A), (C) the character table of the subduced representations from the full irreducible representations of the chosen {\bf k}-vector into the SSG, (D) the decompositions of the representations in (C) into irreps of the SSG and (E) the multiplicities of the decompositions of the induced representations from the irreps of the SSG into irreps at the chosen {\bf k}-vector. According to the Frobenius reciprocity theorem, the table of \cref{dsitesymout}E is the transpose of the table in \cref{dsitesymout}D. 

Given a Wyckoff position and the irreps of its SSG, the systematic application of the site-symmetry approach for all the {\bf k}-vectors that cover the first Brillouin zone, allows the determination of the band representations \cite{Zak1980,ZakBandrep1,ZakBandrep2,bradlyn2017topological,elcoro2017} induced from the irreps 
of the SSG of the Wyckoff position.

Using the tabulation made in LSITESYM, the determination of the band representations induced from every irrep of the site-symmetry group of any Wyckoff position in a given layer group is straightforward. A band representation is described as the direct sum of irreps of the little groups $\mathcal{L}^{\bf k}$ of every {\bf k}-vector in the first Brillouin zone. To construct thus the induced band representation we have just to fix the layer group, the Wyckoff position, the irrep of the site-symmetry group of the Wyckoff position and extract the data from  \cref{dsitesymout}E for all the {\bf k}-vectors of the layer group. Among all the band representations induced from all the Wyckoff positions, those that cannot be expressed as a direct sum of band representations are called \emph{elementary band representations} (EBRs) and are of special importance in the TQC approach (see Ref.~\cite{bradlyn2017topological,elcoro2017,cano2018,vergniory2017} for detailed explanations about the relevance of the EBRs in the analysis of electronic band structures.)

\begin{figure}
	\centering
	\includegraphics[scale=0.7]{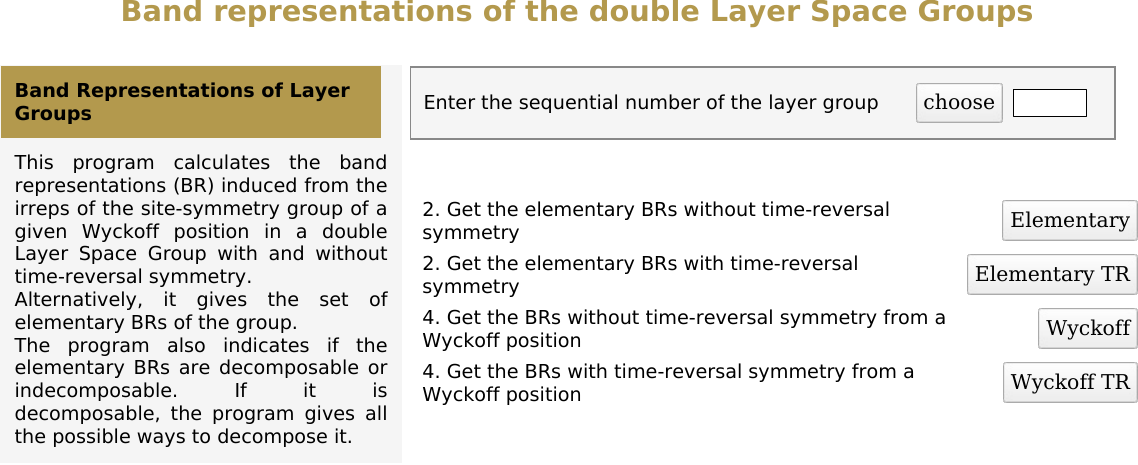}
	\caption{\label{lbandrepmenu} (a) Main menu of the LBANDREP tool. The user introduces the sequential number of the layer group in the box (or after clicking on the \emph{choose} button the group can be selected from the table of symbols of the layer groups) and chooses among the four options given at the bottom: list of elementary band representations (1) with or (2) without time-reversal symmetry or list of induced band representations with (3) or (4) without time-reversal for a Wyckoff position to be selected in a subsequent menu.}
\end{figure}

The recently implemented tool LBANDREP (\href{https://cryst.ehu.es/cryst/lbandrep.html}{cryst.ehu.es/cryst/lbandrep.html}) in the BCS gives all the BRs induced from all irreps of the site-symmetry groups of all the Wyckoff positions. It also identifies the EBRs. The \cref{lbandrepmenu} shows the main menu of LBANDREP. Providing the sequential number of the layer group (or selecting the layer group from the table shown after clicking on the \emph{choose} button), the user has four different options to get a set of band representations. It is possible to get the complete list of EBRs of the layer group or the set of band representations induced from a chosen Wyckoff position. In both cases, the user can choose between layer groups with or without time-reversal symmetry.

\begin{figure}
	\centering
	\includegraphics[scale=0.6]{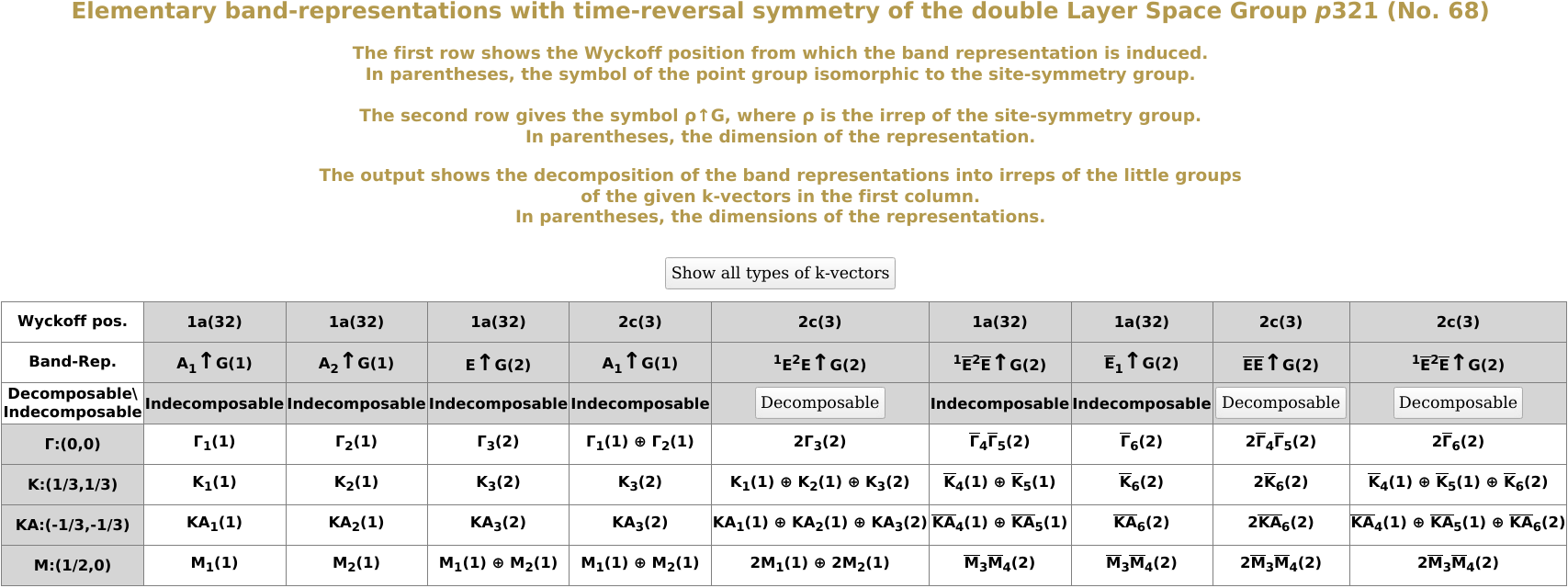}
	\caption{\label{lbandrepout} Default output of LBANDREP for the layer group $p321$ (N. 68) and with the chosen option \emph{Elementary TR}. See the text for an explanation of the information included in the table.}
\end{figure}

\cref{lbandrepout} shows the output of LBANDREP for the layer group $p321$ (N. 68) and with the chosen option \emph{Elementary TR}. The output contains the full list of five single-valued and four double-valued EBRs induced from the two maximal Wyckoff positions $a$ and $c$. The default output gives only the multiplicities of the irreps at the {\bf k}-vectors of maximal symmetry ($\Gamma$, $K$, $KA$, and $M$ for this layer group). This list contains all the information to identify and characterize the EBR. The multiplicities at any other {\bf k}-vector can be obtained from the multiplicities at the maximal {\bf k}-vectors making use of the compatibility relations. Clicking on the box \emph{Show all types of k-vectors} the list of irreps is extended to the {\bf k}-vectors of the whole first Brillouin zone.

\begin{figure}
	\centering
	\includegraphics[scale=0.7]{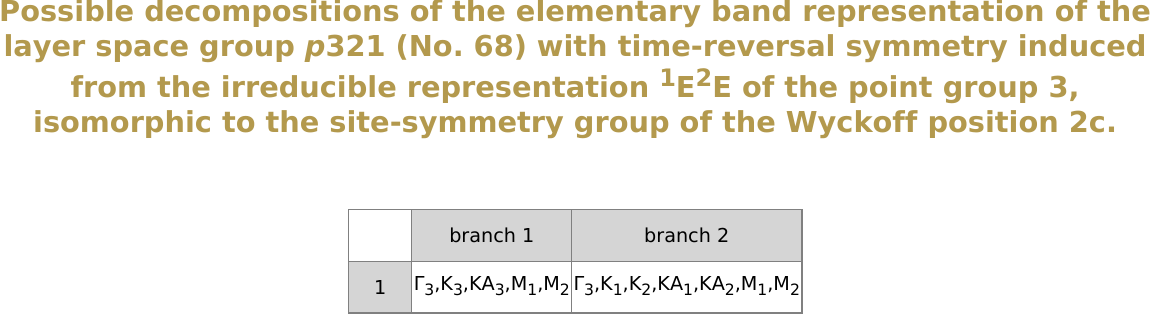}
	\caption{\label{lbandrepdesc} Possible ways to decompose the EBR induced from the single-valued irrep $\,^1E\,^2E$ of the site symmetry group of the Wyckoff position $2c$ in the layer group $p321$ (N. 68) with time-reversal symmetry. This information is accessible by clicking on the box  \emph{Decomposable} below the Wyckoff position $2c$ and irrep $\,^1E\,^2E$ in \cref{lbandrepout}.}
\end{figure}

The first row of the table indicates the Wyckoff position and, in parentheses, the symbol of the point group isomorphic to the site-symmetry group of the Wyckoff position. The second row shows the irrep of the site-symmetry group that induces the EBR and the dimension of the irrep in parentheses. The third row indicates whether the EBR is decomposable or not. An EBR is decomposable whether the set of irreps at every {\bf k}-vector can be divided into two or more subsets of irreps that can be fully connected along the whole reciprocal space. A decomposable EBR thus can be split into several gapped bands. In this layer group, the EBR induced from the single-valued irrep $\,^1E\,^2E$ of the site symmetry group of the Wyckoff position $2c$ is decomposable. Clicking on the box \emph{Decomposable} the program gives the different ways into which the EBR can be decomposed. \cref{lbandrepdesc} gives all the possible decompositions of the EBR into gapped bands. In this particular case, there is a single way to decompose the EBR into two bands defined by the irreps at every maximal {\bf k}-vector. This information is relevant in the classification of the compounds with strong topology into NLC and SEBR (see the main text and refs \cite{bradlyn2017topological,elcoro2017}). The set of bands below the Fermi level of a material tagged as SEBR cannot be expressed as a linear combination of EBRs using positive or negative coefficients (it is thus strong topological) but can be expressed as a linear combination of EBRs and parts of EBRs, as the ones shown in every column of \cref{lbandrepdesc}.

\subsection{Check Topological layer Materials tool in the BCS}
\addcontentsline{stoc}{subsection}{\protect\numberline{\thesubsection}Check Topological layer Materials tool in the BCS}

The tables described in the preceding sections and implemented in the BCS contain all the required information to apply the TQC approach to the (calculated) set of electronic bands of a material. We have implemented the tool \emph{Check Topological Layer Materials} in the BCS (\href{https://cryst.ehu.eus/cryst/checktopologicallayer}{cryst.ehu.eus/cryst/checktopologicallayer}) to make the identification of the topological character of the material based on TQC. This tool has been adapted from the previously implemented \emph{Check Topological Mat.} tool in the BCS for 3-dimensional systems. The use of this tool is explained in detail in the supplementary information (appendices A-E) of Ref.~\cite{Vergniory2019}. The appendix in that reference can also be used as a guide of the tool \emph{Check Topological Layer Materials}. The program needs as input a file in plain text that contains all the required parameters: the number of electrons of the compound, a label that indicates whether SOC has been included in the calculations, the list of symmetry operations of the layer group, the list of coordinates of the {\bf k}-vectors of maximal symmetry, the list of symmetry operations of the little group of every maximal {\bf k}-vector and the traces of the symmetry operations of the little groups of the eigenstates calculated at the {\bf k}-vectors of maximal symmetry (see the details in Ref.~\cite{Vergniory2019}). The whole procedure in the 3D and 2D versions of the program is practically the same except for a subtlety regarding the symmetry operations given. Whereas the (3D) space group type can be unambiguously identified by a set of generators given in any arbitrary setting, this is not true for layer groups. As stressed in \cref{sec:wyckoff}, two different layer groups can correspond to the same reference space group in the monoclinic and orthorhombic crystal systems (for example the layer group types $p112$ (N. 3) and $p211$ (N. 8) which have as reference group the space group type $P2$ (N. 3)). In these cases, the difference between the layer groups is the orientation of their symmetry operations with respect to the 2-dimensional lattice of translations. To avoid any misidentification of the layer group generated by the symmetry operations given in the file to upload in \emph{Check Topological Layer Materials}, in the used setting the ${\bf c}$ unit cell vector must be orthogonal to the lattice of translations of the layer group.

\subsection{RSI-layer: tables of real Space Invariants in the layer groups}
\addcontentsline{stoc}{subsection}{\protect\numberline{\thesubsection}RSI-layer: tables of real Space Invariants in the layer groups}

Following the method introduced in Ref. \cite{fragilesci} to calculate the RSIs in wallpaper groups and subsequently extended to 3D space groups \cite{xu2021three}, we have calculated the RSI in the (double) layer group for both types of irreps: single irreps (for band structure calculations without SOC) and for double irreps  (for band structure calculations with SOC). The definitions of the RSI indices of LGs have been implemented in the BCS (\href{https://cryst.ehu.eus/cryst/RSIlayer}{cryst.ehu.eus/cryst/RSIlayer}).

\section{2D Topological Materials Database}\label{dabase}
\addcontentsline{stoc}{section}{\protect\numberline{\thesection}2D Topological Materials Database}

In this section, we will briefly describe the main features of the \webtwoDTQC\; (\webtwoDTQCAbbr). The database is composed of a search engine, which allows a \emph{basic} and \emph{advanced} search methods, and for each material a detailed description of its crystallographic, electronic, and topological properties.

\begin{figure}
    \centering
    \includegraphics[width=0.8\linewidth]{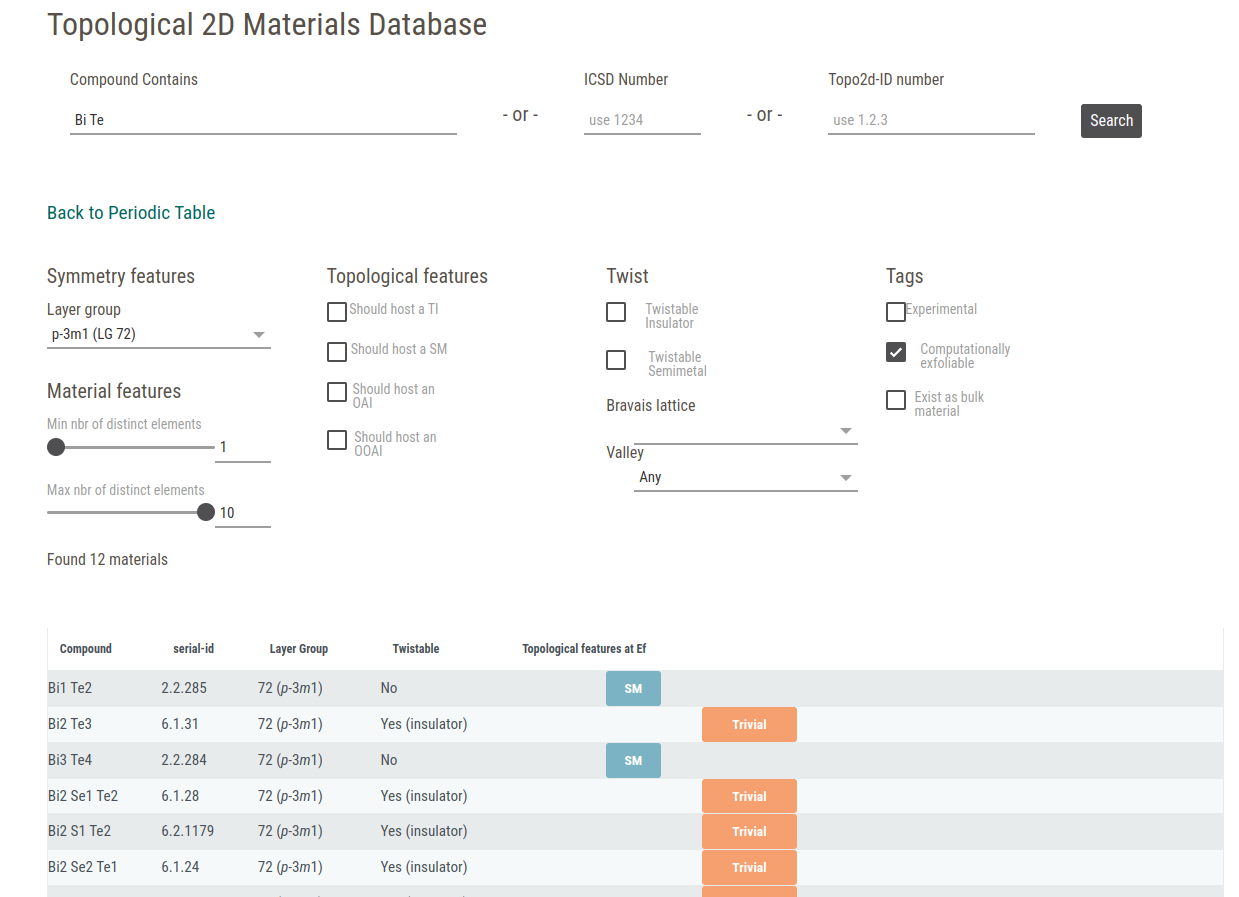}
    \caption{Screenshot of the advanced search interface of the \webtwoDTQCAbbr. The tags for a detailed search involving symmetry, chemical formula, topological, and twist features are shown, alongside the experimental/exfoliable/bulk status.
    }
    \label{fig:db-1}
\end{figure}

The \webtwoDTQCAbbr\; database is largely analogous to the \webTQCAbbr\ or \webTQCphonon. The \emph{basic} search engine page allows searches based on chemical elements or chemical formula, the parent bulk crystal ICSD ID or the serial number assigned to each entry in the \webtwoDTQCAbbr\; database. In addition, the \emph{advanced} search option allows the user to specify more constraints on the material to be searched for, as shown in \cref{fig:db-1}. This includes the layer group (LG) of the structure, the maximum and minimum number of distinct elements present in the formula, and the topological or obstructed class the material belongs to (trivial and not-obstructed, topological insulator (TI), topological semimetal (SM), obstructed atomic insulator (OAI) or orbital-obstructed atomic insulator (OOAI)) at the Fermi level. One can also filter materials with respect to their twist properties, including if it is a twistable insulator or semimetal, the Bravais lattice or the valley type. 
We have also included additional properties of the valence band maxima (VBM) and conduction band minima (CBM) of insulators, including their effective masses in two principal directions, and energy with respect to the vacuum level. 
Finally, a few tags allow to focus on 2D materials that are known to be experimentally realized ((providing references when available)), that exist as bulk or that have been predicted as exfoliable from a bulk material. The search result gives a short overview of each material, including its serial number, LG, twistable properties, and topological features at Fermi energy with SOC.

\begin{figure}
    \centering
    \includegraphics[width=0.9\linewidth]{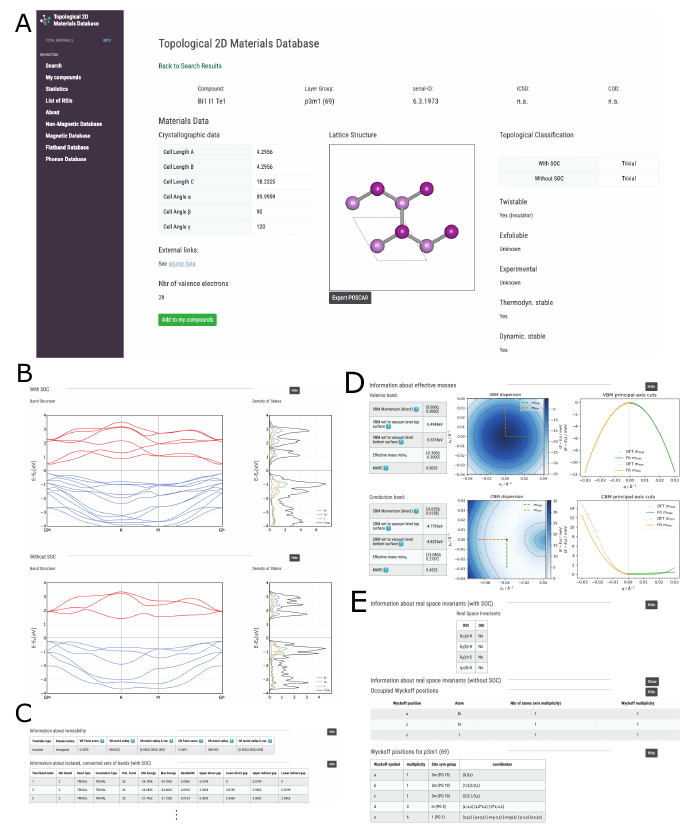}
   \caption{Screenshot of a typical material entry page, using BiTeI in layer group 69 (2D-TQC \serialidweb{6.3.1973}) as an example. The top panel (A) displays key structural details, including unit cell parameters, an interactive 3D structure, and basic properties. The unique serial number identifies the material entry in the \webtwoDTQCAbbr. Links to the original crystal structure sources (either \CtwoDB; or \MCtwoD) are provided, along with topological classifications (with and without SOC), twistable properties, computational exfoliability (from \MCtwoD), experimental status, and thermodynamic and dynamic stability (from \CtwoDB). 
   Panel (B) shows the key electronic properties, including the band structure and atom-projected density of states (PDOS) with and without SOC, presented interactively. Panel (C) provides details on twistability, isolated and connected band sets (with and without SOC) with their topological properties.
    Panel (D) presents the effective masses along the two principal axes for both the valence band maximum (VBM) and conduction band minimum (CBM), along with the band dispersion near the VBM and CBM fitted parabolically along these axes. Additionally, it includes the VBM and CBM energies relative to the vacuum level for both the top and bottom surfaces of the system. 
   Panel (E) shows the real space invariants (RSI) (with and without SOC), occupied Wyckoff positions, and the full list of Wyckoff positions in the corresponding LG. 
    }
    \label{fig:db-stru}
\end{figure}

Once we select a given material, the website displays a large amount of information, as exemplified in \cref{fig:db-stru} for BiTeI in LG 69 (2D-TQC \serialidweb{6.3.1973}), organized as follows. 

At the top of each page, we display detailed crystallographic properties of the material, including unit cell parameters and layer group information. The structure is presented in an interactive 3D plot, enabling magnification and rotation of the unit cell. Additionally, we list both the Wyckoff positions occupied by atoms in the material and the complete set of generic Wyckoff positions for the layer group at the end of the page. Atomic structures used in this work were collected from two established 2D materials databases, \CtwoDB; and \MCtwoD, and all calculations were performed based on these initial structures. Each entry includes a link to the source of the atomic structure file.

For each material, we have calculated the projected density of states (PDOS) and the band structure, both with and without spin-orbit coupling (SOC), as exemplified in \cref{fig:db-stru}B. Both plots are interactive, allowing users to zoom in and explore finer details, similar to the atomic structure visualization. 

We computed the effective masses along two principal axes at the VBM and CBM for insulators, together with the band-edge energies with respect to the vacuum level. In the case of polar materials, the vacuum level of the top and bottom surfaces can be different. Therefore, we provide the band-edge energies referenced to the vacuum level of both surfaces. In \cref{app:sec:table_band_alignment}, we provide a comprehensive table for materials with close vacuum-aligned band edge energies, which are important guidance for future research on twisted homo- and hetero-bilayer engineering.

When the material is predicted as theoretically twistable, we summarize its key twisting properties as studied in Ref.~\cite{jiang2024twist}. For twistable semimetals, we include the moir\'e valley momentum and the twist score. For twistable insulators, we provide the moir\'e valley momentum and twist scores for the conduction and valence bands separately when their respective band edges are twistable. The effective masses and vacuum-aligned energies of the VBM and CBM are particularly useful for the band alignments of (twisted) homo- or hetero-bilayers.

For each entry, we conduct a comprehensive analysis of the topology for all isolated band groups, both with and without spin-orbit coupling (SOC). For each group of bands, we provide: the band index, the number of bands included, the topological classification according to TQC—categorized as SEBR, NLC, FRAGILE, or TRIVIAL—and the cumulative topological classification. Additionally, we report the energy difference to the Fermi level, the minimum and maximum energies of the bands, the upper and lower direct gaps, and the upper and lower indirect gaps.

For topologically trivial materials with a valence band representation, we additionally provide all possible real-space invariants (RSI) consistent with the material’s layer group (LG). For each RSI, we list its index, the associated Wyckoff position, and its specific value in the material. We also indicate whether the index implies that the material is an obstructed atomic insulator (OAI)—a condition met when a nonzero index appears in an empty Wyckoff position.

The \webtwoDTQCAbbr~also provides a statistics page, which offers a convenient option to download key topological properties calculated for each material in CSV format. It includes statistics across various subsets of the database, categorized by topology type, experimental status, and other classifications.

\section{Topological indices}\label{topoindices}
\addcontentsline{stoc}{section}{\protect\numberline{\thesection}Topological indices}

In the main text, we show that the topological quantum chemistry (TQC) formalism can determine if the symmetry data vector of a material corresponds to a topological insulator or not. Materials belonging to the SEBR or NLC type have a strong topology. However, this classification alone does not give further information on the type of topological invariant the material has. Such information can be determined by the symmetry indicators, which we briefly discuss below.

It is instructive to consider a crystal of space group $G$ that undergoes a structural phase transition, so the symmetry of its structure changes into a subgroup of $G$. The loss of one or more symmetry operations may or may not change the (strong) topology of the system. When this happens, we may say that the lost symmetry protects the topology of the original system. A set of topological indices or symmetry indicators (SIs) can be defined in order to express the topology of a material, depending on the symmetry that protects it.
Any given type of strong topology (QSHI, TCI,...) is therefore characterized by a set of SIs.
Each SI allowed in an SG is the generator of an abelian group $\mathbb{Z}_i$, where $i$ is the amount of values the SI can take. The SI group of an SG is the product of the abelian groups generated by each allowed SI \cite{Po2017,Tang2019}, so that each SG has a characteristic SI group and allowed SIs. If at least one allowed SI in a material is not zero, the system will exhibit strong topology.

Given a space group (or layer group), the topological indices or symmetry indicators that it allows can be calculated \cite{Po2017,filanomaly1,Song2018, song2018diagnosis, peng2022topological, elcoro2021}. 
An explicit expression of the topological indices as a function of multiplicities of coreps of the little groups at the maximal $\mathbf{k}$ points of the Brillouin Zone (BZ) can be derived. The expressions of the SI for all SG have been tabulated \cite{Po2017,filanomaly1,Song2018}. On the other hand, the set of possible EBRs of an SG are also fully determined \cite{bradlyn2017topological}. The full information about the SI group of an SG is encoded in its EBR matrix \cite{basicbands2020}. It can be shown that the EBR matrix, which is a non-square matrix with integer coefficients accepts a Smith decomposition,

\begin{equation}\label{eq:smith}
\Delta = L \cdot EBR \cdot R 
\end{equation}

\noindent where $\Delta$ is a diagonal matrix, and $L$ and $R$ are unimodular matrices. Some elements in the diagonal of $\Delta$, $\Delta_{i,i}$, can be zero, since some EBRs may not be independent from the others. These vanishing diagonal coefficients of $\Delta$ establish the conditions for compatibility relations to be satisfied \cite{basicbands2020}. The diagonal coefficients $\Delta_{ii}$ larger than one establish which subspace spanned by EBRs can host non-trivial topology, as we will show below. It was established in Ref.~\cite{bradlyn2017topological} that, if compatibility relations are satisfied, the symmetry data vector $B$ can be expressed as

\begin{equation}
B = EBR \cdot X
\end{equation}

\noindent where $X$ is a vector that can contain integer or fractional elements $X_i$. If $X_i$ are integers, the material is trivial or fragile topological, and if there is at least one fractional coefficient the material is strong topological in the TQC sense (SEBR or NLC). Making use of the Smith decomposition of the EBR matrix in \cref{eq:smith}, we may define a $C$ matrix

\begin{equation}\label{eq:topoin}
C= \Delta \cdot R^{-1} \cdot X = L \cdot B
\end{equation}

If the system does not have strong topology, then $X$ can have only integer coefficients, and the following conditions are obtained for the coefficients of $C$, $c_i$

\begin{equation}\label{eq:ci}
c_i = 0 \mod \Delta_{ii}. 
\end{equation}

The $c_i$ can be nonzero only if $\Delta_{ii}>1$. If $\Delta_{ii}>1$, $c_i$ can have a set of integer values, limited by the value of $\Delta_{ii}$. If $c_i>0$, the system will present strong topology. Therefore, the $\Delta_{ii}$ elements larger than 1 indicate the existence of strong topological classes in the space group considered. Each $\Delta_{ii}>1$ represents an abelian group $\mathbb{Z}_{\Delta_{ii}}$. The topological class of a particular material will be determined by the values of the topological indices, which coincide with the $c_i$ coefficients associated to the $\mathbb{Z}_{\Delta_{ii}}$ groups with $\Delta_{ii}>1$. The information about the topological indices of a system is thus encoded in the $L$ matrix and the symmetry data vector $B$. Therefore, the $\mathbb{Z}_{\Delta_{ii}}$ groups with $\Delta_{ii}>1$ coincide with the abelian groups generated by symmetry indicators.

For example, in the $\Delta$ matrix of SG $P$-1 (no. 2), there are four diagonal elements $>1$, namely (2,2,2,4). Therefore, the SI group of SG no. 2 is $\mathbb{Z}_2\times\mathbb{Z}_2\times\mathbb{Z}_2\times\mathbb{Z}_4$. The $\mathbb{Z}_4$ group \cite{Po2017,Song2018,basicbands2020} is generated by the strong topological index of Fu and Kane. The other three groups have a $\mathbb{Z}_2$ index each, corresponding to the three weak topological indices \cite{topo3D,fu2007topological}.

The $c_i$ can be expressed as a function of EBRs (\cref{eq:topoin}), and thus also as a function of the coreps of the maximal $\mathbf{k}$ points. In general, the unimodular matrices $R$ and $L$ that participate in the Smith decomposition of the EBR matrix are not unique, and so they depend on an arbitrary choice. The compatibility relations impose relationships between the coefficients of the $L$ matrix \cite{Po2017,basicbands2020}. While the Smith normal form of the EBR matrix, and thus the SI group of the SG or LG do not depend on the choice of the $L$ and $R$ matrices, the $c_i$, and therefore the expressions of topological indices, do. In \cite{filanomaly1,Song2018} a convention for the indices was introduced, and all the indices for all SG were calculated and tabulated. This convention allows to establish a unique shape of $L$ for all SG, and under this assumption, the $c_i$ obtained by means of \cref{eq:ci} match the SI indicators tabulated in \cite{Song2018}.

Although we have so far referred to SG in the discussion, all the formal conclusions hold for LG as well. As it is described in more detail in \cref{adap}, in this work we have adapted the TQC machinery to LG. This involved the calculation of all the EBRs of all LG (available in program \LBANDREP), which allows for the construction of their EBR matrices. With the information encoded in the EBR matrices in hand, we have calculated the independent topological indices for all LG, tabulated them, and made them available in the new program \LTOPOINDICES\; in the \webBCSAbbr. The new program \CheckT\; can calculate the values of the topological index for a given material, given its symmetry data vector, using solely TQC recipes. These indices are analogous to the indices for all SG previously calculated in refs \cite{filanomaly1} and \cite{Song2018} in the context of symmetry indicator theory. Also, the SI spaces for all space groups and layer groups were tabulated in \cite{Po2017}. As expected, the SI groups we obtain for all LG ($\Delta_{ii}>1$ elements) by means of TQC match with the abelian SI groups presented in Ref.~\cite{Po2017} for LG. 

\section{Additional details of highlighted materials}\label{section:additional}

\addcontentsline{stoc}{section}{\protect\numberline{\thesection}Additional details of highlighted materials}

\subsection{C$_3$Ta$_4$ in LG \lgsymbnum{72} (2D-TQC \serialidweb{1.1.12})}\label{subsec:c3ta4}
\addcontentsline{stoc}{subsection}{\protect\numberline{\thesubsection}C$_3$Ta$_4$ in LG \lgsymbnum{72} (2D-TQC \serialidweb{1.1.12})}

The band structure of the C$_3$Ta$_4$ in \lgsymbnum{72} is shown in \cref{fig:materials}A in the main text. 
Our calculations reveal that the symmetry data vector of the valence BR can be expressed as 
$$\{4\overline{\Gamma}_4\overline{\Gamma}_5\bigoplus4\overline{\Gamma}_6\overline{\Gamma}_7\bigoplus10\overline{\Gamma}_8\bigoplus10\overline{\Gamma}_9, 10\overline{M}_4\overline{M}_5\bigoplus 18\overline{K}_6, 13\overline{M}_3\overline{M}_4\bigoplus15\overline{M}_5\overline{M}_6\}.$$

According to the new \CheckT\; program, the BR cannot be expressed as a linear combination of EBRs. The EBRs compatible with \lgsymbnum{72} are available in the new program \LBANDREP. The material is a SEBR topological insulator.
Using the symmetry data vector, 
We can also calculate the topological index of the material, using the symmetry indicator approach refs. \cite{Po2017,Song2018}. The SI group of \lgsymbnum{72} is $\mathbb{Z}_2$ \cite{Po2017,Wang2019}, and is associated with the weak TI indicator indices as defined in Ref.~\cite{Po2017}. As a result, the indicator $z_{2w,1}$ determines the topology of the band structure, and the material must be a QSHI (knowing that it is SEBR). This index is present in all LG with inversion symmetry, and refers to one of the weak TI indices of 3D TIs \cite{fu2007topological,topo3D}. In 2D there are no "weak" TIs from the perspective of Refs \cite{fu2007topological} and \cite{topo3D}. This occurs because when we project the BZ to 2D, the expressions of strong and weak TI reduce to the same Fu-Kane formula for the $\mathbb{Z}_2$ invariant topological $\nu$ index of two-dimensional QSHI \cite{fu2007topological}.
In our case we have $z_{2w,1}=1$ as expected. All the topological insulators we found are tabulated in the tables of \cref{sec:ti}, with their corresponding topological indices. 

\subsection{Br$_2$Hg$_3$ in \lgsymbnum{80} (2D-TQC \serialidweb{1.2.133})}
\addcontentsline{stoc}{subsection}{\protect\numberline{\thesubsection}Br$_2$Hg$_3$ in \lgsymbnum{80} (2D-TQC \serialidweb{1.2.133})}

The band structure of the Br$_2$Hg$_3$ in \lgsymbnum{80} is shown in \cref{fig:materials}B in the main text. Our calculations reveal that the symmetry data vector of the valence band representation is 
$$\{6\overline{\Gamma}_7\bigoplus8\overline{\Gamma}_8\bigoplus7\overline{\Gamma}_9\bigoplus\overline{\Gamma}_{10}\bigoplus\overline{\Gamma}_{11}\bigoplus2\overline{\Gamma}_{12},9\overline{K}_7\bigoplus8\overline{K}_88\bigoplus8\overline{K}_9,10\overline{M}_5\bigoplus15\overline{M}_6\}$$. 
With the \CheckT\; program, we determine that the BR cannot be expressed as a linear combination of EBRs, and that it is a SEBR TI. \lgsymbnum{80} has SI group $\mathbb{Z}_6$, as determined in \cite{Po2017}. Following the approach detailed in \cref{topoindices}, we calculate the $\Delta$ matrix, in \cref{eq:smith} of \cref{topoindices}, and obtain a single diagonal element larger than one, $\Delta_{ii}=6$. The topological index $c_i$ corresponding to the SI group in this SG can be obtained in the \LTOPOINDICES\; program in the BCS. We have a single independent index

\begin{equation}
z_{6m,0}=\frac{3}{2}n_{\overline{\Gamma}_7}-\frac{5}{2}n_{\overline{\Gamma}_8}-\frac{1}{2}n_{\overline{\Gamma}_9}-\frac{3}{2}n_{\overline{\Gamma}_{10}}+\frac{5}{2}n_{\overline{\Gamma}_{11}}+\frac{1}{2}n_{\overline{\Gamma}_{12}}+3n_{\overline{K}_7}-5n_{\overline{K}_8}-n_{\overline{K}_9}+\frac{3}{2}n_{\overline{M}_5}-\frac{3}{2}n_{\overline{M}_6} \mod 6
\end{equation}

The topology of the system can thus be fully specified by the mirror Chern number $z_{6m,0}$. This index is protected by the mirror plane parallel to the structure plane, and the $C_6$ axis perpendicular to the structure plane present in \lgsymbnum{80}. There exist other topological indices in LG \lgsymb{80} which can be explicitly obtained from the independent index. One of these indices is the Fu-Kane weak topological index $z_{2w,1}$, protected by time-reversal symmetry. The presence of inversion symmetry allows the calculation of this index \cite{topo3D,fu2007topological}, as we saw in \cref{subsec:c3ta4}. It can be shown that in this LG, the indices $z_{6m,0}$ and $z_{2w,1}$ are related in the following way: if $z_{6m,0}$ is odd (even), then $z_{2w,1}$=1 ($z_{2w,1}$=0). This underscores the importance of non-independent topological indices, since they provide information about the physics of the system that may go unnoticed if only the independent index is considered. In fact, odd values of the mirror Chern number lead to a QSHI in LG \lgsymb{80}, while even values of $z_{6m,0}$ lead to a TCI. This information is encoded in the non-independent index $z_{2w,1}$. When a material in this type of system presents an odd value of the mirror Chern number and therefore also a finite value of $z_{2w,1}$, its topology has been termed as \emph{dual} \cite{dualtopo}. In our system we have $z_{6m,0}=1$,$z_{2w,1}=1$, and $z_{3,0}=2$.

\subsection{Re$_4$S$_8$ in \lgsymbnum{2} (2D-TQC \serialidweb{3.1.1})}
\addcontentsline{stoc}{subsection}{\protect\numberline{\thesubsection}Re$_4$S$_8$ in \lgsymbnum{2} (2D-TQC \serialidweb{3.1.1})}

Re$_4$S$_8$ in \lgsymbnum{2}, is an obstructed atomic insulator (OAI). As it is shown in \cref{table:oaiExperimental} in \cref{oai}, its band gap is 1.34 eV. In \cref{fig:materials}D of the main text we show its band structure. Its non-vanishing RSI are $\delta(b)=1$ and $\delta(d)=1$, calculated using the new program \RSI. Since only $e$ Wyckoff position are occupied in the material, any plane crossing Wyckoff positions $b$ or $d$ and not crossing $e$ Wyckoff positions will present obstructed edge states. The structure is very similar to Re$_4$Se$_8$, also an OAI, the edge states of which are studied in detail in \cref{sec:edgesates} and \cref{comp}.

\begin{figure*}
    \centering
    \includegraphics[width=0.3\linewidth]{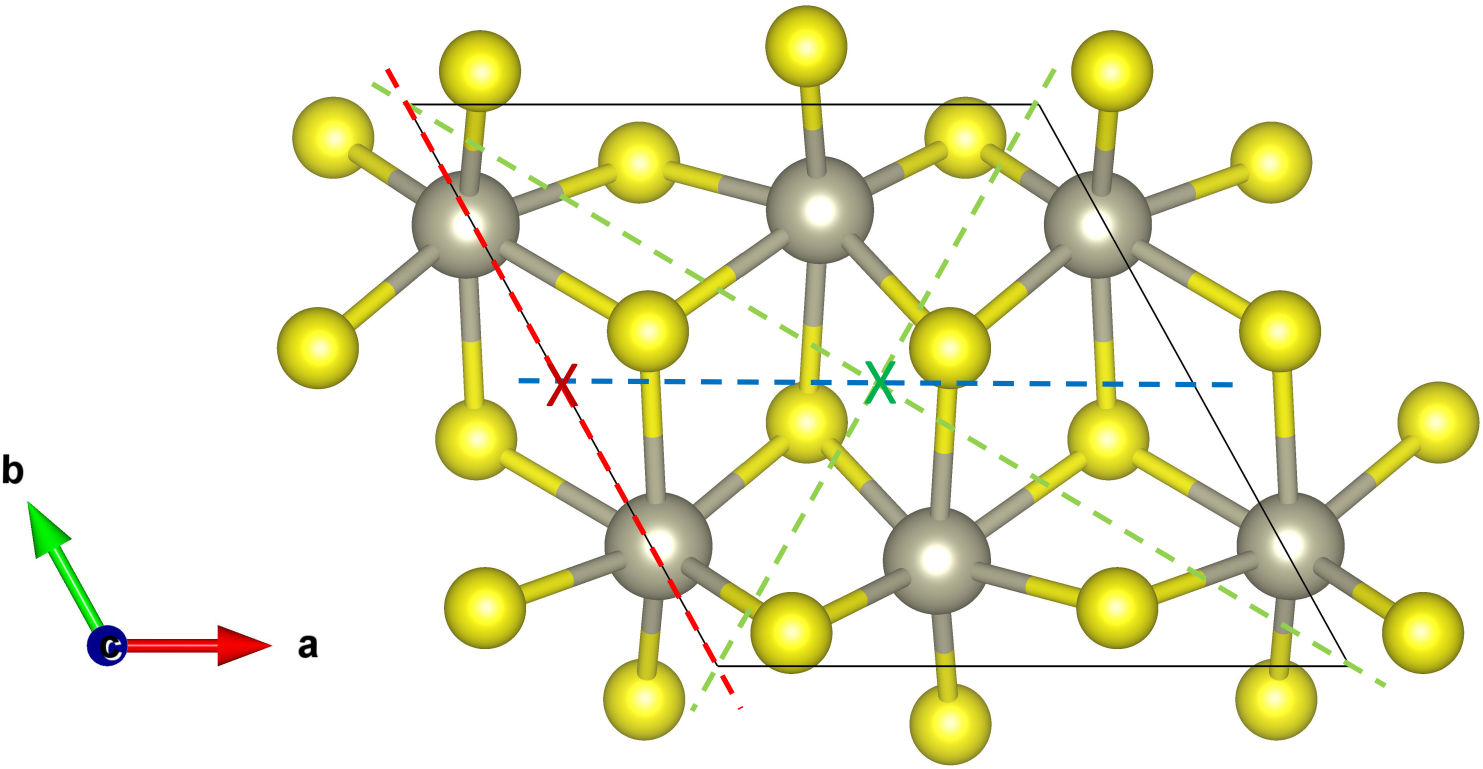}
    \caption{Unit cell of Re$_4$S$_8$, in \lgsymbnum{2}. Re atoms are indicated in grey color, while S atoms are indicated in yellow. The Wyckoff positions cross $b$ (0,1/2,0) and $d$ (1/2,1/2,0) are indicated in dark red and dark green crosses, respectively. With dashed lines, we represent lattice planes perpendicular to the crystal structure that cross Wyckoff positions $b$ and/or $d$ and do not cross any atomic position. The lattice planes are represented in a different color according to the Wyckoff position positions they cross: light red (green) for planes crossing $b$ ($d$) Wyckoff position positions, and blue for planes crossing both Wyckoff position positions.}
    \label{fig:res2}
\end{figure*}

In \cref{fig:res2} Re$_4$S$_8$, we showcase lattice planes crossing the Wyckoff positions with non-vanishing RSI. Ribbons cut along these planes and preserving inversion symmetry are expected to present obstructed edge states due to the filling anomaly.

\subsection{Cu$_2$H$_4$ in \lgsymbnum{72} (TQC \serialidweb{2.4.669}) and Experimental Materials with Fragile Bands near $E_F$}
\addcontentsline{stoc}{subsection}{\protect\numberline{\thesubsection}Cu$_2$H$_4$ in \lgsymbnum{72} (TQC \serialidweb{2.4.669}) and Experimental Materials with Fragile Bands near $E_F$}

Cu$_2$H$_4$ in \lgsymbnum{72} is an odd electron ($N_e=13$) ESFD topological semimetal according to TQC. Kramers degeneracy enforces degenerate irreps to be not filled at the Fermi level, failing to satisfy compatibility relations. The material becomes more interesting if we consider $N_e-1$ electrons.
At $N_e-1$, the material displays fragile cumulative topology. In \cref{table:fragile} we show the lowest energy EBRs of the material, up to $N_e-1$ electrons. Up to $N_e-1$, the symmetry data vector reads $\{\overline{\Gamma_4}\overline{\Gamma_5}\bigoplus4\overline{\Gamma_8}\bigoplus\overline{\Gamma_9},3\overline{K_4}\overline{K_5}\bigoplus3\overline{K_6},5\overline{M_3}\overline{M_4}\bigoplus\overline{M_5}\overline{M_6}\}$.

\begin{table}[htbp]
\centering
\begin{tabular}{c|c|c|c|c|c}
\hline\hline
BR & BR electron no. & BR topology & Cumulative BR & Cumulative electron no. & Cumulative topology \\
\hline
0 0 1 0 1 0 1 0 & 2 & FRAGILE & 0 0 1 0 1 0 1 0 & 2 & FRAGILE \\
\hline
0 0 0 1 0 1 1 0 & 2 & SEBR & 0 0 1 1 1 1 2 0 & 4 & SEBR \\
\hline
0 0 1 0 0 1 0 1 & 2 & SEBR & 0 0 2 1 1 2 2 1 & 6 & TRIVIAL \\
\hline
1 0 0 0 1 0 1 0 & 2 & TRIVIAL & 1 0 2 1 2 2 3 1 & 8 & TRIVIAL \\
\hline
0 0 1 0 0 1 1 0 & 2 & TRIVIAL & 1 0 3 1 2 3 4 1 & 10 & TRIVIAL \\
\hline
0 0 1 0 1 0 1 0 & 2 & FRAGILE & 1 0 4 1 3 3 5 1 & 12 & FRAGILE \\
\end{tabular}
\caption{Details of band representations at electron occupations below 12 in Cu$_2$H$_4$. The vectors in the first and fourth column refer to the following corep multiplicities $\{\overline{\Gamma_4}\overline{\Gamma_5},\overline{\Gamma_6}\overline{\Gamma_7},\overline{\Gamma_8},\overline{\Gamma_9},\overline{K_4}\overline{K_5},\overline{K_6},\overline{M_3}\overline{M_4},\overline{M_5}\overline{M_6}\}$.}
\label{table:fragile}
\end{table}

No real materials with fragile cumulative topology at the Fermi level have been reported, to our knowledge. The main reason behind this is that, in a system with fragile cumulative topology, adding a trivial EBR can turn it into a trivial system \cite{fragile0,basicbands2020}. This is in contrast to systems with SEBR or NLC TI cumulative topology. Thus, it is unlikely that the fragile cumulative topology \emph{survives until} the Fermi level \cite{fragilemonoid}, especially if the material presents many filled bands. 

As stated in the main text, fragile bands at the Fermi level may be connected to the correlated states in twisted bilayer graphene (TBG) without the particle-hole symmetry \cite{fragilegraphene}. We note that in this work, we have encountered a relatively large number of monolayers with fragile EBRs at or near the Fermi level. In \cref{table:fragilemat} we list some examples, corresponding to experimentally existing materials.
Although an exhaustive list of materials is not explicitly listed here, the reader is referred to the \webtwoDTQCAbbr, introduced in this work, for more details.

Considering the Cu$_2$H$_4$, it is not a thermodynamically nor dynamically stable material, thus it is not likely to be fabricated in the future.
Nevertheless, this material may become an interesting starting point to study cumulative fragile topology in real materials.
\begin{table}[htbp]
\centering
\begin{tabular}{c|c|c|c|c|c}
\hline\hline
LG & Formula & 2D-TQC & database & topology  & fragile band location \\
\hline
69 & BiBrTe & \serialidweb{6.1.10} & \CtwoDB & LCEBR & VB\\
\hline
69 & BiClTe & \serialidweb{6.1.11} & \CtwoDB & LCEBR &VB\\
\hline
69 & BiITe & \serialidweb{6.1.13} & \CtwoDB &  LCEBR&VB\\
\hline
72 & Bi$_2$PbSe$_4$ & \serialidweb{6.1.18} & \CtwoDB & LCEBR  &VB\\
\hline
72 & Bi$_2$PbTe$_4$ & \serialidweb{6.1.19} & \CtwoDB & LCEBR  &VB\\
\hline
72 & Bi$_2$SeTe$_2$ & \serialidweb{6.1.27} & \CtwoDB & LCEBR  &VB\\
\hline
72 & Bi$_2$SnTe$_4$ & \serialidweb{6.1.29} & \CtwoDB & LCEBR  &VB\\
\hline
72 & Bi$_2$ & \serialidweb{1.1.9} & \CtwoDB & SEBR  &CB\\
\hline
72 & C$_2$H$_2$ & \serialidweb{3.1.15} & \CtwoDB & OAI  &VB\\
\hline
72 & CO$_2$Ti$_2$ & \serialidweb{6.1.34} & \CtwoDB & LCEBR  &VB\\
\hline
72 & CdBr$_2$ & \serialidweb{6.1.37} & \CtwoDB & LCEBR  &VB\\
\hline
72 & CdI$_2$ & \serialidweb{6.1.37} & \CtwoDB & LCEBR  &VB\\
\hline
72 & HfS$_2$ & \serialidweb{6.1.43} & \CtwoDB & LCEBR  &VB\\
\hline
72 & HfSe$_2$ & \serialidweb{6.1.44} & \CtwoDB & LCEBR  &VB\\
\hline
72 & In$_2$S$_3$ & \serialidweb{6.1.47} & \CtwoDB & LCEBR  &VB\\
\hline
72 & PdTe$_2$ & \serialidweb{6.1.55} & \CtwoDB & LCEBR  &VB,CB\\
\hline
72 & PtS$_2$ & \serialidweb{6.1.56} & \CtwoDB & LCEBR  &VB,CB\\
\hline
72 & PtSe$_2$ & \serialidweb{6.1.58} & \CtwoDB & LCEBR  &VB,CB\\
\hline
72 & S$_2$Zr & \serialidweb{6.1.61} & \CtwoDB & LCEBR  &VB\\
\hline
72 & Sb$_2$Se$_3$ & \serialidweb{6.1.65} & \CtwoDB & LCEBR  &VB\\
\hline
72 & Sb$_2$Te$_3$ & \serialidweb{6.1.64} & \MCtwoD &LCEBR&VB\\
\hline
72 & Se$_2$Zr & \serialidweb{6.1.73} & \CtwoDB & LCEBR  &VB\\
\hline

\end{tabular}
\caption{Materials experimentally existing in monolayer or few-layer form, with a connected group of bands isolated to the rest (indicated as BR in the table) at the top (bottom) of the valence (conduction) band, indicated as VB (CB). The layer group, formula, database, database code, TQC cumulative topology, and location of the fragile BR are given.}
\label{table:fragilemat}
\end{table}

\section{Topological semimetals: general remarks and highlighted materials}\label{suppSM}
\addcontentsline{stoc}{section}{\protect\numberline{\thesection}Topological semimetals: general remarks and highlighted materials}

As it was introduced in \cref{adap}, compatibility relations define how the irreps at high symmetry points (HSPs), lines, and planes of the Brillouin Zone (BZ) can connect to each other, according to the symmetry of the system. 
A small corep of a little group admits a matrix representation, whereby each symmetry operation can be expressed. In general, the matrices of a small corep $\rho^i_{G^{\mathbf{k}}}$ of the little group of an HSP $\mathbf{k}$, also form a representation, which may or may not be irreducible, of the little group ($G^{\mathbf{k}_l}$) of any line (with $\mathbf{k}$ vector $\mathbf{k_l}$) connected to the HSP. This representation can be expressed as a direct sum of the small coreps $\rho^j_{G^{\mathbf{k_l}}}$ of $G^{\mathbf{k}_l}$ \cite{bradlyn2017topological,basicbands2020}:

\begin{equation}
\rho^i_{G^{\mathbf{k}}}\downarrow G^{\mathbf{k}_l} =\bigoplus_{i=1}^{s}m_{ij}^{\mathbf{k},\mathbf{k_l}}\rho^j_{G^{\mathbf{k_l}}}
\end{equation}

Here $s$ the number of irreducible representations of $G^{\mathbf{k}_l}$ and $m_{ij}^{\mathbf{k},\mathbf{k_l}}$ is the multiplicity of $\rho^j_{G^{\mathbf{k}}}$ in the decomposition of $\rho^i_{G^{\mathbf{k}}}$. Irreps $\rho^j_{G^{\mathbf{k}}}$ are termed the subduced irreps of the little group $G^{\mathbf{k}}$ from the irreps $\rho^i_{G^{\mathbf{k}}}$ of the little group of the HSP. If a line (band) connects two HSPs then, these two HSPs define according to what small coreps of the little group of the line the band can transform, if compatibility relations are to be satisfied.
The compatibility relations between two HSP (double) layer groups can be obtained by means of the new \DLCOMPREL\; program. In the Topological Quantum Chemistry (TQC) formalism, the symmetry data vector is constructed with eigenvalues of the $N_e$ lowest energy bands, where $N_e$ is the number of electrons. If bands forming the symmetry data vector do not satisfy compatibility relations, the material is termed a topological semimetal. Bands not satisfying compatibility relations manifest as crossings (but not all Fermi surface crossings necessarily violate compatibility relations) 
of the bands containing the $N_e$th and $N_e+1$th lowest energy electrons. If the $N_e$th and $N_e+1$th electrons belong to the same (degenerate) corep at some point of the BZ, the material will be termed as Enforced Semimetal with Fermi Degeneracy (ESFD). This includes, but is not limited to, time-reversal symmetric (TRS) materials with an odd number of electrons. In some cases, symmetries enforce crossings away from TRS HSP that fail to satisfy compatibility relations (we will see examples below). In these cases, the material will be termed Enforced Semimetal (ES).

In this work, we have found \TwoDDBMaterialsSMWithSOC\; topological semimetals, \TwoDDBMaterialsSMWithSOCAndEvenNbrElectrons\ of which are even-electron materials. All of the topological semimetals we have found are listed in \cref{section:tableSM}. Out of these, we have highlighted some materials. In the main text \cref{fig:materials}C we show the band structure of Eu$_2$I$_8$La$_2$  in \lgsymbnum{7}, while four more TSM are mentioned in the main text. The cases selected correspond to systems where TQC allows for a complete characterization of their non-trivial FS.
In this section we will showcase, in more detail, the topological properties of the crossings of these six materials.

\subsection{Eu$_2$I$_8$La$_2$ in \lgsymbnum{7} (2D-TQC \serialidweb{2.2.30})}\label{subsection:SM2}
\addcontentsline{stoc}{subsection}{\protect\numberline{\thesubsection}Eu$_2$I$_8$La$_2$ in \lgsymbnum{7} (2D-TQC \serialidweb{2.2.30})}

The system, whose band structure is shown in \cref{fig:materials}C of the main text, is a computationally exfoliated compound with 94 electrons. In the band plot, the bands belonging to the lowest energy 94 bands are shown in blue, and the bands in red have higher energy. There are two crossings between the 94th and 95th bands, at high symmetry points $A$ and $B$. From our TQC analysis, performed with the \CheckT\; program, we find that the elementary band representation (EBR) crossed by the Fermi level is formed by the four-dimensional irrep $\overline{A}_2\overline{A}_2$ at the $A$ point. At HKS $B$, the EBR is formed of 4D irrep $\overline{B}_2\overline{B}_2$. Since the EBR contains two electrons, these two irreps are not fully filled, and the material is a topological semimetal, of ESFD type. The crossings at $A$ and $B$ are 4D. If we look at the compatibility relations using \DLCOMPREL\; program, we see both $\overline{A}_2\overline{A}_2$ and $\overline{B}_2\overline{B}_2$ are subduced into a pair of the two-dimensional irrep $\overline{F}_3\overline{F}_4$ in the general k vector of the BZ. Thus the material is a Dirac semimetal.

\begin{figure*}
    \centering
    \includegraphics[width=0.9\linewidth]{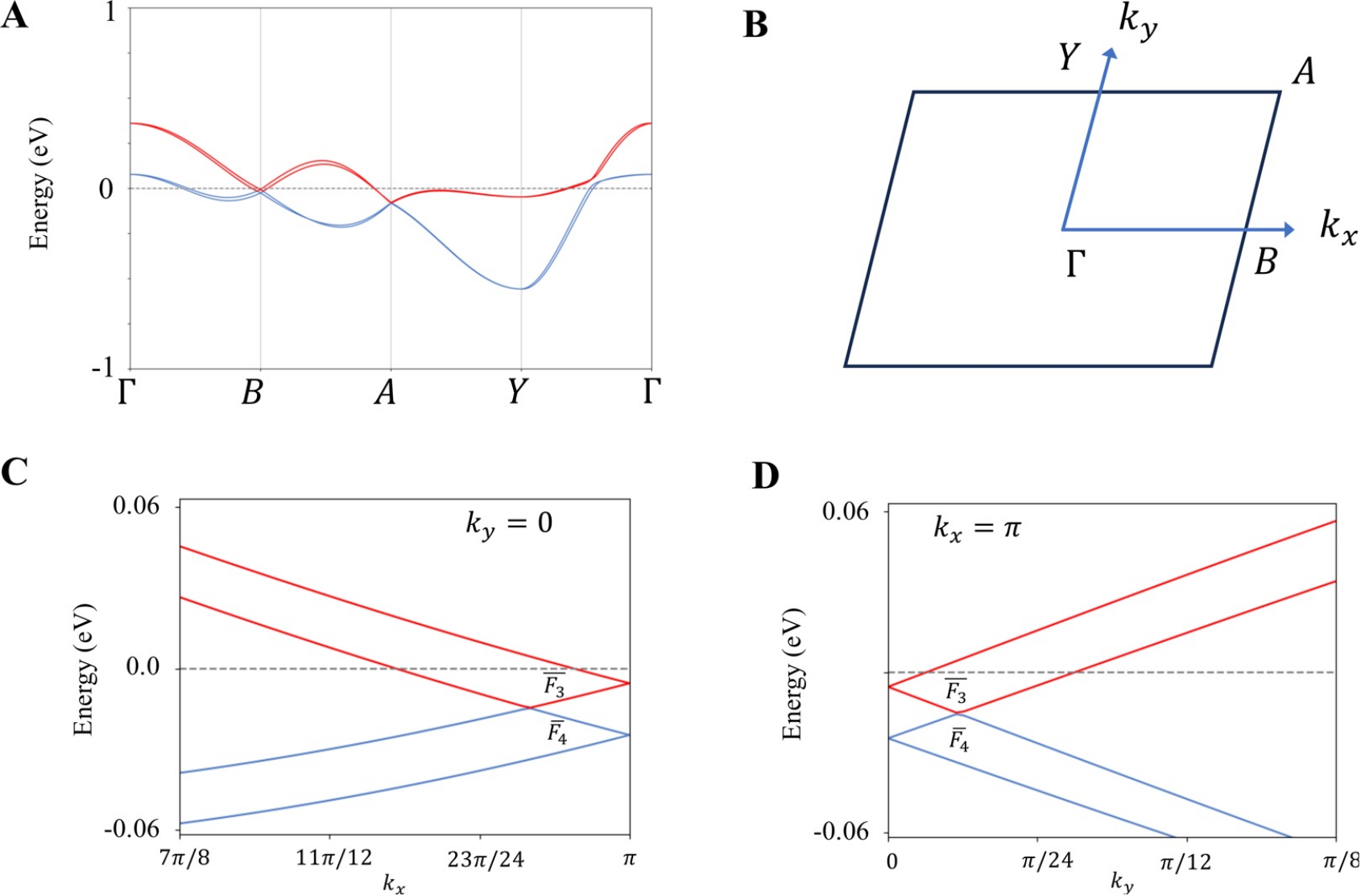}
    \caption{(A) Band structure of O$_6$P$_2$Sn$_2$ in \lgsymbnum{5} (2D-TQC \serialidweb{2.2.25}). (B) Brillouin Zone (BZ) of layer group $P$11$a$ (no 5). The High Symmetry Points are indicated with capital letters. (B) ((C)) Band structure zoomed onto the vicinity of its fermi level band crossing points ($\Gamma-B$ and $B-A$ branches, respectively). Only the highest valence band and lowest conduction bands are shown. $\overline{F}_3$ and $\overline{F}_4$ are the irreps according to which these bands transform in the vicinity of the $B$ point of the BZ.}
    \label{fig:SM1}
\end{figure*}

\subsection{O$_6$P$_2$Sn$_2$  in \lgsymbnum{5} (2D-TQC \serialidweb{2.2.25})}\label{subsection:SM1}
\addcontentsline{stoc}{subsection}{\protect\numberline{\thesubsection}O$_6$P$_2$Sn$_2$  in \lgsymbnum{5} (2D-TQC \serialidweb{2.2.25})}

The system is non-centrosymmetric, with \lgsymbnum{5}, and contains 74 electrons. In our TQC analysis, done with the program \CheckT in the \webBCS, introduced in this work, we find it is a topological semimetal of ES type. The BZ of LG no. 5 is a parallelogram, and is shown in \cref{fig:SM1}A. The band structure presents two crossings around point $B$ in the BZ. In order to inspect the crossings in more detail, we have plotted the zoomed-in band structure near $B$ point in the BZ. In \cref{fig:SM1}B we show the $\Gamma-B$ branch in the BZ, the path going from $k_x=\frac{7\pi}{8}$ to $k_x=\pi$ with $k_y=0$. We can see a crossing point between the points $(23\pi/24,0)$ and $(\pi,0)$. In \cref{fig:SM1}C we show the $B-A$ branch, where $k_y$ goes from 0 to $\frac{\pi}{9}$, while $k_x=\pi$. In this branch, a linear crossing happens between the points $(\pi,0)$ and $(\pi,\pi/24)$. Looking at the eigenvalue spectrum at point $B$ of the BZ, we see the highest occupied Kramers degenerate band pair transforms according to the two-dimensional irrep $\overline{B}_4\overline{B}_4$. The compatibility relations, as it can be seen in the new \DLCOMPREL\; program in the Bilbao Crystallographic Server (BCS), require the subduction of this irrep into a pair of one-dimensional $\overline{F}_4$ irreps away from maximal $\mathbf{k}$ points, due to the lack of inversion symmetry. The lowest non-occupied degenerate band pair at point $B$ transforms according to $\overline{B}_3\overline{B}_3$ irrep. The irrep is further subduced into a pair of $\overline{F}_3$ irreps at generic $\mathbf{k}$ points in the BZ. This implies, as it can be seen in \cref{fig:SM1}B, that the crossing in the figure breaks compatibility relations, yielding a topologically non-trivial crossing. The crossing is of enforced type, and is thus preserved upon any symmetry-preserving perturbation.

\begin{figure*}
    \centering
    \includegraphics[width=0.7\linewidth]{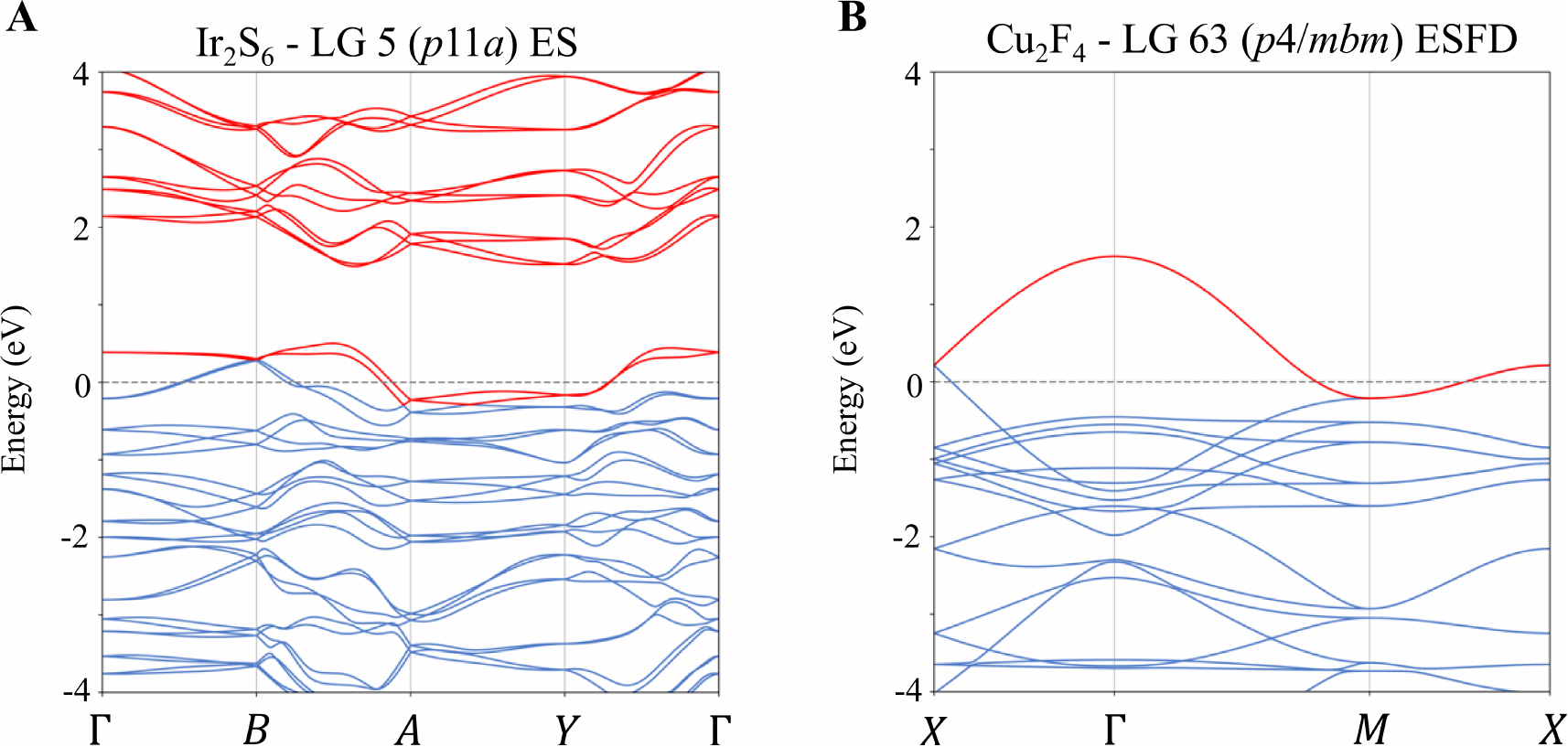}
    \caption{(A) Band structure of dynamically stable Ir$_2$S$_6$, a non-inversion symmetric ES type, topological semimetal with multiple non-trivial crossings near the Fermi level. (B) Computationally exfoliated Cu$_2$F$_4$, an ESFD type nodal line semimetal.
    In all cases, the $N_e$ lowest energy bands, where $N_e$ is the number of electrons in the unit cell of the material, are drawn in blue, and higher energy bands are colored in red. The layer group of the compounds is indicated in the title of each of the plots.}
    \label{fig:SM2}
\end{figure*}

\subsection{Ir$_2$S$_6$ in \lgsymbnum{5} (2D-TQC \serialidweb{2.3.34})}\label{subsection:SM3}
\addcontentsline{stoc}{subsection}{\protect\numberline{\thesubsection}Ir$_2$S$_6$ in \lgsymbnum{5} (2D-TQC \serialidweb{2.3.34})}

In \cref{fig:SM2}A we show the band structure of the dynamically stable Ir$_2$S$_6$ in \lgsymbnum{5}. The system has 54 electrons. In our TQC analysis we find that the Fermi level crosses an EBR of 4 electrons, two of which are occupied. The analysis determines the materials is a topological semimetal of ES type. In the band structure we can see four band crossings at the Fermi surface. At least two of these crossings are non-trivial and enforced by the non-symmorphic glide plane parallel to the structure plane, as we will now demonstrate. The EBR is composed of a pair of two-dimensional $\overline{\Gamma}_3\overline{\Gamma}_4$ irreps at the $\Gamma$ point, each of which is decomposed into the direct sum of $\overline{F}_3$ and $\overline{F}_4$ irreps away from high symmetry points in the BZ. On the other hand, the EBR is formed by the 2D irreps $\overline{B}_3\overline{B}_3$ and $\overline{B}_4\overline{B}_4$ at the $B$ point. In a similar spirit as in the case of Eu$_2$I$_8$La$_2$, $\overline{B}_3\overline{B}_3$ ($\overline{B}_4\overline{B}_4$) is decomposed into a pair of $\overline{F}_3$ ($\overline{F}_4$) irreps in the $\Gamma-B$ line. 
Two crossing bands with $\overline{F}_3$ and $\overline{F}_4$ irreps along $\Gamma-B$ line will give symmetry-protected degeneracy, as they break the compatibility relations. 

In the $A-Y$ line, another non-trivial crossing emerges, driven by the same mechanism as in the $\Gamma-B$ line: while the irreps at $Y$ are subduced into direct products of irreps $\overline{F}_3$ and $\overline{F}_4$, irreps of $A$ are subduced into pairs of identical irreps. There exist two more crossings belonging to the FS, in the $B-A$ line. To depict their topology further analysis is needed, which we leave for future work.

$\Gamma-B$ and $A-Y$, resemble the so-called Hourglass dispersion \cite{hourglass2016}. Surface bands emerging in QSH insulators with hourglass shape protected by non-symmorphic symmetries were introduced in Ref.~\cite{hourglass2016}.
Hourglass shaped dispersions have been observed in the bulk bands of topological semimetals too, including both 3D \cite{hourgla2} and 2D \cite{hour2d} examples.

\subsection{Cu$_2$F$_4$ in \lgsymbnum{63} (2D-TQC \serialidweb{2.2.203})}
\addcontentsline{stoc}{subsection}{\protect\numberline{\thesubsection}Cu$_2$F$_4$ in \lgsymbnum{63} (2D-TQC \serialidweb{2.2.203})}

Cu$_2$F$_4$, in \lgsymbnum{63} is a computationally exfoliated material, with 50 electrons. In \cref{fig:SM2}C we show its band structure with SOC taken into account. We can see the bands containing the 50th and 51th electrons appear degenerated in the $M-X$ line of the BZ. The band close to the Fermi level is formed by the 4D irrep $\overline{M}_6\overline{M}_7$ ($\overline{X}_3\overline{X}_4$) at the $M$ ($X$) point. 
Along the $M-X$ line, the bands at the Fermi level maintain the 4-fold degeneracy. The material is therefore a nodal line semimetal with a 4-fold nodal line along $M-X$. 

\subsection{Ir$_2$O$_4$ in \lgsymbnum{15} (2D-TQC \serialidweb{2.3.134})}
\addcontentsline{stoc}{subsection}{\protect\numberline{\thesubsection}Ir$_2$O$_4$ in \lgsymbnum{15} (2D-TQC \serialidweb{2.3.134})}

\begin{figure*}
    \centering
    \includegraphics[width=0.7\linewidth]{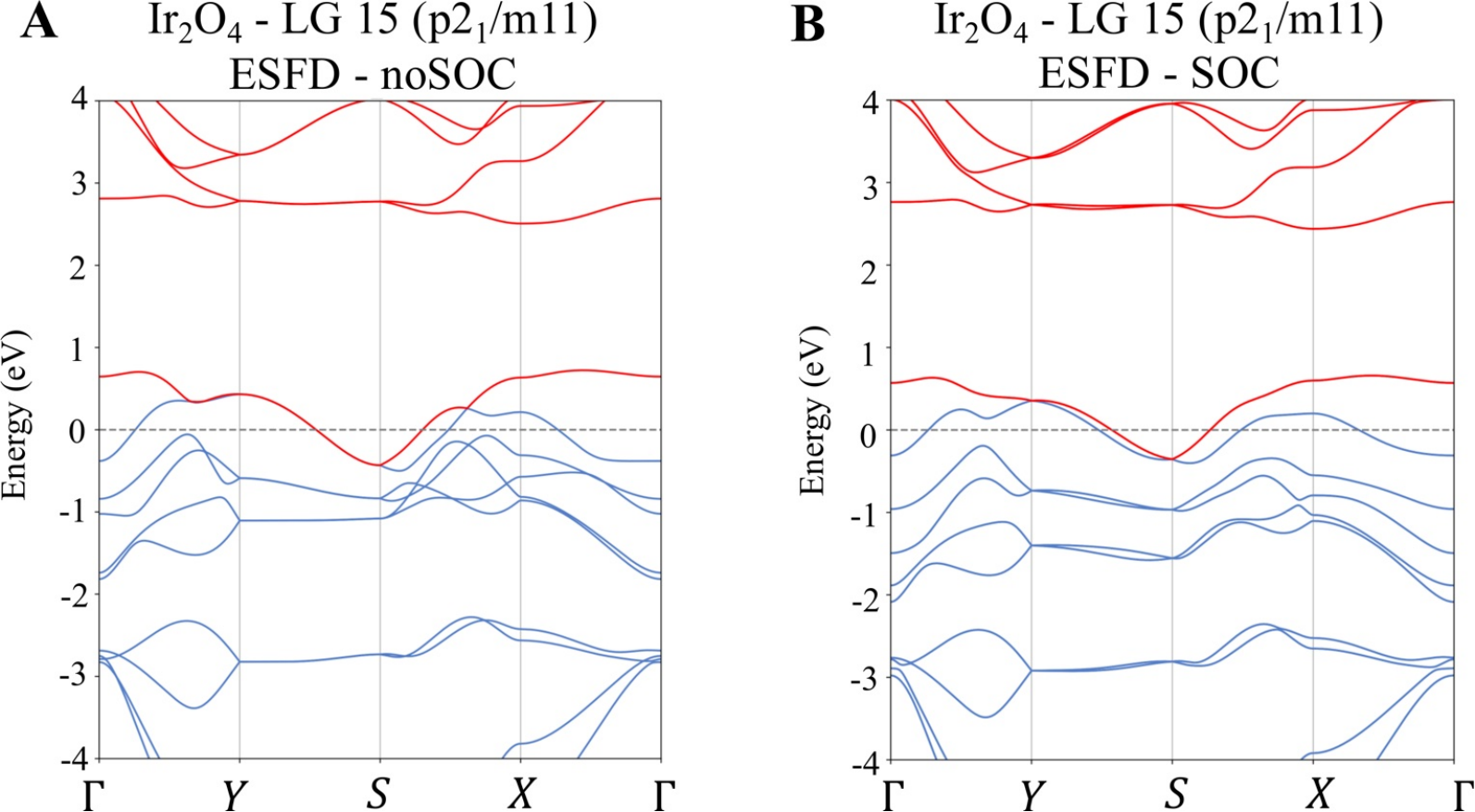}
    \caption{(A) ((B)), the band structure of the dynamically stable material Ir$_2$O$_4$, without (with) SOC included. The material is a nodal line semimetal without SOC and a Dirac semimetal with SOC turned on.
    The $N_e$ lowest energy bands, where $N_e$ is the number of electrons in the unit cell of the material, are drawn in blue, and higher energy bands are colored in red. The layer group and the noSOC/SOC status of the compounds is indicated in the title of each of the plots.}
    \label{fig:ir2s6}
\end{figure*}

Ir$_2$O$_4$ in \lgsymbnum{15}, a dynamically stable material, with 42 electrons. In \cref{fig:ir2s6}A we show the band structure of the system without the presence of spin-orbit coupling (SOC). Our TQC analysis reveals the material is a topological semimetal of ESFD type. The EBR (without SOC) that is crossed by the Fermi level is formed by two bands. 
A nodal line appears along the line $Y-S$.  
This is confirmed by a closer look at the eigenvalue spectra at maximal $k$ points $Y$ and $S$. The EBR at the $S$ ($Y$) is formed of the degenerate irrep $S_1$ ($Y_1$) formed of two Kramers degenerate bands. The irreps of both points subduce into the same (degenerate) irrep, $C_1C_2$ in the $Y-S$ line, thus confirming the nodal line semimetal character of the compound without SOC, protected by the $\{M_x|\frac{1}{2},0,0\}$ and ${C_{2x}|\frac{1}{2},0,0}\cdot\mathcal{T}$ symmetries along the $Y-S$ line. 
When SOC is considered (as shown in \cref{fig:ir2s6}B), the bands at HSP $S$ and $Y$ remain degenerate and are 4-fold as the spin degree of freedom is taken into account. In contrast, SOC splits the states into two 2-fold degenerate bands along the $Y-S$ line, corresponding to irreps $\overline{C}_3\overline{C}_3$ and $\overline{C}_4\overline{C}_4$. Thus, the nodal line is broken and only the two 4D crossing points at $S$ and $Y$ remain, with irrep $\overline{Y}_2\overline{Y}_2$ ($\overline{S}_2\overline{S}_2$) at $Y$ ($S$) point. The material is thus a topological semimetal with 4D crossings. 

\section{Calculation of edge states in selected materials}\label{app:sec:edge-states}
\addcontentsline{stoc}{section}{\protect\numberline{\thesection}Calculation of edge states in selected materials}

\subsection{DFT calculations on finite ribbons}
\addcontentsline{stoc}{subsection}{\protect\numberline{\thesubsection}DFT calculations on finite ribbons}

This section presents a systematic comparison between edge states derived from tight-binding models and those obtained through ab initio calculations of relaxed ribbons. We examine five representative cases: (i) the topological insulators Cu$_4$I$_8$, Bi$_2$Br$_2$, and As$_2$Cl$_2$, and (ii) the obstructed atomic insulators (OAIs) Re$_4$Se$_8$ and C$_6$N$_2$. 

\begin{figure*}
    \centering
    \includegraphics[width=0.99\linewidth]{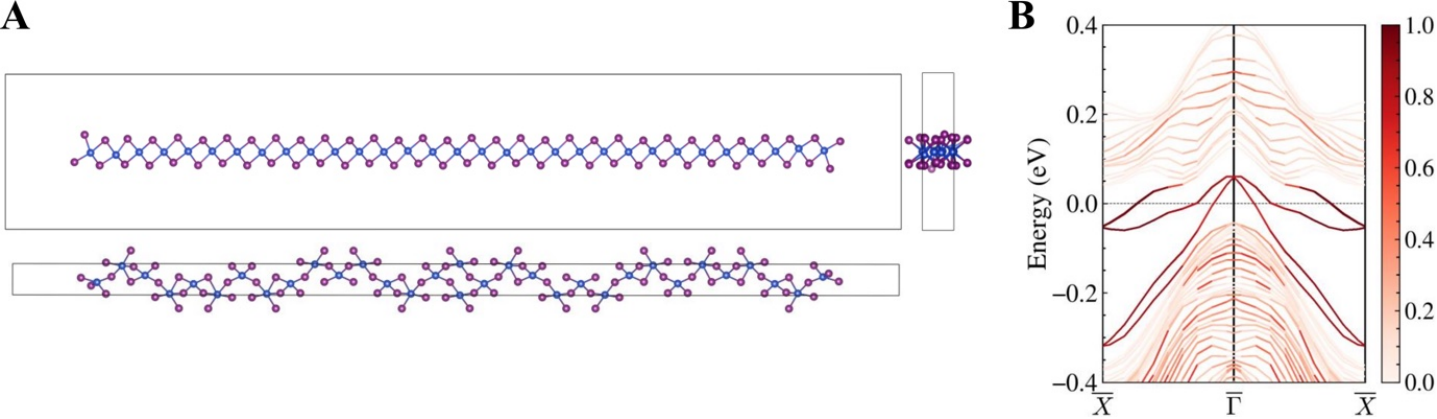}
    \caption{Edge state analysis of finite Cu$_4$I$_8$ ribbons by DFT. (A) Relaxed ribbon structure, projected in the three crystallographic axes. Cu (I) atoms are represented by blue (purple) spheres. (B) Edge-projected band structure of the relaxed ribbon, where a surface Dirac cone centered at $\Gamma$ is observed.} 
    \label{fig:ribbon-dft-cui2}
\end{figure*}

In \cref{fig:ribbon-dft-cui2} 
We show the atomic structure and the edge-projected band structure of a relaxed ribbon of monolayer Cu$_4$I$_8$ in \lgsymbnum{16} (2D-TQC \serialidweb{1.3.26}), with 8 unit cells in the non-periodic in-plane direction. In \cref{fig:ribbon-dft-cui2}A, we show the relaxed ribbon structure projected along the three Cartesian directions, with one periodic direction and a vacuum layer in each of the two open directions. In \cref{fig:ribbon-dft-cui2}B, we display the band structure near the Fermi level of the relaxed ribbon, with atomic weights from atoms within two unit cells near the edge. We find the surface Dirac cone centered at the $\Gamma$ point, in close similarity to the tight-binding results in the main text. In \cref{fig:ribbon-dft-cui2}, the surface Dirac cone centered at the $\Gamma$ point is contributed by the two open edges in the inversion-symmetric ribbon. There also exist trivial edge states at the $X$ point that emerge from the bulk continuum. These trivial edge states are merged into the bulk bands in \cref{fig:ti} of the main text. Thus by adding surface perturbations like hydrogenation, the trivial edge states at $X$ could be pushed into the bulk continuum. 

\begin{figure*}
    \centering
    \includegraphics[width=0.8\linewidth]{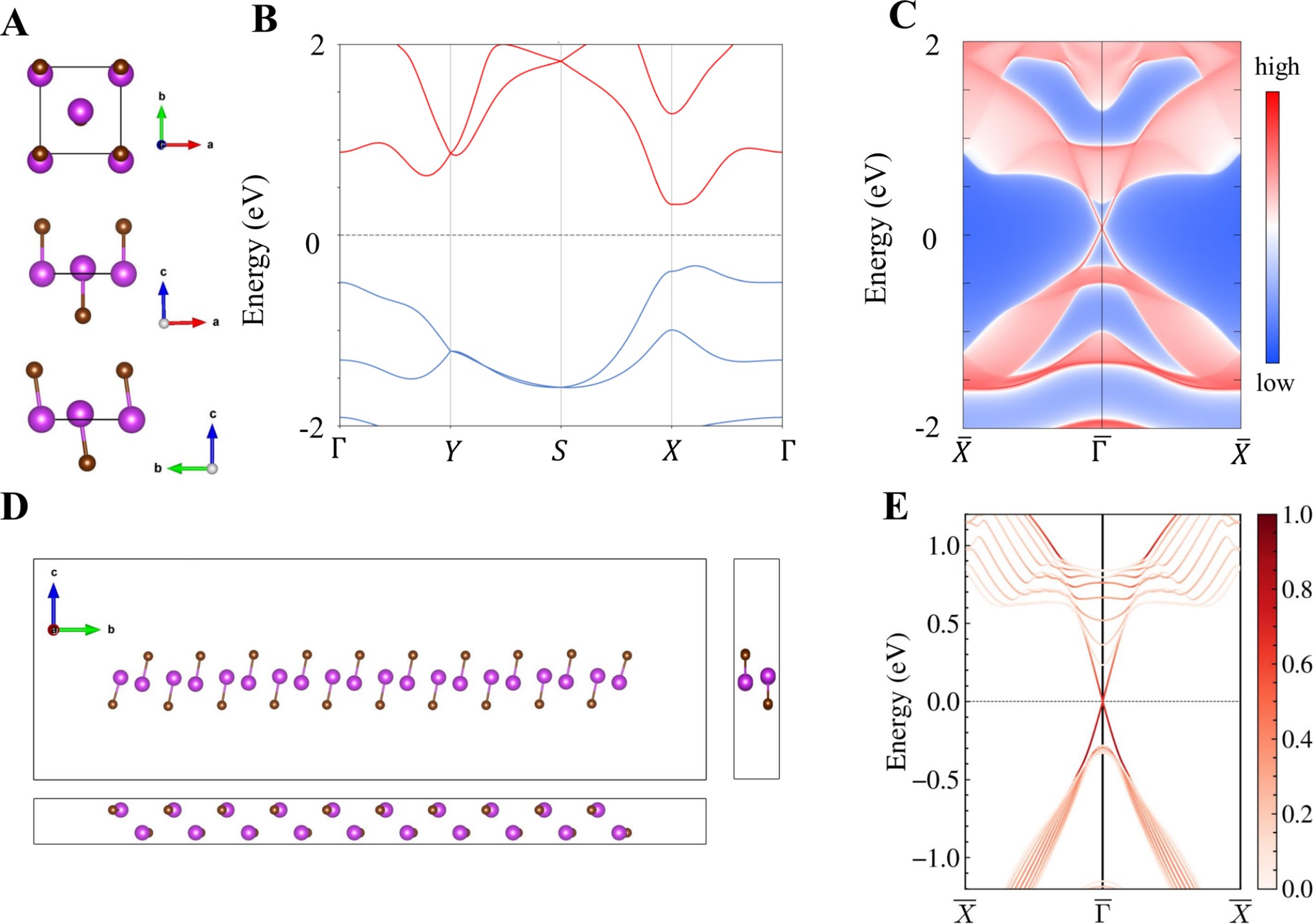}
    \caption{Edge state analysis of Bi$_2$Br$_2$.
    (A) Top and lateral views of the compound. Bi atoms are indicated in purple color, while Br atoms are indicated in brown color. (B) Band structure calculated by DFT with SOC included, where valence bands are represented in red and conduction bands in blue. (C) Spectral function of the tight-binding model on a semi-infinite ribbon, cut along the (100) direction of the structure. Red (blue) color indicates a higher (lower) contribution of edge atoms.
    (D) Relaxed and unrelaxed ribbon structures. The projections onto the three crystallographic axes are shown in the two cases. In \textbf{e}, we present the DFT band structure of a relaxed ribbon of \ch{Bi2Br2} with weights from the edge atoms. A clear surface Dirac cone is observed. 
    } 
    \label{fig:ribbon-dft-bibr}
\end{figure*}

In \cref{fig:ribbon-dft-bibr} we show the atomic and electronic properties of monolayer Bi$_2$Br$_2$ in \lgsymbnum{15} (2D-TQC \serialidweb{1.3.6}). In \cref{fig:ribbon-dft-bibr}A, we show the monolayer structure projected along the three Cartesian directions; in \cref{fig:ribbon-dft-bibr}B, we display its band structure, where valence (conduction) bands are presented in blue (red). In \cref{fig:ribbon-dft-bibr}C, we show the spectral function corresponding to a semi-infinite ribbon of the material, calculated after building a tight-binding model based on the monolayer. In \cref{fig:ribbon-dft-bibr}D, we show the structure of the relaxed (top) and unrelaxed (bottom) 10-unit-cell-wide ribbon of the material, with one periodic direction and a vacuum layer in each of the two open directions. In \cref{fig:ribbon-dft-bibr}D we display the band structure near the Fermi level of the relaxed ribbon, with atomic weights from atoms within 2 unit cells near the edge. We find the surface Dirac cone centered at the $\Gamma$ point, in close similarity to the tight-binding results \cref{fig:ribbon-dft-bibr}C. 
We remark that another \ch{Bi4Br4} structure in \lgsymbnum{18} is reported in literature~\cite{zhou2014large, shumiya2022evidence, yang2022large} with existing bulk form, but is not present in our database. We find the monolayer \ch{Bi4Br4} (\lgsymbnum{18}) phase has a 0.1 eV/atom lower energy compared with the \ch{Bi2Br2} phase (\lgsymbnum{15}). Thus \ch{Bi2Br2} phase is less stable.

\begin{figure*}
    \centering
    \includegraphics[width=0.99\linewidth]{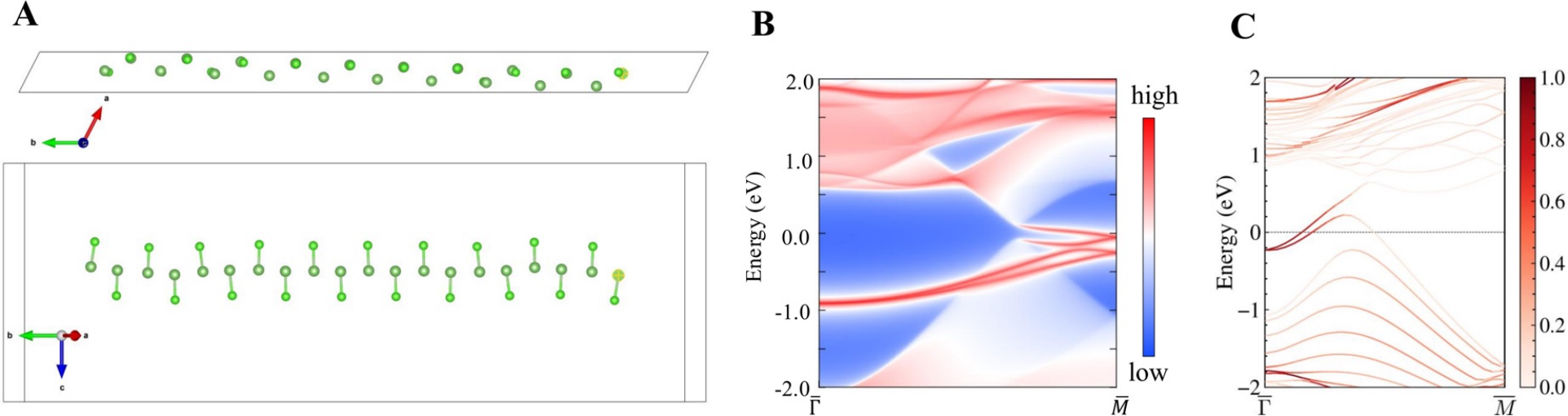}
    \caption{Edge state analysis of finite As$_2$Cl$_2$ ribbons by DFT. (A) Relaxed ribbon structure in two different views. As atoms are indicated in dark green color, while Cl atoms are indicated in light green color. (B) Spectral function of a semi-infinite ribbon cut along the $[1,0]$ direction, calculated on a tight-binding model of the monolayer. (C) Edge-projected band structure of the relaxed ribbon along the same $[1,0]$ cut.
   } 
    \label{fig:ribbon-dft-ascl}
\end{figure*}

In \cref{fig:ribbon-dft-ascl}A, we show a relaxed 10-unit-cell-wide ribbon of As$_2$Cl$_2$ in \lgsymbnum{72} (2D-TQC \serialidweb{1.3.193}), where the ribbon is cut along the $[10]$ direction. After building a tight-binding model based on the monolayer, we calculate the spectral function corresponding to a semi-infinite ribbon. The result is shown in \cref{fig:ribbon-dft-ascl}B. A surface Dirac cone centered at $\bar{M}$ emerges. Another pair of localized states, presumably emerging from dangling bonds, can also be observed below the Dirac cone. In \cref{fig:ribbon-dft-ascl}C, we show the ab initio bands of the relaxed 10-unit-cell-wide ribbon, projected on the edge atoms. A similar surface Dirac emerges as in the tight-binding result.

\begin{figure*}
    \centering
    \includegraphics[width=0.99\linewidth]{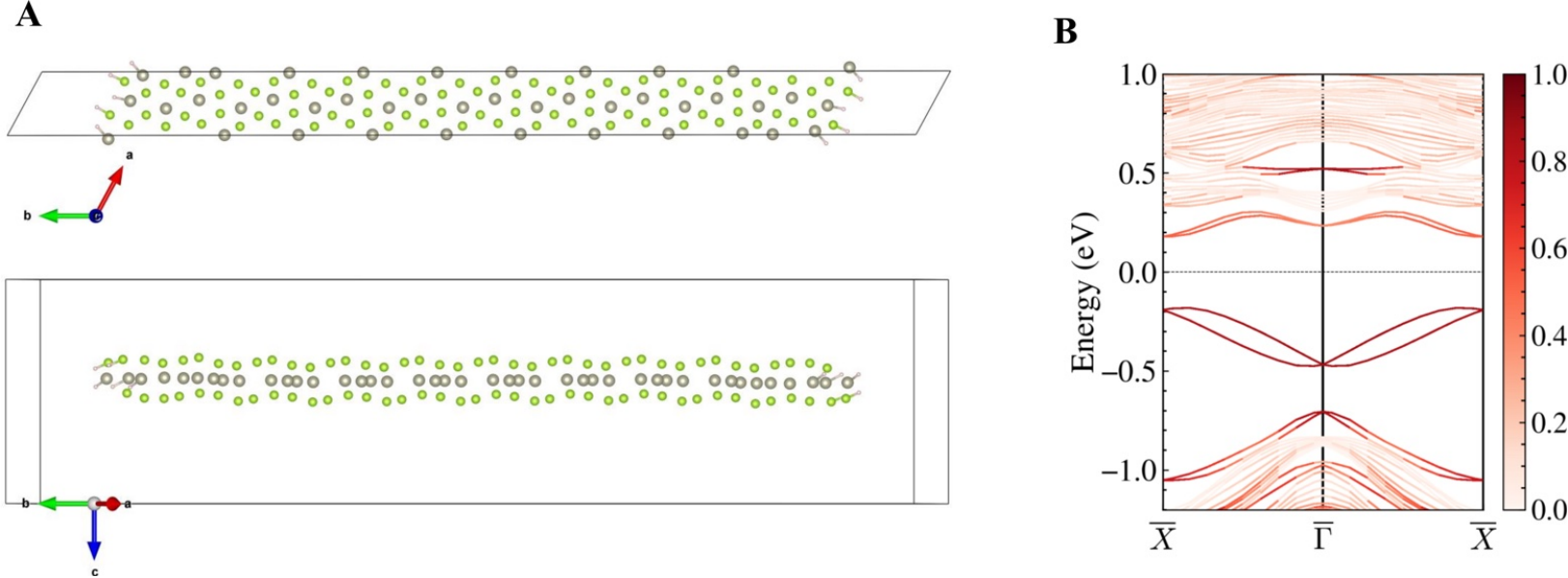}
    \caption{Edge state analysis of a finite inversion-symmetric Re$_4$Se$_8$ ribbon cut along the [10] direction, passivated by H atoms at the edges. (A) Relaxed ribbon structure from ab initio. Re atoms are indicated in grey color, while Se atoms are indicated in green color. H atoms are represented in white spheres. The projections onto two crystallographic axes are shown. (B) Edge-projected band structure of the relaxed ribbon, with eight mid-gap bands, four spinful bands per edge. There also exist other bands with large edge weights that are merged into the bulk continuum. 
    } 
    \label{fig:ribbon-dft-rese2}
\end{figure*}

\begin{figure*}
    \centering
    \includegraphics[width=0.99\linewidth]{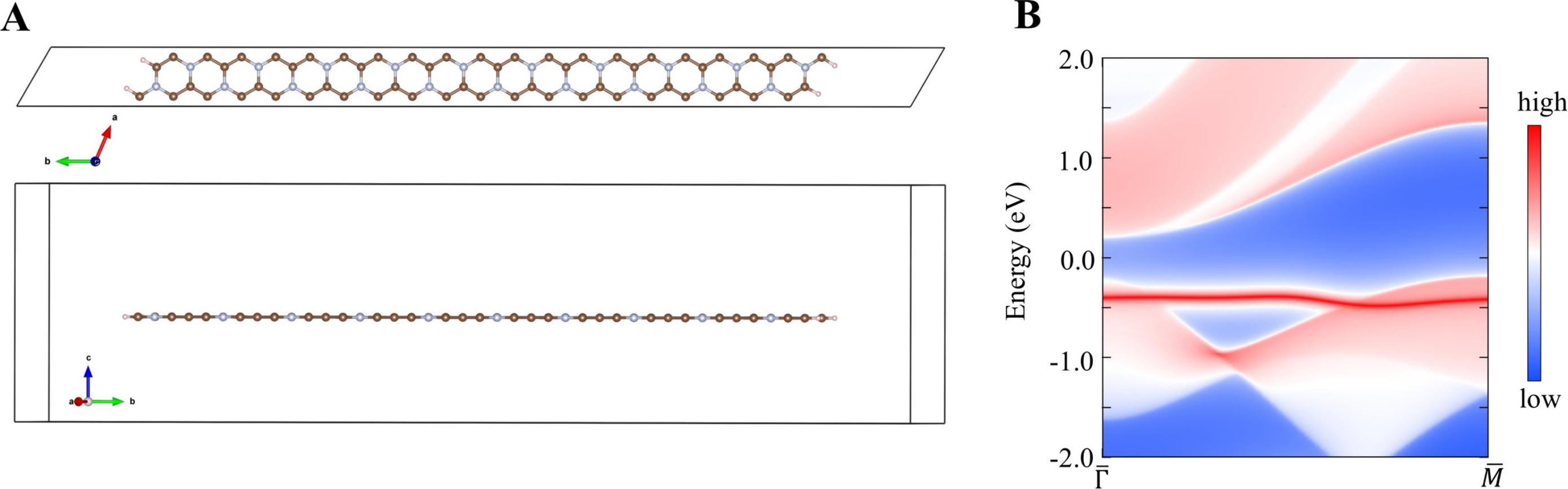}
    \caption{Edge state analysis of a finite C$_6$N$_2$ ribbon cut along the [10] direction, passivated by H atoms at the edges. (A) Relaxed ribbon structure. C atoms are indicated in brown color, while N atoms are indicated in grey color. H atoms are represented by white spheres. The projections onto two crystallographic axes are shown. (B) Spectral function, calculated by the Green's function method of a semi-infinite slab using the tight-binding model for the monolayer.} 
    \label{fig:ribbon-dft-c6n2}
\end{figure*}

\cref{fig:ribbon-dft-rese2}A presents the relaxed H-passivated Re$_4$Se$_8$ ribbon structure cut along the [10] direction, while panel (b) shows the corresponding edge-projected band structure of a nanoribbon. Our ab initio calculations reveal eight mid-gap bands (four spinful bands per edge, two per Wyckoff position cut), along with bands in the bulk continuum that have relatively large surface weights near the Fermi level. Without hydrogenation (see \cref{fig:ribbon-terminations}D), states originating from dangling bonds are present in the gap. Upon hydrogenation, bulk-like bonds are partially restored, pushing the dangling-bond states toward the band edges, while keeping the four obstructed states (associated with the obstructed Wyckoff positions $a$ and $c$) in the mid-gap. This strongly suggests that one obstructed Wannier center is present at each of these Wyckoff positions.

\cref{fig:ribbon-dft-c6n2}A presents the relaxed, hydrogen-passivated C$_6$N$_2$ ribbon structure cut along the [10] direction of the monolayer. Panel (B) displays the spectral function calculated for a semi-infinite ribbon using the Green's function method within our tight-binding model. The obstructed edge states show good agreement with the ab initio ribbon results in the main text \cref{fig:oai}, with only minor energy shifts from the hydrogenation and edge reconstruction effects in the fully relaxed structure.

\subsection{Extended details about Wannier centers in Re$_4$Se$_8$}
\addcontentsline{stoc}{subsection}{\protect\numberline{\thesubsection}Extended details about Wannier centers in Re$_4$Se$_8$}

\begin{figure*}
    \centering
    \includegraphics[width=\linewidth]{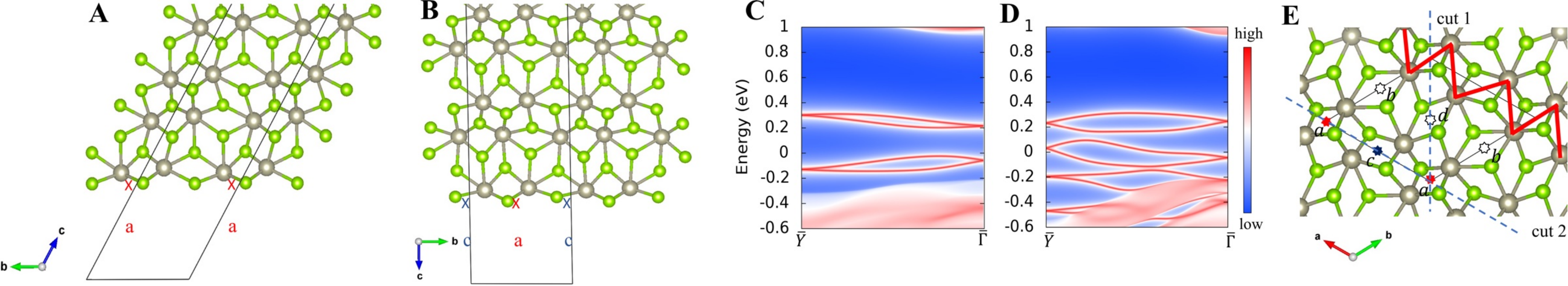}
    \caption{Extended details about edge states of semi-infinite Re$_4$Se$_8$ ribbons. (A) Ribbon termination whose edges cross Wyckoff position $a$ (ribbon A). Re atoms are indicated in grey color, while Se atoms are indicated in green color. (B) Ribbon termination whose edges cross Wyckoff positions $a$ and $c$ (ribbon AC). The $a$ ($c$) Wyckoff position is indicated with a red (blue) cross. (C) Spectral function of a semi-infinite configuration of the A ribbon, calculated using the Green's function method on a tight binding model of the monolayer. (D) Spectral function of the AC ribbon. (E) Unit cell of Re$_4$Se$_8$ and maximal Wyckoff positions in $p\overline{1}$, with the cutting planes defining the two ribbons overlaid. Obstructed Wyckoff positions $a$ and $c$ are shown in red and blue colors, respectively, while non-obstructed Wyckoff positions $b$ and $c$ are filled with white color. The diamond chain~\cite{distorted-diamond} formed by Re atoms is indicated by thick red lines.}
    
    \label{fig:ribbon-terminations}
\end{figure*}

In Re$_4$Se$_8$, when the ribbon cuts only Wyckoff positions $a$ (\cref{fig:ribbon-terminations}A), four mid-gap edge states that are separated into two groups emerge in the spectrum computed from a semi-infinite ribbon using the Wannier tight-binding model~\cite{Wu2018}, as shown in \cref{fig:ribbon-terminations}C. This figure is similar to \cref{fig:oai}B in the main text, but is included here to facilitate comparison with the two-WP case.
We then consider a new cutting plane that crosses both obstructed Wyckoff positions $a$ and $c$ (\cref{fig:ribbon-terminations}B), which gives four obstructed midgap states per edge, in addition to states originating from dangling bonds, as shown in \cref{fig:ribbon-terminations}D. A comparison of the two cuts is shown in \cref{fig:ribbon-terminations}E.

\clearpage
\section{Introduction to the tables}\label{tabintro}
\addcontentsline{stoc}{section}{\protect\numberline{\thesection}Introduction to the tables}

In the following sections, we tabulate the topological materials we found in the high-throughput search. The tables contain the relevant properties of each type of material. The sections include, respectively: topological insulators with a global band gap, topological insulators, topological semimetals, obstructed atomic insulators, and orbital-obstructed atomic insulators

In each section, the materials are divided into four additional categories, namely:
\begin{itemize}
\item \emph{Experimentally existing}: materials that we have manually verified in the literature to have been experimentally fabricated. 

\item \emph{Computationally exfoliated (MC2D)}: materials from the MC2D. In this database, all materials were computationally exfoliated from an experimentally existing material in 3D and then relaxed.

\item \emph{Stable (C2DB)}: materials that are marked as highly thermodynamically stable (the energy above the convex hull $<0.2$ eV/atom, or dynamically stable (all phonon frequencies are real) in the C2DB. 

\item \emph{Not-stable (C2DB)}: materials that do not satisfy the conditions defined above from C2DB.
\end{itemize}

In each table, we list a common set of properties per entry:

\begin{itemize}
\item 2D-TQC: the serial number of the entry, for this work, with a direct link to the corresponding entry of the \webtwoDTQCAbbr.
\item Formula: the chemical formula of the material in the entry. The box includes a link to the external website hosting the publication, showing its experimental realization (if existing).
\item LG: the layer group.
\item Gap: the global gap, given in eV.
\item Gap (Direct): the direct gap, given in eV. 
\item  Source: the database where we obtain the entry, including \CtwoDB~ and \MCtwoD. Notice that these two databases contain some common materials, and we list all of them as separate entries in the table. 

\item Database ID: the identification code of the material in \CtwoDB\; or \MCtwoD, updated in May 2024. We note that this code may be subject to changes as databases are updated by the maintainers.
 
\item Type: the topological classification of the material with SOC, including LCEBR, SEBR, NLC, AccidentalFermi, OAI, OOAI, ES, and ESFD\cite{bradlyn2017topological}. (See main text for a description of each type.)

\item Parent COD (if existing): identification code of the material in the \cod~ database.

\item Parent ICSD (if existing): identification code of the material in the \icsd\; database.

\end{itemize}

Apart from this, there are some features that we show only for certain types of entries. These are

\begin{itemize}

\item Topological indices: index characterizing the topological equivalence class of the Band Representation. (Only for SEBR and NLC.)

\item Electron number: the number of electrons of the material in the DFT calculation, with SOC turned on. (Only for ES and ESFD.)

\item Real space indices (RSI): Non-vanishing RSI of the material at each Wyckoff position. Indices of $\mathbb{Z}$, $\mathbb{Z}_2$, and $\mathbb{Z}_4$ type are labeled as $\delta$, $\eta$ and $\zeta$, respectively. The corresponding Wyckoff position is indicated in parentheses. (Only for OAI and OOAI.)

\end{itemize}

Finally, the tables contain entries from different databases that correspond to the same material (duplicates). We have systematically identified such cases by considering pairs of entries that satisfy a set of conditions. These conditions are as follows:

\begin{itemize}

\item Same layer group
\item Same atomic species
\item Same amount of atoms in the unit cell
\item A maximum difference of 5\% in $a$ lattice parameter length
\item A maximum difference of 5\% in $b$ lattice parameter length
\item A maximum difference of 5\% in $\gamma$ angle of the unit cell

\end{itemize}

We detected 153 pairs of such entries. Of these, 38 pairs are classified as topological insulators (29 SEBR and 9 NLC), 57 as topological semimietals (18 of them with an even electron number), 57 as obstructed atomic insulators and 1 as orbital-selected obstructed atomic insulator. 

Therefore, this work contains 3920 unique topological/obstructed materials. Among these, 867 are TIs, 584 even-electron topological semimetals, 946 obstructed atomic insulators and 35 orbital-selected obstructed atomic insulators. All duplicated entries are listed in \cref{sec:dupli}.

\section{Table of topological insulators with a global band gap}\label{sec:globalti}
\addcontentsline{stoc}{section}{\protect\numberline{\thesection}Table of topological insulators with a global band gap}

\subsection{Summary of results}
\addcontentsline{stoc}{subsection}{\protect\numberline{\thesubsection}Summary of results}

\begin{table}[htbp]
\centering
\begin{tabular}{c|c}
\hline\hline
Type of topological insulator with a full gap & Number of entries\\
\hline

\hyperlink{tigap:experi}{Experimental}& 8\\
\hline
\hyperlink{tigap:exfo}{Computationally exfoliated (MC2D)}& 19\\
\hline
\hyperlink{tigap:stable}{Stable (C2DB)}& 72\\
\hline
\hyperlink{tigap:unstable}{Not-stable (C2DB)}& 40\\
\hline

\hline\hline
\end{tabular}
\caption{\label{tigap:summa}Summary of topological insulators with a global gap in each type.}
\end{table}

\subsection{Experimental}
\addcontentsline{stoc}{subsection}{\protect\numberline{\thesubsection}Experimental}

\begin{longtable}{c|c|c|c|c|c|c|c|c|c|c}
\caption{\label{tigap:experilab}Experimental topological insulators with a global gap}\hypertarget{tigap:experi}\\
\centering
2D-TQC & Formula & LG & Gap & \shortstack{Direct\\ Gap} & Source & Source ID & \shortstack{Topology\\ (SOC)} & \shortstack{Topo.\\ indices} & \shortstack{Parent\\ COD} & \shortstack{Parent\\ ICSD} \\
\endhead
\hline

\serialidweb{1.1.1} & Mo$_{2}$S$_{4}$ & 15 & 0.0483 & 0.0509 & C2DB & \ctwodbpath{2MoS2-1} & NLC & z2w,1=1 & \codpath{4124624} & -\\
\hline
\serialidweb{1.3.24} & S$_{4}$W$_{2}$ & 15 & 0.0351 & 0.1320 & C2DB & \ctwodbpath{2WS2-1} & NLC & z2w,1=1 & - & -\\
\hline
\serialidweb{1.1.5} & Se$_{4}$W$_{2}$ & 15 & 0.0269 & 0.1117 & C2DB & \ctwodbpath{2WSe2-1} & NLC & z2w,1=1 & - & -\\
\hline
\serialidweb{1.1.7} & Hf$_{2}$Te$_{10}$ & 46 & 0.1152 & 0.2918 & C2DB & \ctwodbpath{2HfTe5-1} & SEBR & z2w,1=1 & \codpath{1527386} & -\\
\hline
\serialidweb{1.1.9} & Bi$_{2}$ & 72 & 0.4875 & 0.5386 & C2DB & \ctwodbpath{2Bi-1} & SEBR & z2w,1=1 & \codpath{2310889} & -\\
\hline
\serialidweb{1.1.10} & Bi$_{2}$ & 72 & 0.5294 & 0.5589 & MC2D & \mctwodpath{mc2d-18} & SEBR & z2w,1=1 & - & -\\
\hline
\serialidweb{1.1.15} & Ge$_{2}$ & 72 & 0.0242 & 0.0242 & C2DB & \ctwodbpath{2Ge-1} & SEBR & z2w,1=1 & - & 192507\\
\hline
\serialidweb{1.1.17} & Sn$_{2}$ & 72 & 0.0739 & 0.0739 & C2DB & \ctwodbpath{2Sn-1} & SEBR & z2w,1=1 & - & -\\
\hline

\hline

\hline\hline
\end{longtable}
\clearpage

\subsection{Computationally exfoliated (MC2D)}
\addcontentsline{stoc}{subsection}{\protect\numberline{\thesubsection}Computationally exfoliated (MC2D)}

\begin{longtable}{c|c|c|c|c|c|c|c|c|c|c}
\caption{\label{tigap:exfolab}Computationally exfoliated (MC2D) topological insulators with a global gap}\hypertarget{tigap:exfo}\\
\centering
2D-TQC & Formula & LG & Gap & \shortstack{Direct\\ Gap} & Source & Source ID & \shortstack{Topology\\ (SOC)} & \shortstack{Topo.\\ indices} & \shortstack{Parent\\ COD} & \shortstack{Parent\\ ICSD} \\
\endhead
\hline

\serialidweb{1.2.3} & Fe$_{2}$O$_{12}$Te$_{4}$ & 2 & 0.0142 & 0.0231 & MC2D & \mctwodpath{mc2d-1629} & NLC & z2w,1=1 & - & -\\
\hline
\serialidweb{1.2.5} & H$_{8}$As$_{4}$O$_{16}$Zn & 2 & 0.0285 & 0.0291 & MC2D & \mctwodpath{mc2d-2034} & NLC & z2w,1=1 & - & -\\
\hline
\serialidweb{1.2.13} & Ir$_{2}$Nb$_{2}$Te$_{8}$ & 15 & 0.0302 & 0.0302 & MC2D & \mctwodpath{mc2d-396} & NLC & z2w,1=1 & - & -\\
\hline
\serialidweb{1.2.14} & Ir$_{2}$Ta$_{2}$Te$_{8}$ & 15 & 0.0251 & 0.0251 & MC2D & \mctwodpath{mc2d-706} & NLC & z2w,1=1 & - & -\\
\hline
\serialidweb{1.2.15} & Rh$_{2}$Ta$_{2}$Te$_{8}$ & 15 & 0.0621 & 0.0621 & MC2D & \mctwodpath{mc2d-488} & NLC & z2w,1=1 & - & -\\
\hline
\serialidweb{1.2.25} & Hg$_{8}$O$_{4}$ & 17 & 0.0762 & 0.0762 & MC2D & \mctwodpath{mc2d-769} & NLC & z2w,1=1 & - & -\\
\hline
\serialidweb{1.2.24} & Hg$_{4}$N$_{4}$S$_{4}$ & 17 & 0.0067 & 0.0067 & MC2D & \mctwodpath{mc2d-2366} & NLC & z2w,1=1 & - & -\\
\hline
\serialidweb{1.2.23} & Hg$_{10}$O$_{4}$ & 17 & 0.0625 & 0.0625 & MC2D & \mctwodpath{mc2d-409} & NLC & z2w,1=1 & - & -\\
\hline

\hline

\hline\hline
\end{longtable}
\clearpage

\subsection{Stable (C2DB)}
\addcontentsline{stoc}{subsection}{\protect\numberline{\thesubsection}Stable (C2DB)}

\begin{longtable}{c|c|c|c|c|c|c|c|c|c|c}
\caption{\label{tigap:stablelab}Stable (C2DB) topological insulators with a global gap}\hypertarget{tigap:stable}\\
\centering
2D-TQC & Formula & LG & Gap & \shortstack{Direct\\ Gap} & Source & Source ID & \shortstack{Topology\\ (SOC)} & \shortstack{Topo.\\ indices} & \shortstack{Parent\\ COD} & \shortstack{Parent\\ ICSD} \\
\endhead
\hline

\serialidweb{1.3.1} & As$_{2}$I$_{2}$ & 2 & 1.3298 & 1.4872 & C2DB & \ctwodbpath{2AsI-1} & NLC & z2w,1=1 & - & -\\
\hline
\serialidweb{1.3.3} & P$_{2}$S$_{6}$Ta$_{2}$ & 2 & 0.0434 & 0.0683 & C2DB & \ctwodbpath{2PTaS3-2} & NLC & z2w,1=1 & - & -\\
\hline
\serialidweb{1.3.5} & As$_{2}$F$_{2}$ & 15 & 0.1338 & 0.1376 & C2DB & \ctwodbpath{2AsF-1} & NLC & z2w,1=1 & - & -\\
\hline
\serialidweb{1.3.6} & Bi$_{2}$Br$_{2}$ & 15 & 0.6435 & 0.6710 & C2DB & \ctwodbpath{2BiBr-1} & NLC & z2w,1=1 & - & -\\
\hline
\serialidweb{1.3.7} & Br$_{2}$Sb$_{2}$ & 15 & 0.2826 & 0.2826 & C2DB & \ctwodbpath{2BrSb-1} & NLC & z2w,1=1 & - & -\\
\hline
\serialidweb{1.3.11} & F$_{2}$P$_{2}$ & 15 & 0.0470 & 0.0470 & C2DB & \ctwodbpath{2FP-1} & NLC & z2w,1=1 & - & -\\
\hline
\serialidweb{1.3.12} & Ir$_{2}$Nb$_{2}$Te$_{8}$ & 15 & 0.0374 & 0.0374 & C2DB & \ctwodbpath{2IrNbTe4-1} & NLC & z2w,1=1 & - & 73690\\
\hline
\serialidweb{1.3.13} & Ir$_{2}$Ta$_{2}$Te$_{8}$ & 15 & 0.0276 & 0.0276 & C2DB & \ctwodbpath{2IrTaTe4-1} & NLC & z2w,1=1 & - & 657362\\
\hline

\hline

\hline\hline
\end{longtable}
\clearpage

\subsection{Not-stable (C2DB)}
\addcontentsline{stoc}{subsection}{\protect\numberline{\thesubsection}Not-stable (C2DB)}

\begin{longtable}{c|c|c|c|c|c|c|c|c|c|c}
\caption{\label{tigap:unstablelab}Not-stable (C2DB) topological insulators with a global gap}\hypertarget{tigap:unstable}\\
\centering
2D-TQC & Formula & LG & Gap & \shortstack{Direct\\ Gap} & Source & Source ID & \shortstack{Topology\\ (SOC)} & \shortstack{Topo.\\ indices} & \shortstack{Parent\\ COD} & \shortstack{Parent\\ ICSD} \\
\endhead
\hline

\serialidweb{1.4.7} & H$_{4}$Ag$_{4}$Cl$_{4}$O$_{4}$ & 17 & 0.0113 & 0.0123 & C2DB & \ctwodbpath{4AgClHO-2} & NLC & z2w,1=1 & - & -\\
\hline
\serialidweb{1.4.13} & F$_{2}$Sb$_{2}$ & 18 & 0.3271 & 0.3271 & C2DB & \ctwodbpath{2FSb-1} & SEBR & z2w,1=1 & - & -\\
\hline
\serialidweb{1.4.18} & Mo$_{2}$S$_{2}$ & 18 & 0.1014 & 0.1014 & C2DB & \ctwodbpath{2MoS-2} & SEBR & z2w,1=1 & - & -\\
\hline
\serialidweb{1.4.29} & Bi$_{4}$ & 42 & 0.1076 & 0.1111 & C2DB & \ctwodbpath{4Bi-2} & SEBR & z2w,1=1 & - & -\\
\hline
\serialidweb{1.4.56} & PbSe & 61 & 0.2090 & 0.2090 & C2DB & \ctwodbpath{2PbSe-5} & SEBR & z4m0=2 & - & -\\
\hline
\serialidweb{1.4.55} & PbS & 61 & 0.1054 & 0.1054 & C2DB & \ctwodbpath{2PbS-4} & SEBR & z4m0=2 & - & -\\
\hline
\serialidweb{1.4.57} & PbTe & 61 & 0.2571 & 0.2657 & C2DB & \ctwodbpath{2PbTe-4} & SEBR & z4m0=2 & - & -\\
\hline
\serialidweb{1.4.63} & SSn & 61 & 0.0308 & 0.1674 & C2DB & \ctwodbpath{2SSn-4} & SEBR & z4m0=2 & - & -\\
\hline

\hline

\hline\hline
\end{longtable}

\clearpage

\section{Table of topological insulators}\label{sec:ti}
\addcontentsline{stoc}{section}{\protect\numberline{\thesection}Table of topological insulators}

\subsection{Summary of results}
\addcontentsline{stoc}{subsection}{\protect\numberline{\thesubsection}Summary of results}

\begin{table}[htbp]
\centering
\begin{tabular}{c|c}
\hline\hline
Type of topological insulator & Number of entries\\
\hline

\hyperlink{ti:experi}{Experimental}& 23\\
\hline
\hyperlink{ti:exfo}{Computationally exfoliated (MC2D)}& 135\\
\hline
\hyperlink{ti:stable}{Stable (C2DB)}& 419\\
\hline
\hyperlink{ti:unstable}{Not-stable (C2DB)}& 328\\
\hline

\hline\hline
\end{tabular}
\caption{\label{ti:summa}Summary of topological insulators in each type.}
\end{table}

\subsection{Experimental}
\addcontentsline{stoc}{subsection}{\protect\numberline{\thesubsection}Experimental}

\begin{longtable}{c|c|c|c|c|c|c|c|c|c|c}
\caption{\label{ti:experilab}Experimental topological insulators}\hypertarget{ti:experi}\\
\centering
2D-TQC & Formula & LG & Gap & \shortstack{Direct\\ Gap} & Source & Source ID & \shortstack{Topology\\ (SOC)} & \shortstack{Topo.\\ indices} & \shortstack{Parent\\ COD} & \shortstack{Parent\\ ICSD} \\
\endhead
\hline

\serialidweb{1.1.1} & Mo$_{2}$S$_{4}$ & 15 & 0.0483 & 0.0509 & C2DB & \ctwodbpath{2MoS2-1} & NLC & z2w,1=1 & \codpath{4124624} & -\\
\hline
\serialidweb{1.1.2} & Mo$_{2}$Te$_{4}$ & 15 & 0.00 & 0.1058 & MC2D & \mctwodpath{mc2d-206} & NLC & z2w,1=1 & - & -\\
\hline
\serialidweb{1.1.3} & Mo$_{2}$Te$_{4}$ & 15 & 0.00 & 0.1094 & C2DB & \ctwodbpath{2MoTe2-1} & NLC & z2w,1=1 & - & -\\
\hline
\serialidweb{1.3.24} & S$_{4}$W$_{2}$ & 15 & 0.0351 & 0.1320 & C2DB & \ctwodbpath{2WS2-1} & NLC & z2w,1=1 & - & -\\
\hline
\serialidweb{1.1.5} & Se$_{4}$W$_{2}$ & 15 & 0.0269 & 0.1117 & C2DB & \ctwodbpath{2WSe2-1} & NLC & z2w,1=1 & - & -\\
\hline
\serialidweb{1.1.6} & Te$_{4}$W$_{2}$ & 15 & 0.00 & 0.1419 & C2DB & \ctwodbpath{2WTe2-1} & NLC & z2w,1=1 & \codpath{2310355} & -\\
\hline
\serialidweb{1.1.4} & Te$_{4}$W$_{2}$ & 15 & 0.00 & 0.1423 & MC2D & \mctwodpath{mc2d-255} & NLC & z2w,1=1 & - & -\\
\hline
\serialidweb{1.1.7} & Hf$_{2}$Te$_{10}$ & 46 & 0.1152 & 0.2918 & C2DB & \ctwodbpath{2HfTe5-1} & SEBR & z2w,1=1 & \codpath{1527386} & -\\
\hline

\hline

\hline\hline
\end{longtable}
\clearpage

\subsection{Computationally exfoliated (MC2D)}
\addcontentsline{stoc}{subsection}{\protect\numberline{\thesubsection}Computationally exfoliated (MC2D)}

\begin{longtable}{c|c|c|c|c|c|c|c|c|c|c}
\caption{\label{ti:exfolab}Computationally exfoliated (MC2D) topological insulators}\hypertarget{ti:exfo}\\
\centering
2D-TQC & Formula & LG & Gap & \shortstack{Direct\\ Gap} & Source & Source ID & \shortstack{Topology\\ (SOC)} & \shortstack{Topo.\\ indices} & \shortstack{Parent\\ COD} & \shortstack{Parent\\ ICSD} \\
\endhead
\hline

\serialidweb{1.2.4} & H$_{12}$Co$_{2}$F$_{2}$O$_{12}$P$_{2}$ & 2 & 0.0004 & 0.0025 & MC2D & \mctwodpath{mc2d-905} & NLC & z2w,1=1 & - & -\\
\hline
\serialidweb{1.2.2} & Br$_{4}$Co$_{2}$O$_{6}$Sb$_{4}$ & 2 & 0.00 & 0.0079 & MC2D & \mctwodpath{mc2d-1485} & NLC & z2w,1=1 & - & -\\
\hline
\serialidweb{1.2.1} & As$_{4}$Cu$_{6}$S$_{10}$ & 2 & 0.00 & 0.0085 & MC2D & \mctwodpath{mc2d-2527} & NLC & z2w,1=1 & - & -\\
\hline
\serialidweb{1.2.3} & Fe$_{2}$O$_{12}$Te$_{4}$ & 2 & 0.0142 & 0.0231 & MC2D & \mctwodpath{mc2d-1629} & NLC & z2w,1=1 & - & -\\
\hline
\serialidweb{1.2.5} & H$_{8}$As$_{4}$O$_{16}$Zn & 2 & 0.0285 & 0.0291 & MC2D & \mctwodpath{mc2d-2034} & NLC & z2w,1=1 & - & -\\
\hline
\serialidweb{1.2.6} & Mg$_{3}$O$_{16}$P$_{2}$ & 6 & 0.0030 & 0.0040 & MC2D & \mctwodpath{mc2d-959} & NLC & z2w,1=1 & - & -\\
\hline
\serialidweb{1.2.7} & Cu$_{4}$O$_{14}$P$_{4}$ & 7 & 0.00 & 0.0050 & MC2D & \mctwodpathold{Cu2P2O7} & NLC & z2w,1=1 & - & -\\
\hline
\serialidweb{1.2.8} & Bi$_{16}$I$_{4}$ & 14 & 0.00 & 0.0525 & MC2D & \mctwodpath{mc2d-1335} & NLC & z2w,1=1 & - & -\\
\hline

\hline

\hline\hline
\end{longtable}
\clearpage

\subsection{Stable (C2DB)}
\addcontentsline{stoc}{subsection}{\protect\numberline{\thesubsection}Stable (C2DB)}

\begin{longtable}{c|c|c|c|c|c|c|c|c|c|c}
\caption{\label{ti:stablelab}Stable (C2DB) topological insulators}\hypertarget{ti:stable}\\
\centering
2D-TQC & Formula & LG & Gap & \shortstack{Direct\\ Gap} & Source & Source ID & \shortstack{Topology\\ (SOC)} & \shortstack{Topo.\\ indices} & \shortstack{Parent\\ COD} & \shortstack{Parent\\ ICSD} \\
\endhead
\hline

\serialidweb{1.3.1} & As$_{2}$I$_{2}$ & 2 & 1.3298 & 1.4872 & C2DB & \ctwodbpath{2AsI-1} & NLC & z2w,1=1 & - & -\\
\hline
\serialidweb{1.3.2} & Au$_{2}$F$_{2}$ & 2 & 0.00 & 0.1022 & C2DB & \ctwodbpath{2AuF-1} & NLC & z2w,1=1 & - & -\\
\hline
\serialidweb{1.3.3} & P$_{2}$S$_{6}$Ta$_{2}$ & 2 & 0.0434 & 0.0683 & C2DB & \ctwodbpath{2PTaS3-2} & NLC & z2w,1=1 & - & -\\
\hline
\serialidweb{1.3.5} & As$_{2}$F$_{2}$ & 15 & 0.1338 & 0.1376 & C2DB & \ctwodbpath{2AsF-1} & NLC & z2w,1=1 & - & -\\
\hline
\serialidweb{1.3.6} & Bi$_{2}$Br$_{2}$ & 15 & 0.6435 & 0.6710 & C2DB & \ctwodbpath{2BiBr-1} & NLC & z2w,1=1 & - & -\\
\hline
\serialidweb{1.3.7} & Br$_{2}$Sb$_{2}$ & 15 & 0.2826 & 0.2826 & C2DB & \ctwodbpath{2BrSb-1} & NLC & z2w,1=1 & - & -\\
\hline
\serialidweb{1.3.10} & Cr$_{2}$Te$_{4}$ & 15 & 0.00 & 0.0918 & C2DB & \ctwodbpath{2CrTe2-2} & NLC & z2w,1=1 & - & -\\
\hline
\serialidweb{1.3.11} & F$_{2}$P$_{2}$ & 15 & 0.0470 & 0.0470 & C2DB & \ctwodbpath{2FP-1} & NLC & z2w,1=1 & - & -\\
\hline

\hline

\hline\hline
\end{longtable}
\clearpage

\subsection{Not-stable (C2DB)}
\addcontentsline{stoc}{subsection}{\protect\numberline{\thesubsection}Not-stable (C2DB)}

\begin{longtable}{c|c|c|c|c|c|c|c|c|c|c}
\caption{\label{ti:unstablelab}Not-stable (C2DB) topological insulators}\hypertarget{ti:unstable}\\
\centering
2D-TQC & Formula & LG & Gap & \shortstack{Direct\\ Gap} & Source & Source ID & \shortstack{Topology\\ (SOC)} & \shortstack{Topo.\\ indices} & \shortstack{Parent\\ COD} & \shortstack{Parent\\ ICSD} \\
\endhead
\hline

\serialidweb{1.4.1} & Co$_{2}$I$_{2}$ & 2 & 0.00 & 0.0339 & C2DB & \ctwodbpath{2CoI-1} & NLC & z2w,1=1 & - & -\\
\hline
\serialidweb{1.4.3} & Pb$_{2}$Te$_{2}$ & 14 & 0.00 & 0.0963 & C2DB & \ctwodbpath{2PbTe-5} & NLC & z2w,1=1 & - & -\\
\hline
\serialidweb{1.4.4} & Pd$_{2}$S$_{2}$ & 14 & 0.00 & 0.0495 & C2DB & \ctwodbpath{2PdS-2} & NLC & z2w,1=1 & - & -\\
\hline
\serialidweb{1.4.5} & Cr$_{2}$S$_{4}$ & 15 & 0.00 & 0.0248 & C2DB & \ctwodbpath{2CrS2-1} & NLC & z2w,1=1 & - & -\\
\hline
\serialidweb{1.4.6} & Cr$_{2}$Se$_{4}$ & 15 & 0.00 & 0.0473 & C2DB & \ctwodbpath{2CrSe2-1} & NLC & z2w,1=1 & - & -\\
\hline
\serialidweb{1.4.7} & H$_{4}$Ag$_{4}$Cl$_{4}$O$_{4}$ & 17 & 0.0113 & 0.0123 & C2DB & \ctwodbpath{4AgClHO-2} & NLC & z2w,1=1 & - & -\\
\hline
\serialidweb{1.4.8} & H$_{4}$Ag$_{4}$Cl$_{4}$S$_{4}$ & 17 & 0.00 & 0.0556 & C2DB & \ctwodbpath{4AgClHS-2} & NLC & z2w,1=1 & - & -\\
\hline
\serialidweb{1.4.9} & H$_{4}$Ag$_{4}$F$_{4}$O$_{4}$ & 17 & 0.00 & 0.0921 & C2DB & \ctwodbpath{4AgFHO-1} & NLC & z2w,1=1 & - & -\\
\hline

\hline

\hline\hline
\end{longtable}

\clearpage

\section{Table of topological semimetals}\label{section:tableSM}
\addcontentsline{stoc}{section}{\protect\numberline{\thesection}Table of topological semimetals}

\subsection{Summary of results}
\addcontentsline{stoc}{subsection}{\protect\numberline{\thesubsection}Summary of results}

\begin{table}[htbp]
\centering
\begin{tabular}{c|c}
\hline\hline
Type of topological semimetal & Number of entries\\
\hline

\hyperlink{sm:experi}{Experimental}& 14\\
\hline
\hyperlink{sm:exfo}{Computationally exfoliated (MC2D)}& 340\\
\hline
\hyperlink{sm:stable}{Stable (C2DB)}& 986\\
\hline
\hyperlink{sm:unstable}{Not-stable (C2DB)}& 789\\
\hline
\hline\hline
\end{tabular}
\caption{\label{sm:summa}Summary of topological semimetals in each type.}
\end{table}

\subsection{Experimental}
\addcontentsline{stoc}{subsection}{\protect\numberline{\thesubsection}Experimental}

\begin{longtable}{c|c|c|c|c|c|c|c|c}
\caption{\label{sm:experilab}Experimental topological semimetals}\hypertarget{sm:experi}\\
\centering
2D-TQC & Formula & LG & N. elec.  & Source  & Source ID & \shortstack{Topology\\ (SOC)} & \shortstack{Parent\\ COD} & \shortstack{Parent\\ ICSD} \\
\endhead
\hline

\serialidweb{2.1.1} & IrTe$_{2}$ & 72 & 21 & C2DB & \ctwodbpath{1IrTe2-1} & ESFD & - & -\\
\hline
\serialidweb{2.1.3} & N$_{3}$Ti$_{4}$ & 72 & 63 & C2DB & \ctwodbpath{1N3Ti4-1} & ESFD & - & -\\
\hline
\serialidweb{2.1.2} & Mo$_{2}$N & 72 & 33 & C2DB & \ctwodbpath{1NMo2-1} & ESFD & - & -\\
\hline
\serialidweb{2.1.4} & NV$_{2}$ & 72 & 31 & C2DB & \ctwodbpath{1NV2-1} & ESFD & - & -\\
\hline
\serialidweb{2.1.5} & NW$_{2}$ & 72 & 33 & C2DB & \ctwodbpath{1NW2-1} & ESFD & - & -\\
\hline
\serialidweb{2.1.7} & S$_{2}$Ta & 72 & 23 & C2DB & \ctwodbpath{1TaS2-2} & ESFD & \codpath{1538041} & 52117\\
\hline
\serialidweb{2.1.6} & Se$_{2}$Ta & 72 & 23 & C2DB & \ctwodbpath{1TaSe2-2} & ESFD & \codpath{4031648} & 24318\\
\hline
\serialidweb{2.1.8} & TaTe$_{2}$ & 72 & 23 & C2DB & \ctwodbpath{1TaTe2-2} & ESFD & - & -\\
\hline
\serialidweb{2.1.9} & C$_{2}$Ta$_{3}$ & 78 & 41 & C2DB & \ctwodbpath{1C2Ta3-1} & ESFD & - & -\\
\hline
\serialidweb{2.1.12} & NbS$_{2}$ & 78 & 25 & C2DB & \ctwodbpath{1NbS2-1} & ESFD & - & 645308\\
\hline
\serialidweb{2.1.13} & NbSe$_{2}$ & 78 & 25 & C2DB & \ctwodbpath{1NbSe2-1} & ESFD & \codpath{1539310} & 237740\\
\hline
\serialidweb{2.1.14} & S$_{2}$Ta & 78 & 23 & C2DB & \ctwodbpath{1TaS2-1} & ESFD & - & 651083\\
\hline
\serialidweb{2.1.10} & N$_{3}$W$_{2}$ & 78 & 43 & C2DB & \ctwodbpath{1W2N3-1} & ESFD & - & 186207\\
 \hline
 \serialidweb{2.1.11} & N$_{3}$W$_{2}$ & 78 & 43 & MC2D & \mctwodpath{mc2d-208} & ESFD & - & -\\
\hline

\hline

\hline\hline
\end{longtable}
\clearpage

\subsection{Computationally exfoliated (MC2D)}
\addcontentsline{stoc}{subsection}{\protect\numberline{\thesubsection}Computationally exfoliated (MC2D)}

\begin{longtable}{c|c|c|c|c|c|c|c|c}
\caption{\label{sm:exfolab}Computationally exfoliated (MC2D) topological semimetals}\hypertarget{sm:exfo}\\
\centering
2D-TQC & Formula & LG & N. elec.  &  Source  & Source ID & \shortstack{Topology\\ (SOC)} & \shortstack{Parent\\ COD} & \shortstack{Parent\\ ICSD} \\
\endhead
\hline

\serialidweb{2.2.2} & AgBi$_{2}$F$_{12}$ & 2 & 125 & MC2D & \mctwodpath{mc2d-1545} & ESFD & - & -\\
\hline
\serialidweb{2.2.7} & C$_{2}$AgF$_{6}$O$_{6}$S$_{2}$ & 2 & 109 & MC2D & \mctwodpath{mc2d-2207} & ESFD & - & -\\
\hline
\serialidweb{2.2.3} & AgF$_{12}$Sb$_{2}$ & 2 & 105 & MC2D & \mctwodpath{mc2d-1426} & ESFD & - & -\\
\hline
\serialidweb{2.2.10} & C$_{4}$H$_{4}$CoN$_{14}$ & 2 & 99 & MC2D & \mctwodpath{mc2d-2351} & ESFD & - & -\\
\hline
\serialidweb{2.2.6} & C$_{14}$H$_{14}$CoN$_{2}$O$_{8}$ & 2 & 137 & MC2D & \mctwodpath{mc2d-1803} & ESFD & - & -\\
\hline
\serialidweb{2.2.5} & C$_{12}$H$_{10}$CsN$_{4}$O$_{10}$ & 2 & 147 & MC2D & \mctwodpath{mc2d-834} & ESFD & - & -\\
\hline
\serialidweb{2.2.15} & Cu$_{2}$NO$_{10}$Se$_{2}$ & 2 & 99 & MC2D & \mctwodpath{mc2d-2641} & ESFD & - & -\\
\hline
\serialidweb{2.2.14} & Cl$_{4}$Cu$_{3}$O$_{12}$Te$_{4}$Yb$_{2}$ & 2 & 173 & MC2D & \mctwodpath{mc2d-664} & ESFD & - & -\\
\hline

\hline

\hline\hline
\end{longtable}
\clearpage

\subsection{Stable (C2DB)}
\addcontentsline{stoc}{subsection}{\protect\numberline{\thesubsection}Stable (C2DB)}

\begin{longtable}{c|c|c|c|c|c|c|c|c}
\caption{\label{sm:stablelab}Stable (C2DB) topological semimetals}\hypertarget{sm:stable}\\
\centering
2D-TQC & Formula & LG & N. elec.  & Source  & Source ID & \shortstack{Topology\\ (SOC)} & \shortstack{Parent\\ COD} & \shortstack{Parent\\ ICSD} \\
\endhead
\hline

\serialidweb{2.3.2} & AsBrI$_{5}$Sn & 1 & 61 & C2DB & \ctwodbpath{1AsBrSnI5-1} & ESFD & - & -\\
\hline
\serialidweb{2.3.3} & AsClISSeSn & 1 & 45 & C2DB & \ctwodbpath{1AsClISSeSn-1} & ESFD & - & -\\
\hline
\serialidweb{2.3.1} & AsBr$_{2}$Se & 1 & 25 & C2DB & \ctwodbpath{1AsSeBr2-1} & ESFD & - & -\\
\hline
\serialidweb{2.3.4} & AuBr$_{2}$HfSe$_{2}$ & 1 & 47 & C2DB & \ctwodbpath{1AuHfBr2Se2-1} & ESFD & - & -\\
\hline
\serialidweb{2.3.5} & Br$_{2}$Cu$_{3}$Se$_{2}$ & 1 & 59 & C2DB & \ctwodbpath{1Br2Se2Cu3-1} & ESFD & - & -\\
\hline
\serialidweb{2.3.7} & BrCuIS & 1 & 31 & C2DB & \ctwodbpath{1BrCuIS-1} & ESFD & - & -\\
\hline
\serialidweb{2.3.11} & Cl$_{3}$CuGaS$_{2}$ & 1 & 57 & C2DB & \ctwodbpath{1CuGaS2Cl3-1} & ESFD & - & -\\
\hline
\serialidweb{2.3.12} & FOPd & 1 & 23 & C2DB & \ctwodbpath{1FOPd-1} & ESFD & - & -\\
\hline

\hline

\hline\hline
\end{longtable}
\clearpage

\subsection{Not-stable (C2DB)}
\addcontentsline{stoc}{subsection}{\protect\numberline{\thesubsection}Not-stable (C2DB)}

\begin{longtable}{c|c|c|c|c|c|c|c|c}
\caption{\label{sm:unstablelab}Not-stable (C2DB) topological semimetals}\hypertarget{sm:unstable}\\
\centering
2D-TQC & Formula & LG & N. elec.   & Source  & Source ID & \shortstack{Topology\\ (SOC)} & \shortstack{Parent\\ COD} & \shortstack{Parent\\ ICSD} \\
\endhead
\hline

\serialidweb{2.4.1} & AsGeO$_{2}$ & 1 & 31 & C2DB & \ctwodbpath{1AsGeO2-1} & ESFD & - & -\\
\hline
\serialidweb{2.4.2} & AsO$_{3}$Pd & 1 & 33 & C2DB & \ctwodbpath{1AsPdO3-1} & ESFD & - & -\\
\hline
\serialidweb{2.4.4} & AuCl$_{2}$GaS$_{3}$Ta & 1 & 67 & C2DB & \ctwodbpath{1AuGaTaCl2S3-1} & ESFD & - & -\\
\hline
\serialidweb{2.4.5} & AuS$_{2}$Sn & 1 & 37 & C2DB & \ctwodbpath{1AuSnS2-1} & ESFD & - & -\\
\hline
\serialidweb{2.4.3} & AuBr$_{2}$Sr & 1 & 35 & C2DB & \ctwodbpath{1AuSrBr2-1} & ESFD & - & -\\
\hline
\serialidweb{2.4.10} & BrCaGeO & 1 & 37 & C2DB & \ctwodbpath{1BrCaGeO-1} & ESFD & - & -\\
\hline
\serialidweb{2.4.11} & BrClInOSe & 1 & 39 & C2DB & \ctwodbpath{1BrClInOSe-1} & ESFD & - & -\\
\hline
\serialidweb{2.4.12} & BrInMnSe & 1 & 39 & C2DB & \ctwodbpath{1BrInMnSe-1} & ESFD & - & -\\
\hline

\hline

\hline\hline
\end{longtable}

\clearpage

\section{Table of obstructed atomic insulators}\label{oai}
\addcontentsline{stoc}{section}{\protect\numberline{\thesection}Table of obstructed atomic insulators}

\subsection{Summary of results}
\addcontentsline{stoc}{subsection}{\protect\numberline{\thesubsection}Summary of results}

\begin{table}[htbp]
\centering
\begin{tabular}{c|c}
\hline\hline
Type of obstructed atomic insulator & Number of materials\\
\hline

\hyperlink{oai:experi}{Experimental}& 50\\
\hline
\hyperlink{oai:exfo}{Computationally exfoliated (MC2D)}& 201\\
\hline
\hyperlink{oai:stable}{Stable (C2DB)}& 533\\
\hline
\hyperlink{oai:unstable}{Not-stable (C2DB)}& 220\\
\hline

\hline\hline
\end{tabular}
\caption{\label{oai:summa}Summary of obstructed atomic insulators in each type.}
\end{table}

\subsection{Experimental}\label{table:oaiExperimental}
\addcontentsline{stoc}{subsection}{\protect\numberline{\thesubsection}Experimental}

\begin{longtable}{c|c|c|c|c|c|c|c|c|c|c|c}
\caption{\label{oai:experilab}Experimental obstructed atomic insulators}\hypertarget{oai:experi}\\
\centering
2D-TQC & Formula & LG & Gap & \shortstack{Direct\\ Gap} & Source & Source ID & \shortstack{Topology\\ (SOC)} & RSI & Occ. WP & \shortstack{Parent\\ COD} & \shortstack{Parent\\ ICSD}\\
\endhead
\hline

\serialidweb{3.1.1} & Re$_{4}$S$_{8}$ & 2 & 1.3082 & 1.3082 & C2DB & \ctwodbpath{4ReS2-1} & OAI & \shortstack{$\delta(a)=1$\\$\delta(c)=1$} & \shortstack{$e$} & - & 81814\\
\hline
\serialidweb{3.1.2} & Re$_{4}$S$_{8}$ & 2 & 1.2764 & 1.2764 & MC2D & \mctwodpath{mc2d-869} & OAI & \shortstack{$\delta(b)=1$\\$\delta(d)=1$} & \shortstack{$e$} & - & -\\
\hline
\serialidweb{3.1.3} & Re$_{4}$Se$_{8}$ & 2 & 1.1495 & 1.1495 & C2DB & \ctwodbpath{4ReSe2-1} & OAI & \shortstack{$\delta(a)=1$\\$\delta(c)=1$} & \shortstack{$e$} & \codpath{1539953} & -\\
\hline
\serialidweb{3.1.4} & Re$_{4}$Se$_{8}$ & 2 & 1.1402 & 1.1409 & MC2D & \mctwodpath{mc2d-1579} & OAI & \shortstack{$\delta(a)=1$\\$\delta(c)=1$} & \shortstack{$e$} & - & -\\
\hline
\serialidweb{3.1.5} & Pd$_{2}$S$_{4}$ & 17 & 1.1243 & 1.6124 & C2DB & \ctwodbpath{2PdS2-1} & OAI & \shortstack{$\delta(a)=1$\\$\delta(b)=4$} & \shortstack{$b$\\$c$} & \codpath{2310589} & -\\
\hline
\serialidweb{3.1.6} & Pd$_{2}$Se$_{4}$ & 17 & 1.3036 & 1.4342 & C2DB & \ctwodbpath{2PdSe2-1} & OAI & \shortstack{$\delta(a)=1$\\$\delta(b)=4$} & \shortstack{$b$\\$c$} & \codpath{4310736} & -\\
\hline
\serialidweb{3.1.7} & P$_{4}$ & 42 & 0.9115 & 0.9117 & C2DB & \ctwodbpath{4P-1} & OAI & \shortstack{$\delta(a)=-1$\\$\eta(c,d,e)=1$\\$\zeta(c,d,e)=2$} & \shortstack{$e$} & \codpath{9012486} & 36434\\
\hline
\serialidweb{3.1.8} & P$_{4}$ & 42 & 0.8989 & 0.8992 & MC2D & \mctwodpath{mc2d-220} & OAI & \shortstack{$\delta(a)=-1$\\$\eta(c,d,e)=1$\\$\zeta(c,d,e)=2$} & \shortstack{$e$} & - & -\\
\hline

\hline

\hline\hline
\end{longtable}
\clearpage

\subsection{Computationally exfoliated (MC2D)}
\addcontentsline{stoc}{subsection}{\protect\numberline{\thesubsection}Computationally exfoliated (MC2D)}

\begin{longtable}{c|c|c|c|c|c|c|c|c|c|c|c}
\caption{\label{oai:exfolab}Computationally exfoliated (MC2D) topological insulators}\hypertarget{oai:exfo}\\
\centering
2D-TQC & Formula & LG & Gap & \shortstack{Direct\\ Gap} & Source & Source ID & \shortstack{Topology\\ (SOC)} & RSI & Occ. WP & \shortstack{Parent\\ COD} & \shortstack{Parent\\ ICSD} \\
\endhead
\hline

\serialidweb{3.2.20} & C$_{2}$Ag$_{2}$O$_{4}$ & 2 & 2.3209 & 2.6052 & MC2D & \mctwodpath{mc2d-2678} & OAI & \shortstack{$\delta(b)=1$} & \shortstack{$e$}  & - & -\\
\hline
\serialidweb{3.2.37} & C$_{6}$H$_{12}$Ag$_{2}$N$_{4}$S$_{2}$ & 2 & 2.9953 & 3.3048 & MC2D & \mctwodpath{mc2d-2304} & OAI & \shortstack{$\delta(d)=1$} & \shortstack{$e$}  & - & -\\
\hline
\serialidweb{3.2.36} & C$_{6}$Ag$_{2}$N$_{8}$O$_{10}$ & 2 & 2.0783 & 2.1130 & MC2D & \mctwodpath{mc2d-2094} & OAI & \shortstack{$\delta(b)=1$} & \shortstack{$e$}  & - & -\\
\hline
\serialidweb{3.2.42} & C$_{8}$Ag$_{2}$Br$_{4}$O$_{4}$ & 2 & 2.5679 & 2.8579 & MC2D & \mctwodpath{mc2d-2149} & OAI & \shortstack{$\delta(a)=1$} & \shortstack{$e$}  & - & -\\
\hline
\serialidweb{3.2.1} & Ag$_{2}$Bi$_{2}$P$_{4}$S$_{12}$ & 2 & 1.1285 & 1.1285 & MC2D & \mctwodpath{mc2d-958} & OAI & \shortstack{$\delta(b)=1$\\$\delta(c)=1$} & \shortstack{$e$}  & - & -\\
\hline
\serialidweb{3.2.15} & C$_{14}$H$_{8}$Ag$_{2}$N$_{8}$ & 2 & 1.7248 & 1.7502 & MC2D & \mctwodpath{mc2d-310} & OAI & \shortstack{$\delta(a)=1$} & \shortstack{$e$}  & - & -\\
\hline
\serialidweb{3.2.18} & C$_{16}$H$_{12}$Ag$_{2}$N$_{8}$ & 2 & 1.8653 & 1.8816 & MC2D & \mctwodpath{mc2d-1192} & OAI & \shortstack{$\delta(d)=1$} & \shortstack{$e$}  & - & -\\
\hline
\serialidweb{3.2.24} & C$_{2}$H$_{8}$Ag$_{2}$N$_{8}$ & 2 & 2.8837 & 3.1343 & MC2D & \mctwodpath{mc2d-1090} & OAI & \shortstack{$\delta(b)=1$} & \shortstack{$e$}  & - & -\\
\hline

\hline

\hline\hline
\end{longtable}
\clearpage

\subsection{Stable (C2DB)}
\addcontentsline{stoc}{subsection}{\protect\numberline{\thesubsection}Stable (C2DB)}

\begin{longtable}{c|c|c|c|c|c|c|c|c|c|c|c}
\caption{\label{oai:stablelab}Stable (C2DB) obstructed atomic insulators}\hypertarget{oai:stable}\\
\centering
2D-TQC & Formula & LG & Gap & \shortstack{Direct\\ Gap} & Source & Source ID & \shortstack{Topology\\ (SOC)} & RSI & Occ. WP & \shortstack{Parent\\ COD} & \shortstack{Parent\\ ICSD} \\
\endhead
\hline

\serialidweb{3.3.12} & C$_{6}$Ag$_{2}$Cl$_{2}$F$_{8}$N$_{2}$ & 2 & 1.5745 & 1.6688 & C2DB & \ctwodbpath{2AgClNC3F4-1} & OAI & \shortstack{$\delta(c)=1$} & \shortstack{$e$} & - & -\\
\hline
\serialidweb{3.3.16} & C$_{6}$H$_{8}$Ag$_{2}$Cl$_{2}$N$_{2}$ & 2 & 1.6727 & 1.8673 & C2DB & \ctwodbpath{2AgClNC3H4-1} & OAI & \shortstack{$\delta(c)=1$} & \shortstack{$e$} & - & -\\
\hline
\serialidweb{3.3.1} & As$_{2}$S$_{6}$Si$_{2}$ & 2 & 2.1757 & 2.1843 & C2DB & \ctwodbpath{2AsSiS3-1} & OAI & \shortstack{$\delta(d)=1$} & \shortstack{$e$} & - & -\\
\hline
\serialidweb{3.3.2} & As$_{2}$Se$_{6}$Si$_{2}$ & 2 & 1.7759 & 1.8493 & C2DB & \ctwodbpath{2AsSiSe3-1} & OAI & \shortstack{$\delta(d)=1$} & \shortstack{$e$} & - & -\\
\hline
\serialidweb{3.3.11} & C$_{4}$H$_{4}$Br$_{2}$Cu$_{2}$N$_{2}$ & 2 & 0.6873 & 0.9613 & C2DB & \ctwodbpath{2BrCuNC2H2-1} & OAI & \shortstack{$\delta(b)=1$} & \shortstack{$e$} & \codpath{7052608} & -\\
\hline
\serialidweb{3.3.13} & C$_{6}$Br$_{2}$Cu$_{2}$F$_{8}$N$_{2}$ & 2 & 0.3352 & 0.3423 & C2DB & \ctwodbpath{2BrCuNC3F4-1} & OAI & \shortstack{$\delta(c)=1$} & \shortstack{$e$} & - & -\\
\hline
\serialidweb{3.3.5} & Br$_{2}$Nb$_{4}$S$_{12}$ & 2 & 0.3740 & 0.4044 & C2DB & \ctwodbpath{2BrNb2S6-1} & OAI & \shortstack{$\delta(a)=1$\\$\delta(b)=1$\\$\delta(d)=1$} & \shortstack{$e$} & - & -\\
\hline
\serialidweb{3.3.6} & Br$_{2}$Nb$_{4}$Se$_{12}$ & 2 & 0.6120 & 0.6482 & C2DB & \ctwodbpath{2BrNb2Se6-1} & OAI & \shortstack{$\delta(a)=1$\\$\delta(b)=1$\\$\delta(d)=1$} & \shortstack{$e$} & - & -\\
\hline

\hline

\hline\hline
\end{longtable}
\clearpage

\subsection{Not-stable (C2DB)}
\addcontentsline{stoc}{subsection}{\protect\numberline{\thesubsection}Not-stable (C2DB)}

\begin{longtable}{c|c|c|c|c|c|c|c|c|c|c|c}
\caption{\label{oai:unstablelab}Not-stable (C2DB) obstructed atomic insulators}\hypertarget{oai:unstable}\\
\centering
2D-TQC & Formula & LG & Gap & \shortstack{Direct\\ Gap} & Source & Source ID & \shortstack{Topology\\ (SOC)} & RSI & Occ. WP & \shortstack{Parent\\ COD} & \shortstack{Parent\\ ICSD} \\
\endhead
\hline

\serialidweb{3.4.1} & CdS$_{2}$ & 2 & 1.3877 & 1.3877 & C2DB & \ctwodbpath{1CdS2-1} & OAI & \shortstack{$\delta(c)=1$\\$\delta(d)=5$} & \shortstack{$d$\\$e$}  & - & -\\
\hline
\serialidweb{3.4.2} & F$_{2}$Os$_{2}$ & 10 & 0.00 & 0.0865 & C2DB & \ctwodbpath{2FOs-2} & OAI & \shortstack{$\eta(a)=1$} & \shortstack{$b$}  & - & -\\
\hline
\serialidweb{3.4.3} & GaLiS$_{4}$ & 14 & 0.6825 & 0.8498 & C2DB & \ctwodbpath{1GaLiS4-1} & OAI & \shortstack{$\delta(a)=-5$\\$\delta(b)=-1$\\$\delta(c)=-1$\\$\delta(d)=-1$} & \shortstack{$a$\\$d$\\$g$\\$h$}  & - & -\\
\hline
\serialidweb{3.4.4} & H$_{4}$AlIn & 14 & 0.7369 & 1.0837 & C2DB & \ctwodbpath{2AlInH4-1} & OAI & \shortstack{$\delta(a)=-5$\\$\delta(b)=-1$} & \shortstack{$a$\\$c$\\$g$\\$h$}  & - & -\\
\hline
\serialidweb{3.4.6} & Fe$_{2}$S$_{4}$ & 17 & 0.00 & 0.1668 & C2DB & \ctwodbpath{2FeS2-2} & OAI & \shortstack{$\delta(a)=1$\\$\delta(b)=3$} & \shortstack{$b$\\$c$}  & - & -\\
\hline
\serialidweb{3.4.5} & Cl$_{4}$Cu$_{4}$F$_{4}$S$_{4}$ & 17 & 1.0350 & 1.0900 & C2DB & \ctwodbpath{4ClCuFS-1} & OAI & \shortstack{$\delta(b)=-1$} & \shortstack{$c$}  & - & -\\
\hline
\serialidweb{3.4.7} & H$_{4}$Ag$_{4}$Cl$_{4}$O$_{4}$ & 17 & 0.3071 & 0.3071 & C2DB & \ctwodbpath{4AgClHO-1} & OAI & \shortstack{$\delta(b)=-1$} & \shortstack{$c$}  & - & -\\
\hline
\serialidweb{3.4.18} & O$_{2}$Sb$_{2}$Te & 18 & 0.00 & 0.1773 & C2DB & \ctwodbpath{1TeO2Sb2-1} & OAI & \shortstack{$\delta(a)=-1$\\$\delta(b)=1$\\$\eta(d,e)=1$} & \shortstack{$b$\\$e$}  & - & -\\
\hline

\hline

\hline\hline
\end{longtable}

\clearpage

\section{Table of orbital-selected obstructed atomic insulators}\label{sec:ooai}
\addcontentsline{stoc}{section}{\protect\numberline{\thesection}Table of orbital-selected obstructed atomic insulators}

\subsection{Summary of results}
\addcontentsline{stoc}{subsection}{\protect\numberline{\thesubsection}Summary of results}

\begin{table}[htbp]
\centering
\begin{tabular}{c|c}
\hline\hline
Type of orbital-selected obstructed atomic insulator & Number of materials\\
\hline

\hyperlink{ooai:experi}{Experimental}& 0\\
\hline
\hyperlink{ooai:exfo}{Computationally exfoliated (MC2D)}& 6\\
\hline
\hyperlink{ooai:stable}{Stable (C2DB)}& 22\\
\hline
\hyperlink{ooai:unstable}{Not-stable (C2DB)}& 8\\
\hline

\hline\hline
\end{tabular}
\caption{\label{ooai:summa}Summary of orbital-selected obstructed atomic insulators in each type.}
\end{table}

\subsection{Computationally exfoliated (MC2D)}
\addcontentsline{stoc}{subsection}{\protect\numberline{\thesubsection}Computationally exfoliated (MC2D)}

\begin{longtable}{c|c|c|c|c|c|c|c|c|c|c|c}
\caption{\label{ooai:exfolab}Computationally exfoliated (MC2D) orbital-selected obstructed atomic insulators}\hypertarget{ooai:exfo}\\
\centering
2D-TQC & Formula & LG & Gap & \shortstack{Direct\\ Gap} & Source & Source ID & \shortstack{Topology\\ (SOC)} & RSI & Occ. WP & \shortstack{Parent\\ COD} & \shortstack{Parent\\ ICSD} \\
\endhead
\hline

\serialidweb{4.2.2} & FeSe$_{2}$ & 61 & 0.00 & 0.0076 & MC2D & \mctwodpath{mc2d-2649} & OOAI & \shortstack{$\delta(b)=-1$\\$\delta(b)=-3$\\$\delta(d)=2$\\$\delta(e)=-1$\\$\eta(d,e,f,g,h,i)=1$\\$\zeta(d,e,f,g,h,i,j,k,l,m)=3$} & \shortstack{$b(d)$\\$b(s)$\\$d(p)$\\$d(s)$}  & - & -\\
\hline
\serialidweb{4.2.1} & Bi$_{2}$Pd & 61 & 0.00 & 0.5591 & MC2D & \mctwodpath{mc2d-857} & OOAI & \shortstack{$\delta(a)=-1$\\$\delta(a)=-3$\\$\delta(d)=-1$\\$\delta(e)=1$\\$\zeta(d,e,f,g,h,i,j,k,l,m)=2$} & \shortstack{$a(d)$\\$a(s)$\\$e(d)$\\$e(p)$\\$e(s)$}  & - & -\\
\hline
\serialidweb{4.2.3} & Cl$_{2}$NiO$_{8}$ & 66 & 0.00 & 0.0048 & MC2D & \mctwodpath{mc2d-942} & OOAI & \shortstack{$\delta(a)=-2$\\$\delta(a)=-2$\\$\delta(b)=-1$} & \shortstack{$a(d)$\\$a(s)$\\$c(p)$\\$c(s)$\\$e(p)$\\$e(s)$}  & - & -\\
\hline
\serialidweb{4.2.4} & Ba$_{2}$Pt & 72 & 0.00 & 0.1663 & MC2D & \mctwodpath{mc2d-2558} & OOAI & \shortstack{$\delta(a)=-3$\\$\delta(a)=-2$\\$\delta(c)=1$\\$\eta(e,f)=1$} & \shortstack{$a(d)$\\$a(s)$\\$c(d)$\\$c(p)$\\$c(s)$}  & - & -\\
\hline
\serialidweb{4.2.5} & Nb$_{2}$NiS$_{4}$ & 72 & 0.00 & 0.0282 & MC2D & \mctwodpath{mc2d-1934} & OOAI & \shortstack{$\delta(a)=-2$\\$\delta(a)=-2$\\$\delta(b)=-1$\\$\delta(c)=3$\\$\eta(e,f)=1$} & \shortstack{$a(d)$\\$a(s)$\\$c(d)$\\$c(p)$\\$c(s)$}  & - & -\\
\hline
\serialidweb{4.2.6} & Te$_{2}$Zn & 72 & 0.00 & 0.0488 & MC2D & \mctwodpath{mc2d-1903} & OOAI & \shortstack{$\delta(a)=-3$\\$\delta(a)=-1$\\$\delta(c)=2$} & \shortstack{$a(d)$\\$a(s)$\\$c(p)$\\$c(s)$}  & - & -\\
\hline

\hline

\hline\hline
\end{longtable}
\clearpage

\subsection{Stable (C2DB)}
\addcontentsline{stoc}{subsection}{\protect\numberline{\thesubsection}Stable (C2DB)}

\begin{longtable}{c|c|c|c|c|c|c|c|c|c|c|c}
\caption{\label{ooai:stablelab}Stable (C2DB) orbital-selected obstructed atomic insulators}\hypertarget{ooai:stable}\\
\centering
2D-TQC & Formula & LG & Gap & \shortstack{Direct\\ Gap} & Source & Source ID & \shortstack{Topology\\ (SOC)} & RSI & Occ. WP & \shortstack{Parent\\ COD} & \shortstack{Parent\\ ICSD} \\
\endhead
\hline

\serialidweb{4.3.1} & Bi$_{2}$Ni & 61 & 0.00 & 0.4370 & C2DB & \ctwodbpath{1NiBi2-1} & OOAI & \shortstack{$\delta(b)=-1$\\$\delta(b)=-3$\\$\delta(d)=1$\\$\delta(e)=-1$\\$\zeta(d,e,f,g,h,i,j,k,l,m)=2$} & \shortstack{$b(d)$\\$b(s)$\\$d(d)$\\$d(p)$\\$d(s)$} & - & -\\
\hline
\serialidweb{4.3.2} & Bi$_{2}$Pd & 61 & 0.00 & 0.5685 & C2DB & \ctwodbpath{1PdBi2-1} & OOAI & \shortstack{$\delta(b)=-1$\\$\delta(b)=-3$\\$\delta(d)=1$\\$\delta(e)=-1$\\$\zeta(d,e,f,g,h,i,j,k,l,m)=2$} & \shortstack{$b(d)$\\$b(s)$\\$d(d)$\\$d(p)$\\$d(s)$} & \codpath{9012857} & -\\
\hline
\serialidweb{4.3.4} & PdSb$_{2}$ & 61 & 0.00 & 0.1769 & C2DB & \ctwodbpath{1PdSb2-1} & OOAI & \shortstack{$\delta(b)=-1$\\$\delta(b)=-3$\\$\delta(d)=2$\\$\delta(e)=-1$\\$\eta(d,e,f,g,h,i)=1$\\$\zeta(d,e,f,g,h,i,j,k,l,m)=3$} & \shortstack{$b(d)$\\$b(s)$\\$d(p)$\\$d(s)$} & - & -\\
\hline
\serialidweb{4.3.3} & Bi$_{2}$Pt & 61 & 0.00 & 0.1121 & C2DB & \ctwodbpath{1PtBi2-1} & OOAI & \shortstack{$\delta(b)=-1$\\$\delta(b)=-3$\\$\delta(d)=1$\\$\delta(e)=-1$\\$\zeta(d,e,f,g,h,i,j,k,l,m)=2$} & \shortstack{$b(d)$\\$b(s)$\\$d(d)$\\$d(p)$\\$d(s)$} & - & -\\
\hline
\serialidweb{4.3.10} & CdGa$_{2}$Se$_{4}$ & 72 & 0.0184 & 0.0199 & C2DB & \ctwodbpath{1CdGa2Se4-1} & OOAI & \shortstack{$\delta(a)=-4$\\$\delta(a)=-1$\\$\delta(b)=1$\\$\delta(c)=2$\\$\eta(e,f)=1$} & \shortstack{$a(d)$\\$a(s)$\\$c(d)$\\$c(p)$\\$c(s)$} & - & -\\
\hline
\serialidweb{4.3.11} & CdGa$_{2}$Te$_{4}$ & 72 & 0.00 & 0.0209 & C2DB & \ctwodbpath{1CdGa2Te4-1} & OOAI & \shortstack{$\delta(a)=-4$\\$\delta(a)=-1$\\$\delta(b)=1$\\$\delta(c)=2$\\$\eta(e,f)=1$} & \shortstack{$a(d)$\\$a(s)$\\$c(d)$\\$c(p)$\\$c(s)$} & - & -\\
\hline
\serialidweb{4.3.17} & HgIn$_{2}$Se$_{4}$ & 72 & 0.0152 & 0.0266 & C2DB & \ctwodbpath{1HgIn2Se4-1} & OOAI & \shortstack{$\delta(a)=-4$\\$\delta(a)=-1$\\$\delta(b)=1$\\$\delta(c)=2$\\$\eta(e,f)=1$} & \shortstack{$a(d)$\\$a(s)$\\$c(d)$\\$c(p)$\\$c(s)$} & - & -\\
\hline

\hline

\hline\hline
\end{longtable}
\clearpage

\subsection{Not-stable (C2DB)}
\addcontentsline{stoc}{subsection}{\protect\numberline{\thesubsection}Not-stable (C2DB)}

\begin{longtable}{c|c|c|c|c|c|c|c|c|c|c|c}
\caption{\label{ooai:unstablelab}Not-stable (C2DB) orbital-selected obstructed atomic insulators}\hypertarget{ooai:unstable}\\
\centering
2D-TQC & Formula & LG & Gap & \shortstack{Direct\\ Gap} & Source & Source ID & \shortstack{Topology\\ (SOC)} & RSI & Occ. WP & \shortstack{Parent\\ COD} &  \shortstack{Parent\\ ICSD} \\
\endhead
\hline

\serialidweb{4.4.1} & NiSb$_{2}$ & 61 & 0.00 & 0.1322 & C2DB & \ctwodbpath{1NiSb2-1} & OOAI & \shortstack{$\delta(b)=-1$\\$\delta(b)=-3$\\$\delta(d)=2$\\$\delta(e)=-1$\\$\eta(d,e,f,g,h,i)=1$\\$\zeta(d,e,f,g,h,i,j,k,l,m)=3$} & \shortstack{$b(d)$\\$b(s)$\\$d(p)$\\$d(s)$}  & - & -\\
\hline
\serialidweb{4.4.3} & CdS$_{2}$ & 72 & 0.00 & 0.0123 & C2DB & \ctwodbpath{1CdS2-3} & OOAI & \shortstack{$\delta(a)=-3$\\$\delta(a)=-1$\\$\delta(c)=2$} & \shortstack{$a(d)$\\$a(s)$\\$c(p)$\\$c(s)$}  & - & -\\
\hline
\serialidweb{4.4.4} & F$_{2}$NiO$_{2}$ & 72 & 0.00 & 0.0584 & C2DB & \ctwodbpath{1NiF2O2-1} & OOAI & \shortstack{$\delta(a)=-2$\\$\delta(a)=-2$\\$\delta(b)=-1$} & \shortstack{$a(d)$\\$a(s)$\\$c(p)$\\$c(s)$}  & - & -\\
\hline
\serialidweb{4.4.2} & Al$_{2}$O$_{4}$Pd & 72 & 0.00 & 0.0195 & C2DB & \ctwodbpath{1PdAl2O4-1} & OOAI & \shortstack{$\delta(a)=-2$\\$\delta(a)=-2$\\$\delta(b)=-1$\\$\delta(c)=1$} & \shortstack{$a(d)$\\$a(s)$\\$c(p)$\\$c(s)$}  & - & -\\
\hline
\serialidweb{4.4.5} & F$_{2}$O$_{2}$Pd & 72 & 0.00 & 0.0601 & C2DB & \ctwodbpath{1PdF2O2-1} & OOAI & \shortstack{$\delta(a)=-2$\\$\delta(a)=-2$\\$\delta(b)=-1$} & \shortstack{$a(d)$\\$a(s)$\\$c(p)$\\$c(s)$}  & - & -\\
\hline
\serialidweb{4.4.7} & H$_{2}$O$_{2}$Pd & 72 & 0.00 & 0.0339 & C2DB & \ctwodbpath{1PdH2O2-1} & OOAI & \shortstack{$\delta(a)=-2$\\$\delta(a)=-2$\\$\delta(b)=-1$\\$\eta(e,f)=1$} & \shortstack{$a(d)$\\$a(s)$\\$c(p)$\\$c(s)$}  & - & -\\
\hline
\serialidweb{4.4.6} & F$_{2}$Pt & 72 & 0.00 & 0.0321 & C2DB & \ctwodbpath{1PtF2-1} & OOAI & \shortstack{$\delta(a)=-2$\\$\delta(a)=-2$\\$\delta(b)=-1$\\$\eta(e,f)=1$} & \shortstack{$a(d)$\\$a(s)$\\$c(p)$\\$c(s)$}  & - & -\\
\hline
\serialidweb{4.4.8} & I$_{2}$Pt & 72 & 0.00 & 0.0496 & C2DB & \ctwodbpath{1PtI2-2} & OOAI & \shortstack{$\delta(a)=-3$\\$\delta(a)=-1$\\$\delta(c)=2$} & \shortstack{$a(d)$\\$a(s)$\\$c(p)$\\$c(s)$}  & - & -\\
\hline

\hline

\hline\hline

\end{longtable}



\clearpage

\clearpage

\section{Band alignment of twistable materials}\label{app:sec:table_band_alignment}
\addcontentsline{stoc}{section}{\protect\numberline{\thesection}Band alignment of twistable materials}

In this section, we analyze the band alignment of theoretically twistable 2D materials~\cite{jiang2024twist}. For each material, we identify and tabulate other materials with closely matched band-edge energies under the following conditions:
\begin{enumerate}
\item Valence band maxima (VBM) aligned with VBM.
\item Conduction band minima (CBM) aligned with CBM.
\item VBM aligned with CBM.
\item CBM aligned with VBM.
\end{enumerate}

We apply the following constraints to ensure the table is compact:
\begin{itemize}
\item Energy matching criterion: Band edges must align within 150 meV.
\item Momentum matching: The k-points of the aligned bands edge must coincide within a tolerance of 0.05 (for direct coordinates). 
\item Material stability: thermodynamically unstable materials are excluded.
\item Compositional simplicity: Materials with more than 4 atoms or more than 3 distinct elements are excluded.
\item Lattice compatibility: Only materials sharing the same Bravais lattice are aligned.
\item Band edge simplicity: We only consider band edges that are theoretically twistable~\cite{jiang2024twist}. 
\item For polar materials, we only consider the vacuum-aligned energies of the top surface for simplicity. 
\end{itemize}

For clarity in the following tables, we adopt the notation BE$_1$-BE$_2$ to indicate band edge alignment between materials. Here, BE$_1$ represents the band edge energy of the reference material (row entry), while BE$_2$ denotes the band edge energy of the comparison material (column entry).

\subsection{Rectangular lattice}
\addcontentsline{stoc}{subsection}{\protect\numberline{\thesubsection}Rectangular lattice}

\begin{longtable}{c|c|c|c|c|c|c}
    \caption{\label{app:table_band_align_rectangular} Band alignment of materials in rectangular lattice.}\\
    \centering
    Formula & LG & Mat. type & VBM-VBM & CBM-CBM & VBM-CBM & CBM-VBM\\
    \endhead
    \hline
    \ch{Ge2SSe}(\twodtqclink{6.3.478}) & 11 & Stable & \makecell{\ch{SnS}(\twodtqclink{6.1.2})\\ \ch{GeSe}(\twodtqclink{6.1.1})\\ \ch{As}(\twodtqclink{3.3.152})} & \makecell{\ch{SnSe}(\twodtqclink{6.1.4})\\ \ch{Sn2SSe}(\twodtqclink{6.3.508})\\ \ch{GeS}(\twodtqclink{6.2.721})\\ \ch{GeS}(\twodtqclink{6.3.1288})} & \makecell{\ch{AgTe}(\twodtqclink{6.3.575})\\ \ch{AuSe}(\twodtqclink{6.3.580})} & / \\\hline
\ch{SnSe}(\twodtqclink{6.1.4}) & 32 & \href{https://www.nature.com/articles/s41598-023-46092-1}{Exp.M.Exfo} & \ch{Sn2SSe}(\twodtqclink{6.3.508}) & \makecell{\ch{Ge2SSe}(\twodtqclink{6.3.478})\\ \ch{Sn2SSe}(\twodtqclink{6.3.508})\\ \ch{GeS}(\twodtqclink{6.2.721})\\ \ch{GeS}(\twodtqclink{6.3.1288})} & \ch{AuTe}(\twodtqclink{6.3.581}) & / \\\hline
\ch{SSi}(\twodtqclink{6.3.1327}) & 32 & Stable & \ch{AuCuBr2}(\twodtqclink{6.3.937}) & / & / & / \\\hline
\ch{AgGaI2}(\twodtqclink{6.3.425}) & 11 & Stable & / & / & \ch{HgF2S}(\twodtqclink{6.3.548}) & / \\\hline
\ch{SnS}(\twodtqclink{6.1.2}) & 32 & \href{https://www.nature.com/articles/s41598-023-46092-1}{Exp.M.Exfo} & \makecell{\ch{Ge2SSe}(\twodtqclink{6.3.478})\\ \ch{GeSe}(\twodtqclink{6.1.1})\\ \ch{As}(\twodtqclink{3.3.152})} & \ch{GeS}(\twodtqclink{6.2.721}) & \makecell{\ch{AgTe}(\twodtqclink{6.3.575})\\ \ch{AuSe}(\twodtqclink{6.3.580})} & / \\\hline
\ch{GeSiTe2}(\twodtqclink{6.3.483}) & 11 & Stable & / & / & \ch{As}(\twodtqclink{3.3.152}) & / \\\hline
\ch{GaSe}(\twodtqclink{3.3.120}) & 18 & Stable & \ch{InSe}(\twodtqclink{3.3.130}) & \makecell{\ch{InS}(\twodtqclink{3.3.129})\\ \ch{GaS}(\twodtqclink{3.3.119})\\ \ch{InTe}(\twodtqclink{3.3.131})\\ \ch{InSe}(\twodtqclink{3.3.130})} & \ch{AgS}(\twodtqclink{6.3.573}) & / \\\hline
\ch{InS}(\twodtqclink{3.3.129}) & 18 & Stable & \ch{GaS}(\twodtqclink{3.3.119}) & \makecell{\ch{GaSe}(\twodtqclink{3.3.120})\\ \ch{GaS}(\twodtqclink{3.3.119})\\ \ch{InTe}(\twodtqclink{3.3.131})\\ \ch{InSe}(\twodtqclink{3.3.130})} & / & / \\
\hline

\hline
\end{longtable}

\subsection{Square lattice}
\addcontentsline{stoc}{subsection}{\protect\numberline{\thesubsection}Square lattice}

    \begin{longtable}{c|c|c|c|c|c|c}
    \caption{\label{app:table_band_align_square} Band alignment of materials in square lattice.}\\
    \centering
    Formula & LG & Mat. type & VBM-VBM & CBM-CBM & VBM-CBM & CBM-VBM\\
    \endhead
    \hline
    \ch{AuI}(\twodtqclink{6.3.1728}) & 64 & Stable & \makecell{\ch{ZnTe}(\twodtqclink{6.3.1814})\\ \ch{TiTe2}(\twodtqclink{6.3.1693})} & \makecell{\ch{CdBr2}(\twodtqclink{6.3.1630})\\ \ch{CuCl}(\twodtqclink{6.3.1751})} & \ch{HgBr2}(\twodtqclink{6.3.1632}) & / \\\hline
\ch{AgBr}(\twodtqclink{6.3.1721}) & 64 & Stable & / & \makecell{\ch{ZnTe}(\twodtqclink{6.3.1814})\\ \ch{HgI2}(\twodtqclink{6.3.1669})} & / & \makecell{\ch{CuI}(\twodtqclink{6.3.1761})\\ \ch{PbO}(\twodtqclink{6.3.1804})} \\\hline
\ch{PbTe}(\twodtqclink{6.3.1809}) & 64 & Stable & \makecell{\ch{PbS}(\twodtqclink{6.3.1807})\\ \ch{PbSe}(\twodtqclink{6.3.1808})} & / & / & / \\\hline
\ch{PbS}(\twodtqclink{6.3.1807}) & 64 & Stable & \makecell{\ch{PbTe}(\twodtqclink{6.3.1809})\\ \ch{PbSe}(\twodtqclink{6.3.1808})} & / & / & / \\\hline
\ch{CdBr2}(\twodtqclink{6.3.1630}) & 59 & Stable & / & \makecell{\ch{AuI}(\twodtqclink{6.3.1728})\\ \ch{CuCl}(\twodtqclink{6.3.1751})} & / & / \\\hline
\ch{PbSe}(\twodtqclink{6.3.1808}) & 64 & Stable & \makecell{\ch{PbTe}(\twodtqclink{6.3.1809})\\ \ch{PbS}(\twodtqclink{6.3.1807})} & / & / & / \\\hline
\ch{SiTe2}(\twodtqclink{6.3.1691}) & 59 & Stable & / & \ch{S2Si}(\twodtqclink{6.3.1684}) & \ch{SnTe2}(\twodtqclink{6.3.1692}) & / \\\hline
\ch{Br2Ge}(\twodtqclink{6.3.1631}) & 59 & Stable & / & \ch{PbBr2}(\twodtqclink{6.3.1634}) & / & / \\
\hline

\hline
\end{longtable}

\subsection{Hexagonal lattice}
\addcontentsline{stoc}{subsection}{\protect\numberline{\thesubsection}Hexagonal lattice}

    \begin{longtable}{c|c|c|c|c|c|c}
    \caption{\label{app:table_band_align_hexagonal} Band alignment of materials in hexagonal lattice.}\\
    \centering
    Formula & LG & Mat. type & VBM-VBM & CBM-CBM & VBM-CBM & CBM-VBM\\
    \endhead
    \hline
    \ch{PbTe}(\twodtqclink{6.3.2560}) & 72 & Stable & \makecell{\ch{PbSe}(\twodtqclink{6.3.2559})\\ \ch{TiClI}(\twodtqclink{3.3.274})\\ \ch{WTe2}(\twodtqclink{3.1.47})\\ \ch{TiCl2}(\twodtqclink{3.3.494})\\ \ch{Ge}(\twodtqclink{1.1.15})} & \makecell{\ch{WSSe}(\twodtqclink{3.3.292})\\ \ch{WSe2}(\twodtqclink{3.1.46})\\ \ch{WTe2}(\twodtqclink{3.1.47})\\ \ch{GeO}(\twodtqclink{6.3.2066})\\ \ch{SnF2}(\twodtqclink{6.3.2432})\\ \ch{AlO}(\twodtqclink{3.3.475})\\ \ch{MoTe2}(\twodtqclink{3.1.43})} & \makecell{\ch{SnBr2}(\twodtqclink{6.3.2331})\\ \ch{CrTe2}(\twodtqclink{3.3.499})\\ \ch{CrSe2}(\twodtqclink{3.3.498})\\ \ch{SnCl2}(\twodtqclink{6.3.2415})\\ \ch{GaSe}(\twodtqclink{6.2.1106})\\ \ch{GaP}(\twodtqclink{6.3.2065})\\ \ch{SnCl2}(\twodtqclink{6.3.2635})\\ \ch{MoSeTe}(\twodtqclink{3.3.287})\\ \ch{GaAs}(\twodtqclink{6.3.1940})\\ \ch{CdBr}(\twodtqclink{3.3.380})\\ \ch{Cl2Ge}(\twodtqclink{6.3.2402})\\ \ch{InN}(\twodtqclink{6.3.2654})\\ \ch{BrSb}(\twodtqclink{1.3.212})} & \makecell{\ch{ZrBrCl}(\twodtqclink{3.3.267})\\ \ch{ZrBr2}(\twodtqclink{3.3.483})\\ \ch{HfCl2}(\twodtqclink{3.3.493})\\ \ch{TiI2}(\twodtqclink{3.3.506})} \\
\hline

\hline
\end{longtable}

\clearpage
\section{Materials with multiple entries in the tables (duplicates)}\label{sec:dupli}
\addcontentsline{stoc}{section}{\protect\numberline{\thesection}Materials with multiple entries in the tables (duplicates)}

\begin{longtable}{c|c|c|c|c}
\caption{Pairs of entries in the tables above that correspond to a single material}\hypertarget{duplitable}\\
\centering
LG & Formula & 2D-TQC1 & 2D-TQC2 & Topology \\
\endhead
\hline

15 & Ir$_{2}$Nb$_{2}$Te$_{8}$ & \serialidweb{1.3.12} & \serialidweb{1.2.13} & NLC \\
\hline
15 & Ir$_{2}$Ta$_{2}$Te$_{8}$ & \serialidweb{1.3.13} & \serialidweb{1.2.14} & NLC \\
\hline
15 & Mo$_{2}$Te$_{4}$ & \serialidweb{1.1.3} & \serialidweb{1.1.2} & NLC \\
\hline
15 & Rh$_{2}$Ta$_{2}$Te$_{8}$ & \serialidweb{1.3.23} & \serialidweb{1.2.15} & NLC \\
\hline
15 & Se$_{12}$Ta$_{4}$ & \serialidweb{1.3.25} & \serialidweb{1.2.16} & NLC \\
\hline
15 & Te$_{4}$W$_{2}$ & \serialidweb{1.1.4} & \serialidweb{1.1.6} & NLC \\
\hline
17 & Cd$_{4}$I$_{4}$O$_{4}$ & \serialidweb{1.3.29} & \serialidweb{1.2.19} & NLC \\
\hline
41 & Bi$_{4}$Cu$_{4}$ & \serialidweb{1.3.54} & \serialidweb{1.2.49} & NLC \\
\hline

\hline

\hline\hline
\end{longtable}


\clearpage 

\end{document}